%% file: ANA-HIGG-2021-12-PAPER.tex
\documentclass[cernpreprint, texlive=2016, UKenglish, texmf]{atlasdoc}
 
\usepackage[subcaption]{atlaspackage}
\usepackage{atlasbiblatex}
 
\usepackage{atlasphysics}
\usepackage{atlasprocess}
 
\usepackage{bm}
\usepackage{todonotes}
\usepackage{enumitem}
\usepackage{multirow}
 
\graphicspath{{logos/}{figures/}}

\AtlasTitle{Direct constraint on the Higgs--charm coupling from a search for Higgs boson
decays into charm quarks with the ATLAS detector}
 
\AtlasAbstract{
A search for the Higgs boson decaying into a pair of charm quarks
is presented. The analysis uses proton--proton collisions to target the production of a Higgs
boson in association with a leptonically decaying $W$ or $Z$ boson.
The dataset delivered by the LHC at a
centre-of-mass energy of $\sqrt{s}=13~\TeV$ and recorded by the ATLAS
detector corresponds to an integrated luminosity of 139~\ifb .
Flavour-tagging algorithms are used to identify jets originating from the
hadronisation of charm quarks.
 
The analysis method is validated with the simultaneous measurement
of $WW,WZ$ and $ZZ$ production, with observed (expected) significances of
$2.6$ ($2.2$) standard deviations above the background-only prediction for the $(W/Z)Z(\to c\bar{c})$ process and
$3.8$ ($4.6$) standard deviations for the $(W/Z)W(\to cq)$ process. The
$(W/Z)H(\to c \bar{c})$ search yields an observed (expected) upper limit of $26$
($31$) times the predicted Standard Model cross-section times
branching fraction for a Higgs
boson with a mass of $125~\GeV$, corresponding to an observed (expected) constraint
on the charm Yukawa coupling modifier $|\kappa_c| < 8.5~(12.4)$, at
the 95\% confidence level.
 
A combination with the ATLAS $(W/Z)H, H\to b\bar{b}$ analysis is
performed, allowing the ratio $\kappa_c / \kappa_b$ to be
constrained to less than 4.5 at the 95\% confidence level, smaller
than the ratio of the $b$- and $c$-quark masses, and therefore
determines the Higgs--charm coupling to be weaker than the
Higgs--bottom coupling at the 95\% confidence level.  }
 
\author{The ATLAS Collaboration}
 
\AtlasRefCode{HIGG-2021-12}
 
\PreprintIdNumber{CERN-EP-2021-251}
 
\AtlasJournalRef{Eur. Phys. J. C 82 (2022) 717}
\AtlasDOI{10.1140/epjc/s10052-022-10588-3}

\hypersetup{pdftitle={ATLAS document},pdfauthor={The ATLAS Collaboration}}

\begin{document}
 
\maketitle
 
\clearpage
\section{Introduction}
 
Since the discovery of a new particle, $H$, with a mass of approximately 125~\GeV\ by
the ATLAS~\cite{HIGG-2012-27} and CMS~\cite{CMS-HIG-12-028} collaborations at the
LHC~\cite{Evans:2008zzb}, studies of its properties have indicated that it is
consistent with the Standard Model (SM) Higgs boson~\cite{PhysRevLett.13.321,HIGGS1964132,PhysRevLett.13.508,PhysRevLett.13.585}.
The interactions between the Higgs boson and the charged fermions of the third
generation have been observed by both the
ATLAS~\cite{HIGG-2017-07,HIGG-2018-04,HIGG-2018-13} and
CMS~\cite{CMS-HIG-16-043,CMS-HIG-17-035,CMS-HIG-18-016} collaborations, and CMS has
reported evidence for the decay of the Higgs boson into a pair of
muons~\cite{CMS-HIG-19-006}, while ATLAS reported a $2\sigma$ excess over the
background-only prediction~\cite{HIGG-2019-14}. Direct searches by the ATLAS and
CMS collaborations for Higgs boson decays into a charm quark--antiquark
pair, $H\to c\bar{c}$~\cite{HIGG-2017-01,CMS-HIG-18-031}, decays into an
electron--positron pair~\cite{HIGG-2018-58,CMS-HIG-13-007}, exclusive decays
into mesons~\cite{HIGG-2016-13,HIGG-2016-23,CMS-HIG-19-012,CMS-SMP-17-012,PhysRevD.88.053003,PhysRevLett.114.101802},
and reinterpretations of the Higgs \pt spectrum~\cite{Soreq:2016rae,HIGG-2019-13}, have
not yet found experimental evidence for Higgs boson couplings to the first-generation fermions
or second-generation quarks.
Taken together, the results of these measurements and searches
are consistent with the prediction that the coupling strength of the
Higgs boson to each fermion scales proportionally to the fermion's mass.
 
In the SM the branching fraction of $H\to c\bar{c}$ is
2.89\%~\cite{Djouadi:1997yw}, approximately 20 times smaller than the branching
fraction of the Higgs boson to a bottom quark--antiquark pair, $H \to b\bar{b}$.
Physics beyond the Standard Model can significantly enhance or reduce the
coupling of the Higgs boson to the charm quark, and in turn the $H \to c\bar{c}$
branching
fraction~\cite{PhysRevD.89.033014,PhysRevD.92.033016,Botella:2016krk,PhysRevD.98.055001,GHOSH2016504,Egana-Ugrinovic:2018znw,Egana-Ugrinovic:2019dqu}.
Direct searches for $H \to c\bar{c}$ in proton--proton ($pp$) collisions have set upper limits
on the cross-section times branching fraction for the production
of a $W$ or $Z$ boson in association with
a Higgs boson decaying into $c\bar{c}$. The ATLAS Collaboration has performed a
search in the $Z(\to \ell\ell)H (\to c\bar{c})$ channel, where $\ell=e,\mu$,
using $36.1~\ifb$ of
$pp$ collision data recorded at $\sqrt{s}=13~\TeV$~\cite{HIGG-2017-01}, setting
an observed (expected) upper limit at 110 (150) times the SM prediction, at 95\%
confidence level (CL). The CMS Collaboration has also performed a search using
$35.9~\ifb$ of $pp$ collision data recorded at $13~\TeV$~\cite{CMS-HIG-18-031};
the search was conducted in three channels based on the number of charged leptons,
namely the 0-, 1- and
2-lepton channels, targeting the $ZH\to\nu\nu c\bar{c}$,
$WH\to\ell\nu c\bar{c}$ and $ZH \to\ell\ell c\bar{c}$ signatures, respectively. These
were combined to set an observed (expected) upper limit of 70 (37) times the SM
prediction, at 95\% CL.
 
This paper presents a new search for $VH (\to c\bar{c})$, where
$V=W$ or $Z$, using
$139~\ifb$ of $pp$ collision data collected at a centre-of-mass energy of
13~\TeV\ by the ATLAS detector from 2015 to 2018. Events are selected in
the 0-, 1- and 2-lepton channels and are
categorised according to the transverse momentum, \pt, of the vector boson and the number of
jets.
 
Higgs boson candidates are constructed from the two jets with the highest \pt.
One of the main challenges in searching for $H \to c\bar{c}$ is to recognise jets
originating from the hadronisation of charm quarks. To
identify these jets, a multivariate charm-jet tagging algorithm is used. Additionally, a
bottom-jet identification algorithm is used to veto bottom jets, ensuring this analysis is
orthogonal to the recent ATLAS $VH(\to b\bar{b})$
measurement~\cite{HIGG-2018-51}.
Events are selected if one or both of the two highest-\pt jets are
$c$-tagged.
 
In order to search for the $H \to c\bar{c}$ signal the distributions of the
dijet invariant mass, $m_{cc}$, in all event categories are used simultaneously in a
binned maximum-likelihood fit, which allows the signal yield and the main
background normalisations to be extracted. The analysis strategy is validated by
the simultaneous measurement of the diboson processes in which one of the bosons
decays to at least one charm quark, $VW(\to cq)$ and $VZ(\to c\bar{c})$, where
$q$ is a down-type quark. The result is interpreted in the kappa
framework~\cite{Heinemeyer:2013tqa,deFlorian:2016spz} in terms of $\kappa_c$,
the modifier of the coupling between the Higgs boson and the charm quark.
 
The analysis is combined with the ATLAS $VH, H \to b\bar{b}$
measurement~\cite{HIGG-2018-51} and the results are interpreted in the kappa
framework in terms of both $\kappa_b$ and $\kappa_c$, and in terms of the ratio
$\kappa_c / \kappa_b$.
\newcommand{\AtlasCoordFootnote}{
ATLAS uses a right-handed coordinate system with its origin at the nominal
interaction point (IP) in the centre of the detector and the \(z\)-axis along
the beam pipe. The \(x\)-axis points from the IP to the centre of the LHC ring,
and the \(y\)-axis points upwards. Cylindrical coordinates \((r,\phi)\) are used
in the transverse plane, \(\phi\) being the azimuthal angle around the
\(z\)-axis. The pseudorapidity is defined in terms of the polar angle \(\theta\)
as \(\eta = -\ln \tan(\theta/2)\). Angular distance is measured in units of
\(\Delta R \equiv \sqrt{(\Delta\eta)^{2} + (\Delta\phi)^{2}}\).}

\section{ATLAS detector}
 
The ATLAS experiment~\cite{PERF-2007-01} at the LHC is a multipurpose particle
detector with a forward--backward symmetric cylindrical geometry and a near
\(4\pi\) coverage in solid angle.\footnote{\AtlasCoordFootnote} It consists of
an inner tracking detector (ID) surrounded by a thin superconducting solenoid
providing a \SI{2}{\tesla} axial magnetic field, electromagnetic and hadron
calorimeters, and a muon spectrometer. The inner tracking detector covers the
pseudorapidity range \(|\eta| < 2.5\). It consists of silicon pixel, silicon
microstrip, and transition radiation tracking detectors. Lead/liquid-argon (LAr)
sampling calorimeters provide electromagnetic (EM) energy measurements with high
granularity. A steel/scintillator-tile hadron calorimeter covers the central
pseudorapidity range (\(|\eta| < 1.7\)). The endcap and forward regions are
instrumented with LAr calorimeters for both the EM and hadronic energy
measurements up to \(|\eta| = 4.9\). The muon spectrometer surrounds the
calorimeters and is based on three large superconducting air-core toroidal
magnets with eight coils each. The field integral of the toroids ranges between
\num{2.0} and \SI{6.0}{\tesla\metre} across most of the detector. The muon
spectrometer includes a system of precision chambers for tracking and fast detectors
for triggering. A two-level trigger system is used to select events. The
first-level trigger is implemented in hardware and uses a subset of the detector
information to accept events at a rate below \SI{100}{\kHz}~\cite{TRIG-2019-04}.
This is followed by a software-based trigger that reduces the accepted event
rate to \SI{1}{\kHz} on average depending on the data-taking conditions.
An extensive software suite~\cite{ATL-SOFT-PUB-2021-001} is used in the
reconstruction and analysis of real and simulated data, in detector operations,
and in the trigger and data acquisition systems of the experiment.
\section{Dataset and simulated event samples}
 
This analysis uses data recorded by the ATLAS detector during Run 2 of
the LHC, which took place from 2015 to 2018 at a centre-of-mass energy
of 13 TeV. Data were collected using a combination of missing
transverse momentum triggers~\cite{TRIG-2019-01},
in the 0- and 1-lepton channels, and
single-lepton triggers~\cite{TRIG-2018-05,TRIG-2018-01},
in the 1- and 2-lepton channels. Events are
required to be of good quality and recorded while all
relevant detector components were in operation~\cite{DAPR-2018-01}.
The dataset corresponds to an integrated luminosity of
$139.0\pm2.4~\ifb$~\cite{ATLAS-CONF-2019-021}.
 
The Monte Carlo (MC) simulation samples used in this analysis are largely the
same as those used in the ATLAS $VH(\to b\bar{b})$ analysis~\cite{HIGG-2018-51}, and are
summarised in Table~\ref{tab:mc_samples}. Samples of simulated events were
generated for $VH$ production with a Higgs boson mass, $m_H$, of $125~\GeV$,
for both $H \to c\bar{c}$ and $H \to b\bar{b}$
decays, with branching fractions of 2.89\% and 58.2\%, respectively, and for the
main background processes ($\ttbar$, single-top, $V+\,$jets and diboson). The samples are
used to optimise the analysis and perform the statistical analysis of the data.
 
The background from multi-jet events is negligible in the 0- and 2-lepton
channels after applying the selection criteria detailed in
Section~\ref{sec:selection}. In the 1-lepton channel, it is estimated using a
data-driven method. All samples of simulated events are initially normalised to the most
accurate theoretical cross-section predictions currently
available.
Samples produced using alternative event generators are used to assess
systematic uncertainties in the modelling of the signal and background
processes, and are discussed in Section~\ref{sec:systematics}.
 
All samples of MC events were passed through the ATLAS detector response
simulation~\cite{SOFT-2010-01} based on
\GEANT~\cite{Agostinelli:2002hh} and were
reconstructed with the same algorithms as used for data.
The effect of multiple interactions in the same
and neighbouring bunch crossings (pile-up) was modelled by overlaying the simulated hard-scattering event
with inelastic $pp$ events generated with
\PYTHIA[8.186]~\cite{Sjostrand:2007gs} using the \NNPDF[2.3lo] set of parton
distribution functions (PDF)~\cite{Ball:2012cx} and the A3
set of tuned parameters (A3 tune)~\cite{ATL-PHYS-PUB-2016-017}.
The simulated events were weighted to
reproduce the distribution of the average number of interactions per bunch crossing seen
in data. The \EVTGEN[1.6.0] program~\cite{Lange:2001uf} was used to
describe the decays of bottom and charm hadrons in all samples, except those
generated with
\SHERPA~\cite{Gleisberg:2008ta,Hoeche:2009rj,Cascioli:2011va}.
 
\begin{table*}
\scriptsize
\caption{Signal and background processes and their corresponding MC
generators used in this analysis. The acronyms ME and PS stand for matrix
element and parton shower, respectively. The cross-section order refers to
the order of the cross-section calculation used for process normalisation in
QCD, unless otherwise stated, with ((N)N)LO and ((N)N)LL standing for
((next-to-)next-to-)leading order and ((next-to-)next-to-)leading log,
respectively.}
\centering
\begin{tabular}{llllll}
\toprule
\multirow{2}{*}{Process} & \multirow{2}{*}{ME generator} & \multirow{2}{*}{ME PDF} & PS and & \multirow{2}{*}{Tune} & Cross-section \\
&  &  & hadronisation & & order \\
\midrule
\multirow{3}{*}{\shortstack[l]{$qq \to VH$ \\ \phantom{qq}$(H \to c\bar{c}/b\bar{b})$}}          & \POWHEGBOX[v2]~\cite{Frixione:2007vw,Alioli:2010xd}
& \multirow{3}{*}{\NNPDF[3.0nlo]~\cite{Ball:2014uwa}}
& \multirow{3}{*}{\PYTHIA[8.212]~\cite{Sjostrand:2014zea}}
& \multirow{3}{*}{\AZNLO~\cite{STDM-2012-23} }
& \multirow{3}{*}{\shortstack[l]{NNLO(QCD)\\+NLO(EW)~\cite{Ciccolini:2003jy,Brein:2003wg,Brein:2011vx,Denner:2014cla,Brein:2012ne}}} \\
& + \GOSAM~\cite{Cullen:2011ac}  & & & & \\
& + \MINLO~\cite{Hamilton:2012np,Luisoni:2013cuh} \\
\addlinespace
$gg \to ZH$                           & \multirow{2}{*}{\POWHEGBOX[v2]}
& \multirow{2}{*}{\NNPDF[3.0nlo]}
& \multirow{2}{*}{\PYTHIA[8.212]}
& \multirow{2}{*}{\AZNLO}
& \multirow{2}{*}{NLO+NLL~\cite{Altenkamp:2012sx,Harlander:2014wda}} \\
\phantom{gg}$(H \to c\bar{c}/b\bar{b})$ \\
\addlinespace
\multirow{2}{*}{\ttbar}               & \multirow{2}{*}{\POWHEGBOX[v2]~\cite{Frixione:2007nw}}
& \multirow{2}{*}{\NNPDF[3.0nlo]}
& \multirow{2}{*}{\PYTHIA[8.230]}
& \multirow{2}{*}{A14~\cite{ATL-PHYS-PUB-2014-021}}
& NNLO \\
& & & & &                              +NNLL~\cite{Beneke:2011mq,Cacciari:2011hy,Baernreuther:2012ws,Czakon:2012zr,Czakon:2012pz,Czakon:2013goa,Czakon:2011xx} \\
\addlinespace
$t/s$-channel                         & \multirow{2}{*}{\POWHEGBOX[v2]~\cite{Alioli:2009je}}
& \multirow{2}{*}{\NNPDF[3.0nlo]}
& \multirow{2}{*}{\PYTHIA[8.230]}
& \multirow{2}{*}{A14}
& \multirow{2}{*}{NLO~\cite{Aliev:2010zk,Kant:2014oha}} \\
single top \\
\addlinespace
$Wt$-channel                          & \multirow{2}{*}{\POWHEGBOX[v2]~\cite{Re:2010bp}}
& \multirow{2}{*}{\NNPDF[3.0nlo]}
& \multirow{2}{*}{\PYTHIA[8.230]}
& \multirow{2}{*}{A14}
& Approx.\\
single top & & & & &  NNLO~\cite{Kidonakis:2010ux,Kidonakis:2013zqa} \\
\addlinespace
$V$+\,jets                              & \SHERPA[2.2.1]~\cite{Gleisberg:2008ta,Hoeche:2009rj,Cascioli:2011va}
& \NNPDF[3.0nnlo]~\cite{Ball:2014uwa}
& \SHERPA[2.2.1]
& Default
& NNLO~\cite{Anastasiou:2003ds} \\
\addlinespace
$qq \to VV$                           & \SHERPA[2.2.1]
& \NNPDF[3.0nnlo]
& \SHERPA[2.2.1]
& Default
& NLO \\
\addlinespace
$gg \to VV$
& \SHERPA[2.2.2]
& \NNPDF[3.0nnlo]
& \SHERPA[2.2.2]
& Default
& NLO \\
\bottomrule
\end{tabular}
\label{tab:mc_samples}
\end{table*}
\section{Object and event selection}
\label{sec:selection}
 
\subsection{Object selection}
 
Interaction vertices are reconstructed from tracks in the ID. The vertex with
the highest sum of squared transverse momenta of associated tracks is used as
the primary vertex~\cite{ATL-PHYS-PUB-2015-026}.
 
Electrons are reconstructed by matching ID tracks with energy clusters in the
EM calorimeter~\cite{EGAM-2018-01}. Electrons are required to have $\pt > 7~\GeV$
and $|\eta| < 2.47$. They must satisfy the \textit{loose} identification criterion,
based on a likelihood discriminant combining observables related to the shower
shape in the calorimeter and to the track matching the energy cluster, and
are required to be isolated in both the ID and calorimeter using \pt-dependent
criteria.
In the 1-lepton channel, more stringent requirements
are placed on the identification and isolation of electrons. These electrons,
called \textit{tight} electrons, are required to satisfy the \textit{tight}
likelihood criterion and a stricter calorimeter-based isolation.
 
Muons are reconstructed within the acceptance of the muon
spectrometer, $|\eta| < 2.7$~\cite{MUON-2018-03}. They
are required to have $\pt > 7~\GeV$,
to satisfy the \textit{loose} identification criteria and to be isolated in the ID
using \pt-dependent criteria. As with electrons, in
the 1-lepton channel more stringent requirements are placed on the identification
and isolation of muons. These muons, called \textit{tight} muons, must satisfy
the \textit{medium} identification criteria and a stricter track-based isolation,
and have $|\eta| < 2.5$.
 
Hadronically decaying $\tau$-leptons~\cite{PERF-2014-06,ATLAS-CONF-2017-029},
identified with a \textit{medium} quality
criterion~\cite{ATLAS-CONF-2017-029}, are required to have $\pt > 20~\GeV$ and
$|\eta| < 2.5$, excluding the transition region of $1.37 < |\eta| < 1.52$ between the barrel
and endcap sections of the electromagnetic calorimeter. Reconstructed
$\tau$-leptons are not directly used in the event selection, but are
used in the calculation of missing transverse momentum and to avoid
double-counting reconstructed $\tau$-leptons as other objects.

Jets are reconstructed from topological clusters of energy deposits in
the calorimeter~\cite{PERF-2016-04,PERF-2014-07,ATL-PHYS-PUB-2015-036}
by using the anti-\kt algorithm~\cite{Cacciari:2008gp,Fastjet} with a radius parameter of $R=0.4$.
The jets are classified as \textit{central} or
\textit{forward} jets depending on their pseudorapidity. Jets are classified as
forward jets if they have $2.5 < |\eta| < 4.5$ and $\pt > 30~\GeV$. Jets
are classified as central jets if they have $\pt>20~\GeV$ and $|\eta| < 2.5$.
Additionally, central jets with $\pt <
120~\GeV$ are required to be identified as originating from
the primary vertex using a jet-vertex tagging algorithm~\cite{PERF-2014-03}.
To improve the measurement of each
jet's energy and direction, and consequently the measurement of $m_{cc}$,
if any muons are found within a cone of jet-\pt-dependent size around the jet axis, the
four-momentum of the muon closest to the jet is added to the jet four-momentum,
following the procedure described in Ref.~\cite{HIGG-2018-51}.
An overlap removal procedure is applied to avoid double-counting between
electrons, muons, hadronically decaying
$\tau$-leptons and jets.
 
Central jets are tagged as containing either $b$- or $c$-hadrons using two
discriminants resulting from multivariate tagging algorithms, MV2 and
DL1~\cite{FTAG-2018-01}. Jets are $b$-tagged using the MV2 discriminant,
configured to select $b$-jets with 70\% efficiency in simulated \ttbar events.
A $c$-tagging configuration of the DL1 discriminant, $\mathrm{DL1_c}$, was
optimised for this analysis, and includes a veto on jets $b$-tagged by the MV2
algorithm.
This configuration gives an average
efficiency of 27\% to tag $c$-jets in simulated \ttbar events, and $b$- and
light-jet misidentification rates of 8\% and 1.6\%, respectively.
The efficiencies in simulation are calibrated to match those in data using
control samples of $\ttbar$ and \Zjets events with a precision of 5\%--10\%, using methods identical
to those applied to $b$-tagging
algorithms~\cite{FTAG-2018-01,ATLAS-CONF-2018-001,ATLAS-CONF-2018-006}.
Jets in simulated events are labelled using
information from the MC generator's event `truth' record,
exclusively as $b$-, $c$-, or $\tau$-jets, in this order,
according to whether they contain a $b$-hadron, $c$-hadron, or $\tau$-lepton
with $\pt > 5~\GeV$ within a cone of size $\Delta R = 0.3$ around their
axis. Jets not labelled as $b$-, $c$- or $\tau$-jets are labelled as light jets.
Diboson, $V+$\,jets and top-quark backgrounds are classified according to the
flavour labels of the jets that form the Higgs boson candidate in those
selected events.
 
To maximise the statistical power of the available MC samples, the $c$-tagging
requirement is not applied to the diboson, $V+$\,jets or top-quark samples.
Instead, events are weighted by the probabilities for each jet to be $c$-tagged,
based on its flavour label and as a function of the jet \pt and $|\eta|$,
to obtain predictions for events with either one or two
$c$-tagged jets. This is referred to as \textit{truth-flavour tagging} since it uses
information from the MC generator's event `truth' record.
In the $V+$\,jets samples, differences
of up to 20\% are observed between the two methods and are corrected for
by weights assigned to each jet, dependent on the $\Delta R$ to the closest other jet and
on the flavour labels of the jet and the closest other jet.
Finally, to correct for any residual non-closure in the truth-flavour tagging procedure,
small normalisation corrections are applied to the diboson, $V+$\,jets and top-quark
predictions such that the number of events for each process in
each analysis region (defined in Section~\ref{sec:event_selection}) matches that
obtained when directly applying $c$-tagging.
These normalisation corrections vary between 0.9 and 1.05.
 
The missing transverse momentum, $\bm{E_{\mathrm{T}}^{\mathrm{miss}}}$ is
reconstructed as the negative of the vector sum of the transverse momenta of electrons,
muons, hadronically decaying $\tau$-leptons, jets, and a `soft term' which is
constructed from tracks associated with the primary vertex but not with any
reconstructed object~\cite{PERF-2016-07}. The magnitude of the
$\bm{E_{\mathrm{T}}^{\mathrm{miss}}}$ is referred to as
$E_{\mathrm{T}}^{\mathrm{miss}}$.
The track-based missing transverse momentum,
$\bm{p^{\mathrm{miss}}_{\mathrm{T}}}$, is constructed using all ID tracks
associated with the primary vertex and satisfying the quality criteria detailed in
Ref.~\cite{PERF-2015-08}, with its magnitude denoted by
$p^{\mathrm{miss}}_{\mathrm{T}}$.
 
\subsection{Event selection and categorisation}
\label{sec:event_selection}
Events are categorised into 0-, 1- and 2-lepton
channels based on the number of \textit{loose} electrons and muons they contain.
Events with at least two central jets are selected, and they are further categorised as 2- or
3-jet events according to the total number of jets.
Events with more than three jets are rejected in the 0- and
1-lepton channels to reduce the \ttbar background. In the 2-lepton
channel, events with more than three jets are included in the 3-jet category.
 
Since the signal-to-background ratio increases for large transverse
momentum of the vector boson, $\pt^V$, events with
reconstructed $\pt^V > 75~\GeV$ are selected~\cite{Butterworth:2008iy}. Two
$\pt^V$ regions are used: $75~\GeV < \pt^V < 150~\GeV$
(only in the 2-lepton channel) and $\pt^V > 150~\GeV$. The
$\pt^V$ corresponds to \met in the 0-lepton channel, the magnitude of the vector sum
of the $\bm{E^{\mathrm{miss}}_{\mathrm{T}}}$ and the lepton
$\bm{p_{\mathrm{T}}}$ in the 1-lepton channel, and the magnitude of the vector sum of the two
lepton transverse momenta in the 2-lepton channel.
 
The main discriminating variable in this analysis is the invariant mass, $m_{cc}$, of
the two central jets with the highest \pt, hereafter referred to as \textit{signal}
jets. At least one signal jet must have $\pt > 45~\GeV$.
\textit{Signal regions} are composed of events in which one or both of
these jets are $c$-tagged, with the two cases defining separate categories,
referred to as 1-$c$-tag and 2-$c$-tag, respectively.
Furthermore, any additional non-signal jet must \textit{not} be $b$-tagged.
This requirement means that events in the signal regions can contain at most one $b$-tagged jet.
Combined with an identical jet selection, this
ensures that selected events are orthogonal to those selected in the ATLAS $VH(\to b\bar{b})$
analysis~\cite{HIGG-2018-51}.
Events selected in the control regions are not completely orthogonal with those selected in the $VH(\to b\bar{b})$ analysis,
and the impact of this is discussed in Section~\ref{sec:combination}.
In total, 16 signal regions are defined, arising from the combination of three
lepton channels, two $\pt^V$ categories (in the 2-lepton channel), two
number-of-jets categories and two number-of-$c$-tagged-jets categories.
 
To reduce the background contamination in all channels, the $\Delta R$ between
the two signal jets is required to be $\Delta R < 2.3$ in events with $75 <
\pt^V < 150~\GeV$, $\Delta R < 1.6$ in events with $150 < \pt^V < 250~\GeV$ and
$\Delta R < 1.2$ in events with $\pt^V> 250~\GeV$. The $\Delta R$ selection
criteria, optimised for each $\pt^V$ range, are approximately $80\%$ efficient for signal events.
For each signal region, a corresponding control region is defined as containing
events failing the $\Delta R$ selection, up to a maximum $\Delta R$ of 2.5. These control
regions, hereafter referred to as the high-$\Delta R$ control regions, are
designed to constrain the $V+$\,jets background normalisations and shapes.
 
In addition to the high-$\Delta R$ control regions, further control regions are
defined to constrain the modelling of the most important other background processes.
Top control regions, enriched in \ttbar and single-top events, are defined in
all lepton channels. In the 0- and 1-lepton channels, events
with three jets are selected, and in these events exactly one of the signal jets is $c$-tagged and the non-signal jet is
$b$-tagged, resulting in one control region for each of these lepton channels. In the 2-lepton
channel a pure sample of \ttbar events is selected by requiring the two
leptons to have different flavours ($e\mu$) and opposite electric charges.
One control region is
defined for each number-of-jets category and $\pt^V$ region combination, resulting in a total of four 2-lepton top
control regions, each containing exactly one $c$-tagged jet.
 
Finally, in the 1- and 2-lepton channels, events in which neither of the two
signal jets are $c$-tagged and no non-signal jets are $b$-tagged
are used to constrain the normalisation of the $V+$\,light-jets backgrounds.
One control region is defined for each number-of-jets category and
$\pt^V$ region combination, for a total of six 0-$c$-tag control regions.
 
A total of 44 analysis
regions are defined: 16 signal regions, 16 high-$\Delta R$ control regions, 6
top control regions and 6 0-$c$-tag control regions.
The signal-region selection criteria
specific to each channel are described below, and summarised in
Table~\ref{tab:evselTable}.

\begin{description}[leftmargin=0cm]
\item[0-lepton channel]
Data were collected using \met triggers with thresholds ranging from
70~\GeV\ in 2015 to 110~\GeV\ in 2018~\cite{TRIG-2019-01}. Events must not contain any
\textit{loose} electrons or muons, and are required to have $\met >
150~\GeV$. At 150~\GeV\ the \met triggers are approximately 75\%--90\% efficient,
depending on the year, reaching a full efficiency plateau at about 200~\GeV. A
requirement on the scalar sum of jet transverse momenta,
$H_{\mathrm{T}}$, of 120 (150)~\GeV\ in 2-jet (3-jet) events is imposed to
remove a small region of phase space where the trigger efficiency
depends on the number of jets. To remove non-collision backgrounds,
$p^{\mathrm{miss}}_{\mathrm{T}}$ is required
to exceed $30~\GeV$. Background multi-jet events with high \met typically
arise from mismeasured jet energies in the calorimeter and can be
rejected using angular separation requirements (detailed in
Table~\ref{tab:evselTable}) between the jets,
$\bm{E^{\mathrm{miss}}_{\mathrm{T}}}$ and
$\bm{p^{\mathrm{miss}}_{\mathrm{T}}}$.
\item[1-lepton channel]
Events must contain exactly one \textit{loose} lepton, that is then
required to also be \textit{tight}. If the lepton is an electron (muon),
it must have $\pt > 27~(25)~\GeV$ and $|\eta| < 2.47~(2.5)$. In the muon
sub-channel, data were collected with the same \met triggers as in the
0-lepton channel. The online \met calculation does not include muons,
so these triggers effectively select on $\pt^V$ and perform
more efficiently than single-muon triggers in the analysis phase space.
In the electron sub-channel, single-electron triggers were used to
collect data with thresholds starting at 24~\GeV\ for data collected in
2015 and 26~\GeV\ for data collected between 2016 and 2018~\cite{TRIG-2018-05}.
In the
electron sub-channel, there is a significant background from events with
jets misidentified as electrons at low \met. These events are rejected
by requiring $\met > 30~\GeV$. The transverse mass of the reconstructed
$W$ boson,\footnote{$m_{\mathrm{T}}^W = \sqrt{2 \pt^{\ell} \pt^{\nu} (1 -
\cos(\phi^{\ell} - \phi^{\nu}))}$, where
$\bm{E^{\mathrm{miss}}_{\mathrm{T}}}$ is used as an approximation for
$\bm{p^{\nu}_{\mathrm{T}}}$.} $m_{\mathrm{T}}^W$, is required to be
less than 120~\GeV.
The number of multi-jet background events that survive the 1-lepton
channel event selection is estimated in each analysis region by
performing a template fit to the $m_{\mathrm{T}}^W$ distribution, which
offers good discrimination between multi-jet and simulated backgrounds.
The multi-jet $m_{\mathrm{T}}^W$ templates are extracted from control
regions defined by inverting the tight isolation requirements on the
leptons, after subtraction of the simulated backgrounds. The shapes of
the multi-jet $m_{cc}$ distributions are obtained using the same
procedure. More details of the template-fit method can be found
in Ref.~\cite{HIGG-2018-04}.
\item[2-lepton channel]
The 2-lepton channel must contain exactly two same-flavour \textit{loose} leptons. At
least one of the leptons must have $\pt > 27~\GeV$ to be consistent with
the online single-lepton trigger selection. In the dimuon case, they must have
opposite electric charges. This requirement is not imposed in the
dielectron sub-channel due to a higher probability of charge misidentification.
The invariant mass of the two leptons, $m_{\ell\ell}$, is required to be
consistent with the mass of the $Z$~boson, $81 < m_{\ell\ell} <  101~\GeV$.
\end{description}

Following the event selection, the \Zjets, \Wjets and \ttbar processes
constitute the main backgrounds in the 0-lepton channel. In the 1-lepton
channel, the main backgrounds arise from the \Wjets and \ttbar processes. For
both the 0- and 1-lepton channels, the relative background composition depends
substantially on the analysis region. In the 2-lepton channel the main
background is \Zjets in all regions. The efficiency to select the $VH(\to
c\bar{c})$ signal, in which the $V$ decays to leptons, is $\approx1\mbox{--}2$\%, and the
expected signal-to-background ratio in the mass range $100~\GeV < m_{cc} <
150~\GeV$ is $(1\mbox{--}7) \times 10^{-4}$ in the 1-$c$-tag signal regions and
$(0.6\mbox{--}8) \times 10^{-3}$ in the 2-$c$-tag signal regions.
 
\begin{table}[htb]
\centering
\small
\caption{Summary of the signal region event selection in the 0-, 1- and 2-lepton channels.
Jet1 and jet2 refer to the two signal jets and $H$ refers to the jet1--jet2 system.}
\begin{tabular}{ll}
\toprule
\multicolumn{2}{c}{Common selections}\\
\midrule
Central jets & $\geq$  2  \\
Signal jet \pt & $\geq 1$ signal jet with $\pt > 45~\GeV$ \\
$c$-jets &  One or two $c$-tagged signal jets \\
$b$-jets & No $b$-tagged non-signal jets \\
Jets & $2, 3$ (0- and 1-lepton); $2,~\geq 3$ (2-lepton) \\
\addlinespace
\multirow{2}{*}{$\pt^V$ regions}           & 75--150~\GeV~(2-lepton) \\
& $> 150~\GeV$  \\
\addlinespace
\multirow{3}{*}{$\Delta R(\textbf{jet1},~\textbf{jet2})$} & $75 < \pt^V < 150~\GeV$: $\Delta R \leq 2.3$ \\
& $150 < \pt^V < 250~\GeV$: $\Delta R \leq 1.6$   \\
& $\pt^V > 250~\GeV$: $\Delta R \leq 1.2$   \\
\midrule
\multicolumn{2}{c}{0-lepton channel} \\
\midrule
Trigger             & $\met$ \\
Leptons             & No \textit{loose} leptons \\
$\met$              & $> 150~\GeV$  \\
$p_{\mathrm{T}}^{\mathrm{miss}}$      & $> 30~\GeV$  \\
$H_{\mathrm{T}}$ & $ >120~\GeV$ (2 jets)$,~>150~\GeV$ (3 jets)  \\
$\min|\Delta\phi(\bm{E^{\textbf{miss}}_{\mathrm{T}}}, \textbf{jet})|$   & $> 20\ensuremath{^\circ}$ (2 jets)$,~> 30\ensuremath{^\circ}$ (3 jets) \\
$|\Delta\phi(\bm{E^{\textbf{miss}}_{\mathrm{T}}}, \bm{H})|$             & $> 120\ensuremath{^\circ}$ \\
$|\Delta\phi(\textbf{jet1, jet2})|$    & $< 140\ensuremath{^\circ}$ \\
$|\Delta\phi(\bm{E^{\textbf{miss}}_{\mathrm{T}}},\bm{p_{\mathrm{T}}^{\textbf{miss}}})|$   & $< 90\ensuremath{^\circ}$ \\
\midrule
\multicolumn{2}{c}{1-lepton channel} \\
\midrule
\multirow{2}{*}{Trigger} &  $e$ sub-channel: single electron \\
&  $\mu$ sub-channel: \met \\
Leptons & One \textit{tight} lepton and no additional \textit{loose} leptons \\
$\met$   & $> 30~\GeV$ ($e$ sub-channel) \\
$m_\mathrm{T}^W$   & $< 120~\GeV$ \\
\midrule
\multicolumn{2}{c}{2-lepton channel}\\
\midrule
Trigger & Single lepton\\
\multirow{2}{*}{Leptons} & Exactly two \textit{loose} leptons \\
&  Same flavour, opposite charge for $\mu\mu$ \\
$m_{\ell\ell}$       & $81 < m_{\ell\ell} <  101~\GeV$ \\
\bottomrule
\end{tabular}
\label{tab:evselTable}
\end{table}
\section{Systematic uncertainties}
\label{sec:systematics}
 
The sources of systematic uncertainties affecting this analysis can be broadly
divided into two groups: those related to experimental effects and those due to
the theoretical modelling of signal and background processes.
The estimation
of these uncertainties closely follows the procedures outlined in
Ref.~\cite{HIGG-2018-51}.
 
\subsection{Experimental uncertainties}
\label{sec:exp_uncerts}
 
The leading experimental uncertainties in this analysis are due to imperfect calibration of
the $c$-tagging efficiency, jet energy scales and jet energy resolutions.
Correction factors for
$c$- and $b$-tagging are determined from the difference between tagging
efficiencies in data and simulation and are derived separately for $c$-jets, $b$-jets and
light-flavour jets~\cite{FTAG-2018-01,ATLAS-CONF-2018-001,ATLAS-CONF-2018-006}.
The uncertainties in the correction factors originate from
multiple sources and are decomposed into independent components that are
correlated between different analysis regions.
Two additional uncertainties are included in MC samples where truth-flavour
tagging, described in Section~\ref{sec:selection}, is used.
An uncertainty is included in the $V$+\,jets samples, equal to the $\Delta R$ correction
that is applied to improve agreement between the truth-tagged and direct-tagged
simulation samples, and for each MC prediction an uncertainty is
included in the overall normalisation correction between direct and
truth-flavour tagging.
 
The uncertainties in the calibration of jet energy scales and resolutions are
estimated from multiple measurements~\cite{PERF-2016-04}. Uncertainties in the
jet energy scale and resolution are combined into
independent components that are correlated between
different analysis regions. An additional
uncertainty in the calibration of $b$- and $c$-jets is also included.
 
Uncertainties in the reconstruction, identification, isolation and trigger
efficiencies of electrons and muons, and the uncertainties in their energy scale
and resolution, have been measured in data and found to be negligible compared
to other uncertainties~\cite{EGAM-2018-01,MUON-2018-03}.
These uncertainties, along with the jet energy scale and
resolution uncertainties, are propagated to the calculation of \met following the
method described in Ref.~\cite{HIGG-2018-51}.
The \met calculation has additional uncertainties associated with the \pT scale, \pT resolution and
reconstruction efficiency of the tracks used to build the \met soft term,
and with the modelling of the underlying event~\cite{PERF-2016-07}.
An uncertainty in the \met trigger efficiency is also included.

The uncertainty in the combined 2015--2018 integrated luminosity is
1.7\%~\cite{ATLAS-CONF-2019-021}, obtained using the LUCID-2 detector~\cite{LUCID2}
for the primary luminosity measurements. The average number of interactions per
bunch-crossing in simulation is scaled by 1.03 to improve agreement with data,
with an uncertainty corresponding to the full size of the correction ($\pm3\%$).

\subsection{Signal and background modelling uncertainties}
 
Modelling uncertainties are evaluated using samples of simulated events. For
each process, four categories of uncertainty are considered:
cross-section and acceptance uncertainties, which account for the overall normalisation of
backgrounds that are not allowed to float freely in the global likelihood fit;
flavour-fraction uncertainties, which account for uncertainties in the make-up
of subcomponents of each background; relative acceptance uncertainties, which account for the
relative normalisations of backgrounds in cases where the overall normalisation
is considered correlated across two or more analysis regions;
and shape uncertainties, accounting for uncertainties
in the shape of the $m_{cc}$ distribution in each region.
Unless stated otherwise, normalisation and shape effects of each systematic
uncertainty are treated as uncorrelated with one another
to ensure that modelling uncertainties are not artificially constrained.
The decision to correlate or decorrelate normalisation and shape effects is made
for each modelling uncertainty considered and is based on studies
of different correlation models, and in cases where the impact of the choice is
non-negligible, the model giving the most conservative
uncertainty is used.
Background modelling uncertainties considered in this analysis are summarised in
Table~\ref{tab:syst_summary}.
 
\begin{description}[leftmargin=0cm]
\item[$VH$]
Uncertainties in the $VH(\to c\bar{c})$ signal are evaluated following the
recommendations of the LHC Higgs Working
Group~\cite{Dittmaier:2011ti,Dittmaier:2012vm}, and include uncertainties in the
cross-section of $VH$ production and in the $H\to c\bar{c}$ branching fraction.
In addition, acceptance and shape uncertainties are
evaluated by comparing the
nominal $VH(\to c\bar{c})$ samples with alternatives generated using
\POWHER{7}~\cite{Bellm:2015jjp}, and by independently varying the renormalisation
(\muR) and factorisation (\muF) scales by factors of one-half and two
in the nominal generator.
Comparisons are made separately
for the three production processes; $q\bar{q} \to ZH$, $q\bar{q} \to WH$, and
$gg \to ZH$. Uncertainties in the total acceptance are found to be similar between
lepton channels, 4\%--6\% in
the quark-initiated processes and 31\%--35\% for $gg \to ZH$.
Similarly, uncertainties in the ratio of 3-jet to 2-jet events are found to be similar
between channels for the quark-initiated processes, 6\%--12\%,
while uncertainties in the $gg \to ZH$ processes are larger, 19\%--56\%.
In the 2-lepton channel an uncertainty is included to account for differences
in acceptance between the two $\pt^V$ regions, and is 2\% for $q\bar{q} \to ZH$
and 5\% for $gg \to ZH$.
Uncertainties in the shapes of the $m_{cc}$ distributions for each of the three
production processes are evaluated in a similar way, and good agreement is found
between lepton channels, allowing the use of one shape uncertainty per
production process.
 
Uncertainties in the normalisation of the $WH(\to b\bar{b})$ and
$ZH(\to b\bar{b})$ background are
taken from the recent ATLAS measurements~\cite{HIGG-2018-51}. Uncertainties in
the number of jets and $\pt^V$ acceptance ratios are set to be the
same as those derived for the $H\to c\bar{c}$ signal. Shape uncertainties in the
$H \to b\bar{b}$ background are evaluated in an equivalent way to the
$H\to~c\bar{c}$ signal shape uncertainties.
 
\item[Diboson] Uncertainties in the diboson prediction are evaluated by
comparing the nominal diboson MC samples, generated using \SHERPA, with
alternative samples generated using \POWPY{8} and
by independently varying \muR and \muF in the
nominal samples by factors of one-half and two. Inclusive acceptance
uncertainties are assigned to the $WW$, $WZ$ and $ZZ$
processes, and uncertainties in the ratios of 3-jet to 2-jet
events and the ratios of events in different $\pt^V$ regions are also included.
 
Uncertainties in the $m_{cc}$ shape of the diboson signal, $VW(\to cq)$ and
$VZ(\to c\bar{c})$, and background components are evaluated separately by
comparing MC events from the two generator set-ups in each lepton channel. These
comparisons are made inclusively in the number of jets and, in the 2-lepton
channel, split into $\pt^V$ regions. The shape uncertainties are found to be
consistent between channels.
 
\item[$V+$\,jets production]
The largest backgrounds in this analysis originate from $V+$\,jets, with \Wjets
and \Zjets being the largest background in the 1-lepton and 2-lepton
channels, respectively, and a combination of the two making up the majority of
the background in the 0-lepton channel. In all channels, the $V+$\,jets background is
divided into three subsamples based on the flavours of the two signal jets:
$V+\,$\textit{hf}, $V+\,$\textit{mf} and $V+\,$\textit{lf}, where \textit{hf}, \textit{mf}
and \textit{lf} stand
for heavy-flavour, mixed-flavour and light-flavour, respectively. The
$V+\,$\textit{hf} background consists of the $V{+}\,bb$ and $V{+}\,cc$ contributions,
the $V+\,$\textit{mf} background consists of the $V{+}\,bl$, $V{+}\,cl$ and $V{+}\,bc$ contributions,
where $l$ refers to a \textit{light} jet that doesn't contain a heavy-flavour hadron
or $\tau$-lepton,
and the $V+\,$\textit{lf} background consists of the remaining contribution where
neither of the signal jets contain a heavy-flavour hadron. In the 0-lepton channel
a non-negligible component of events contains a $W \to \tau \nu$ decay in which
the $\tau$-lepton decays hadronically and is selected as one of the two signal jets.
These jets are considered as light jets for the purpose of assigning these events to the
$W+\,$\textit{mf} and $W+\,$\textit{lf} background components.
 
Uncertainties in the $V+$\,jets backgrounds due to normalisation, acceptance,
flavour-fractions and shapes are evaluated using alternative Monte Carlo generator
set-ups. These include taking the same \SHERPA set-up used to generate the
nominal $V+$\,jets samples but varying \muR and \muF
by factors of one-half and two independently, and separate samples generated
using \MGNLO~\cite{Alwall:2014hca} at leading order in QCD with up to
four additional partons in the matrix element calculation,
interfaced to \PYTHIA[8].
 
The normalisations of all the \Wjets and  \Zjets components are free to
float in the likelihood fit to data. The $W+\,$\textit{lf} and $Z\,+\,$\textit{lf}
components float independently in the 2-jet and 3-jet signal regions.
The $Z\,+\,$\textit{hf}, $Z\,+\,$\textit{mf} and $Z\,+\,$\textit{lf} components float
independently in the two $\pt^V$ regions.
The normalisations of the $V+\,$\textit{lf} backgrounds are constrained by the
0-$c$-tag control regions, while the $V+\,$\textit{hf} and $V+\,$\textit{mf}
components are constrained by the signal regions and high-$\Delta R$ control
regions. Uncertainties are included to account for acceptance effects between
number-of-jet categories and lepton channels.
Uncertainties in the relative contributions of the components of the
$V+\,$\textit{hf}, $V+\,$\textit{mf} and $V+\,$\textit{lf} backgrounds
are found to be consistent between lepton channels, so one uncertainty
per component is used, taken from the lepton channel which offers the most precise
estimate.
 
Shape uncertainties are derived in each analysis region and channel,
separately for \Wjets and \Zjets, with an
uncertainty being included for each of the three comparisons performed, namely
the comparisons between the nominal \SHERPA sample and the \muR and
\muF variations, and the comparison with \MGNLOPY{8}. The
comparison between \SHERPA and \MGNLOPY{8} is performed in
a two-step process. First, a comparison is made in the high-$\Delta R$ control
region, with differences in both the shape and normalisation between the two
models propagated to the signal regions in a correlated way. Second, the
\SHERPA model is weighted such that the two models agree in the
high-$\Delta R$ control region and any residual difference between the two models in
the signal regions is included as a shape-only uncertainty.
 
\item[Top-quark background]
 
The top-quark background comprises \ttbar and single-top-quark events. In
the 2-lepton channel, the normalisation of this background in each signal region
is determined using the corresponding top control region described in
Section~\ref{sec:event_selection}. Due to the small size of this background, no
shape uncertainties are considered.
 
In the 0- and 1-lepton channels, where this background is larger, \ttbar and
single-top $Wt$-channel events are divided into two components based on the
flavours of the two signal jets: top($b$) events, in which at least one of the signal
jets originates from a $b$-quark; and top(other) events. For the latter component,
the two signal jets are mostly produced in the decay of a $W$ boson and
therefore their invariant mass peaks at the $W$-boson mass,
while for the former component it does not. The
normalisations of these two components are free to float separately in the
global likelihood fit, with information from the top control regions
contributing significantly. The background from $t$- and $s$-channel single-top-quark
production is small and is considered separately, with uncertainties in
the $t$-channel and $s$-channel production cross-sections included.

Acceptance and shape uncertainties are derived separately for each component by
comparing the nominal \POWPY{8} \ttbar and single-top MC
samples with alternative samples, generated using
\POWHER{7} and \MGNLOPY{8}~\cite{Alwall:2014hca}. In addition, the impact of additional
radiation is assessed using \POWPY{8} samples with modified
parameter values. Uncertainties are
included to cover differences in the normalisation between lepton channels,
between number-of-jets categories, and between the signal regions and top control
regions. The dominant single-top contribution comes from $Wt$ production.
The estimated uncertainty in the relative contributions of \ttbar and $Wt$ events
to the total top-quark background is included, as is an uncertainty due to
the interference between \ttbar and $Wt$ events, evaluated using an alternative
$Wt$ MC sample in which the $\ttbar/Wt$ interference is dealt with using the
diagram subtraction scheme instead of the diagram removal scheme used in the
nominal $Wt$ sample~\cite{Frixione:2008yi,White:2009yt}.
 
Shape differences in each of the various MC sample comparisons are
considered as separate shape uncertainties. Shape uncertainties are derived for
each component of the total top-quark background, top($b$) and top(other), separately
for \ttbar and $Wt$ events.

\item[Multi-jet]
Uncertainties in the multi-jet background are evaluated separately in the electron and
muon sub-channels. Normalisation and shape uncertainties are derived by changing the
definition of the multi-jet control region and
by modifying the normalisation of the \ttbar and \Wjets backgrounds in
this control region by up to 25\%. Additionally, the impact of using an alternative
variable instead of $m_{\mathrm{T}}^W$, namely the azimuthal angle between the charged lepton and
$\bm{E_{\mathrm{T}}^{\mathrm{miss}}}$, $\Delta \phi (\ell,
\bm{E_{\mathrm{T}}^{\mathrm{miss}}})$, in the template fit is
considered as an uncertainty.
 
\end{description}
 
\begin{table}[!htb]
\small
\centering
\caption{Summary of the background modelling systematic uncertainties
considered. The values given refer to the size of the uncertainty affecting
the yield of each background. Where the size of an acceptance
systematic uncertainty varies between analysis regions, a range is
displayed. Uncertainties in the shapes of the $m_{cc}$ distributions are not
shown below, but are taken into account for all backgrounds.
CR and SR stand for control region and signal region.}
\begin{tabular}{lc}
\toprule
\textbf{$VH(\to b\bar{b})$} \\
$WH(\to b\bar{b})$ normalisation & 27\% \\
$ZH(\to b\bar{b})$ normalisation & 25\% \\
\midrule
\textbf{Diboson} \\
$WW/ZZ/WZ$ acceptance & 10\%/5\%/12\% \\
$\pt^V$ acceptance & 4\% \\
$N_{\mathrm{jet}}$ acceptance & 7\%--11\% \\
\midrule
\textbf{\Zjets} \\
$Z\,+\,$\textit{hf} normalisation & Floating \\
$Z\,+\,$\textit{mf} normalisation & Floating \\
$Z\,+\,$\textit{lf} normalisation & Floating \\
$Z\,{+}\,bb$ to $Z\,{+}\,cc$ ratio & 20\% \\
$Z\,{+}\,bl$ to $Z\,{+}\,cl$ ratio & 18\%\\
$Z\,{+}\,bc$ to $Z\,{+}\,cl$ ratio & ~~6\%\\
$\pt^V$ acceptance & 1\%--8\% \\
$N_{\mathrm{jet}}$ acceptance & 10\%--37\%\\
High-$\Delta R$ CR to SR & 12\%--37\% \\
0- to 2-lepton ratio & 4\%--5\% \\
\midrule
\textbf{\Wjets} \\
$W+\,$\textit{hf} normalisation & Floating \\
$W+\,$\textit{mf} normalisation & Floating \\
$W+\,$\textit{lf} normalisation & Floating \\
$W{+}\,bb$ to $W{+}\,cc$ ratio & ~~4\%--10\% \\
$W{+}\,bl$ to $W{+}\,cl$ ratio & 31\%--32\% \\
$W{+}\,bc$ to $W{+}\,cl$ ratio & 31\%--33\% \\
$W\to\tau \nu (+ c)$ to $W{+}\,cl$ ratio & 11\% \\
$W\to\tau \nu (+ b)$ to $W{+}\,cl$ ratio & 27\% \\
$W\to\tau \nu (+ l)$ to $W{+}\,l$ ratio & ~~8\% \\
$N_{\mathrm{jet}}$ acceptance & ~~8\%--14\%\\
High-$\Delta R$ CR to SR & 15\%--29\% \\
$W\to \tau \nu$ SR to high-$\Delta R$ CR ratio & ~~5\%--18\% \\
0- to 1-lepton ratio & 1\%-- 6\% \\
\midrule
\textbf{Top quark (0- and 1-lepton)} \\
Top$(b)$ normalisation & Floating \\
Top(other) normalisation & Floating \\
$N_{\mathrm{jet}}$ acceptance & 7\%--9\% \\
0- to 1-lepton ratio & ~~4\% \\
SR/top CR acceptance (\ttbar) & ~~9\% \\
SR/top CR acceptance ($Wt$) & 16\% \\
$Wt$ / \ttbar ratio & 10\% \\
\midrule
\textbf{Top quark (2-lepton)} \\
Normalisation & Floating \\
\midrule
\textbf{Multi-jet (1-lepton)} \\
Normalisation & ~~20\%--100\% \\
\bottomrule
\end{tabular}
\label{tab:syst_summary}
\end{table}
 
\clearpage
\section{Statistical analysis and results}
\label{sec:stat_analysis}
 
A binned maximum-likelihood fit to the
$m_{cc}$ distribution is performed across the 44 analysis
regions to extract three parameters of interest (POI), $\bm{\mu}$.
The parameters of
interest, $\mu_{VH(c\bar{c})}$, $\mu_{VW(cq)}$ and $\mu_{VZ(c\bar{c})}$, correspond to signal
strengths that multiply the SM cross-sections times branching fractions of the
$VH(\to c\bar{c})$, $VW(\to cq)$ and $VZ(\to c\bar{c})$ processes,
and are extracted by
maximising the likelihood function with respect to both $\bm{\mu}$ and nuisance
parameters, 
which account for the systematic uncertainties
discussed in Section~\ref{sec:systematics}. Uncertainties are constrained with
Gaussian or log-normal distributions in the likelihood function with the exception of
the normalisations of the $V+$\,jets and top-quark backgrounds,
which are allowed to float freely in the fit.
The uncertainties due to the limited number of events in the simulated samples
used in the fit are included using the Beeston--Barlow technique~\cite{BARLOW1993219}
for the total MC prediction, excluding the $VH(c\bar{c})$ signal.
Systematic uncertainties that exhibit large fluctuations are smoothed and
uncertainties with negligible impact on the final results are `pruned'
following the procedures outlined in Ref.~\cite{HIGG-2013-23}.
 
The $VH(\to b\bar{b})$ background, expected to be about eight (two) times larger
than the SM $H \to c\bar{c}$ signal in the 1-$c$-tag (2-$c$-tag) signal regions,
is included in the likelihood function with uncertainties;
however, at the present level of signal
sensitivity it does not significantly impact the search  for $VH(\to c\bar{c})$.
 
The $m_{cc}$ resolution is studied using simulation in the 2-lepton channel and
its value is 10--20~\GeV\ depending on the signal region.
The resolution is better in the 2-jet signal regions than those with more than two jets,
and better in the 2-$c$-tag signal regions than the 1-$c$-tag signal regions.
In the 2-lepton channel, the resolution in the $\pt^V > 150~\GeV$ signal regions is better than in the
$75~\GeV < \pt^V < 150~\GeV$ signal regions.
The following $m_{cc}$ ranges and
uniform binnings are used in the various signal and control regions:
\begin{itemize}
\item 16 bins from 50 to 210~\GeV\ in the signal regions and 0- and 1-lepton
top control regions, with the exception of the 2-lepton, 2-$c$-tag, $\pt^V >
150~\GeV$ signal regions.
\item 9 (10) bins from 50 to $185~(200)~\GeV$ in the 2-lepton, 2-$c$-tag,
$\pt^V > 150~\GeV$, 2-jet (3-jet) signal region.
\item 13 bins from 80 to 340~\GeV\ in each of the high-$\Delta R$ control
regions, with the exception of the 2-lepton, 2-$c$-tag, $\pt^V > 150~\GeV$
high-$\Delta R$ regions, where 9 bins from 80 to 350~\GeV\ are used.
\item A single bin from 50 to 210~\GeV\ in each of the 0-$c$-tag control
regions and 2-lepton top control regions.
\end{itemize}

\begin{figure*}
\centering{\includegraphics[width=0.32\textwidth]{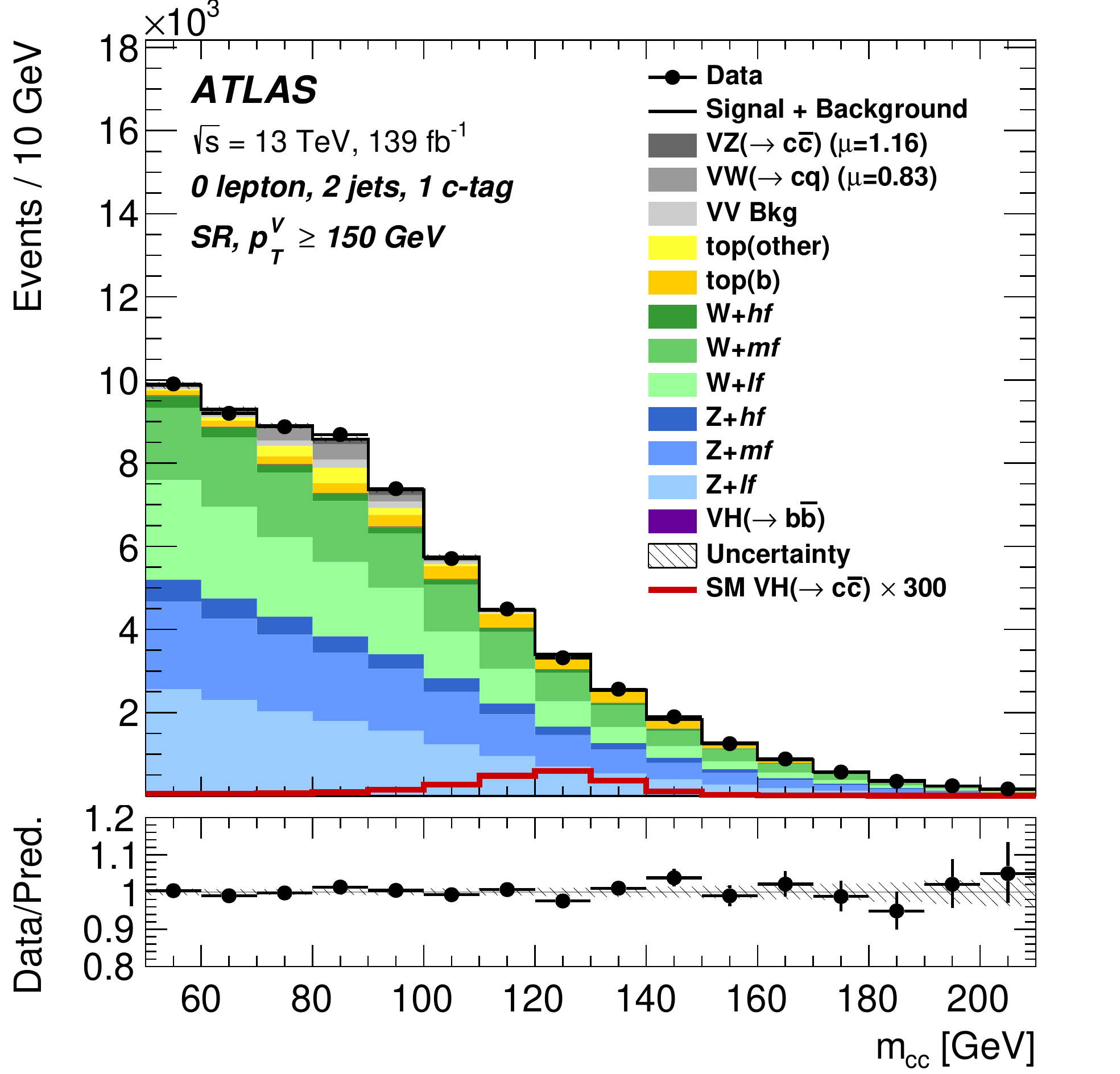}}{\includegraphics[width=0.32\textwidth]{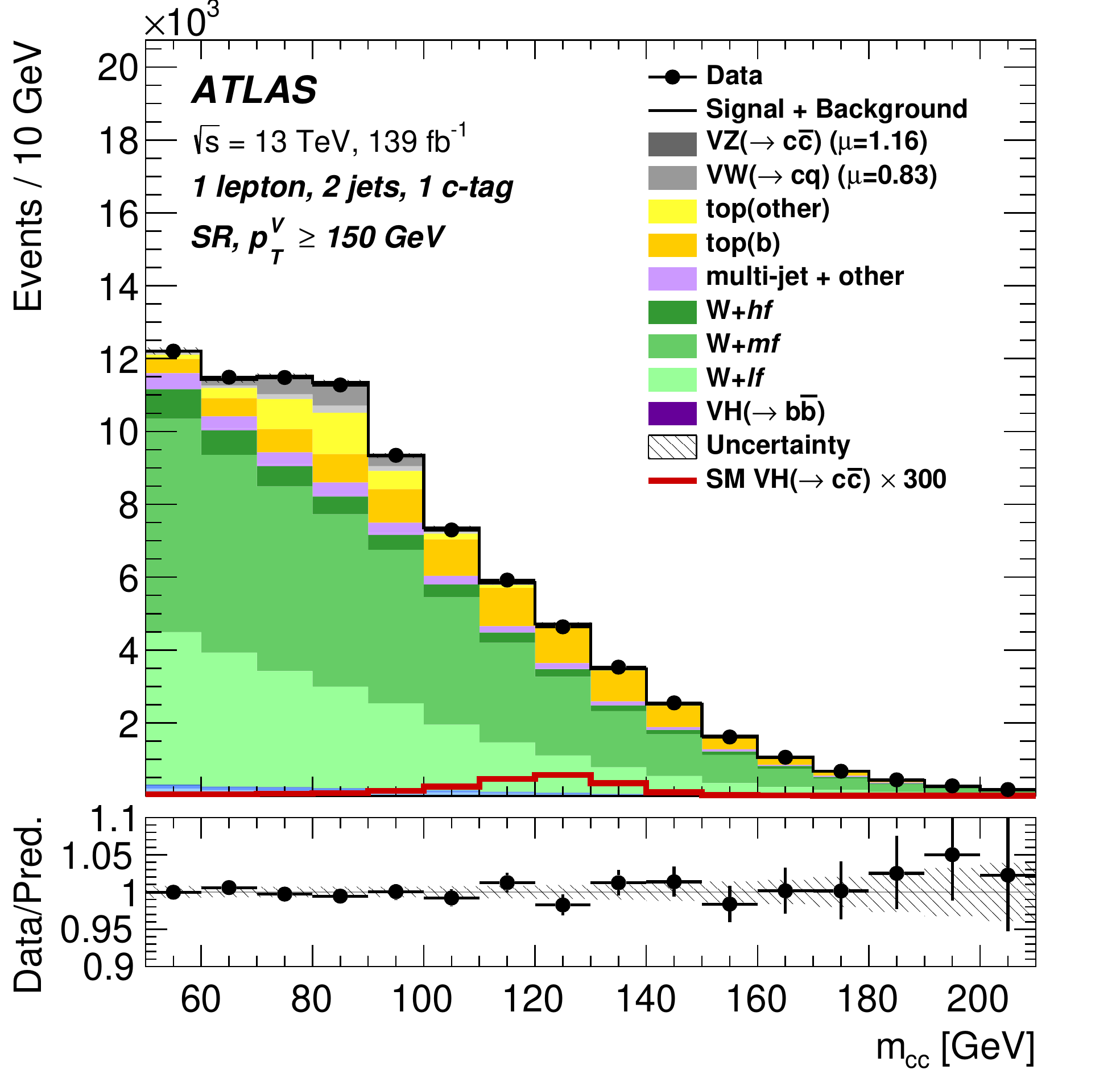}}{\includegraphics[width=0.32\textwidth]{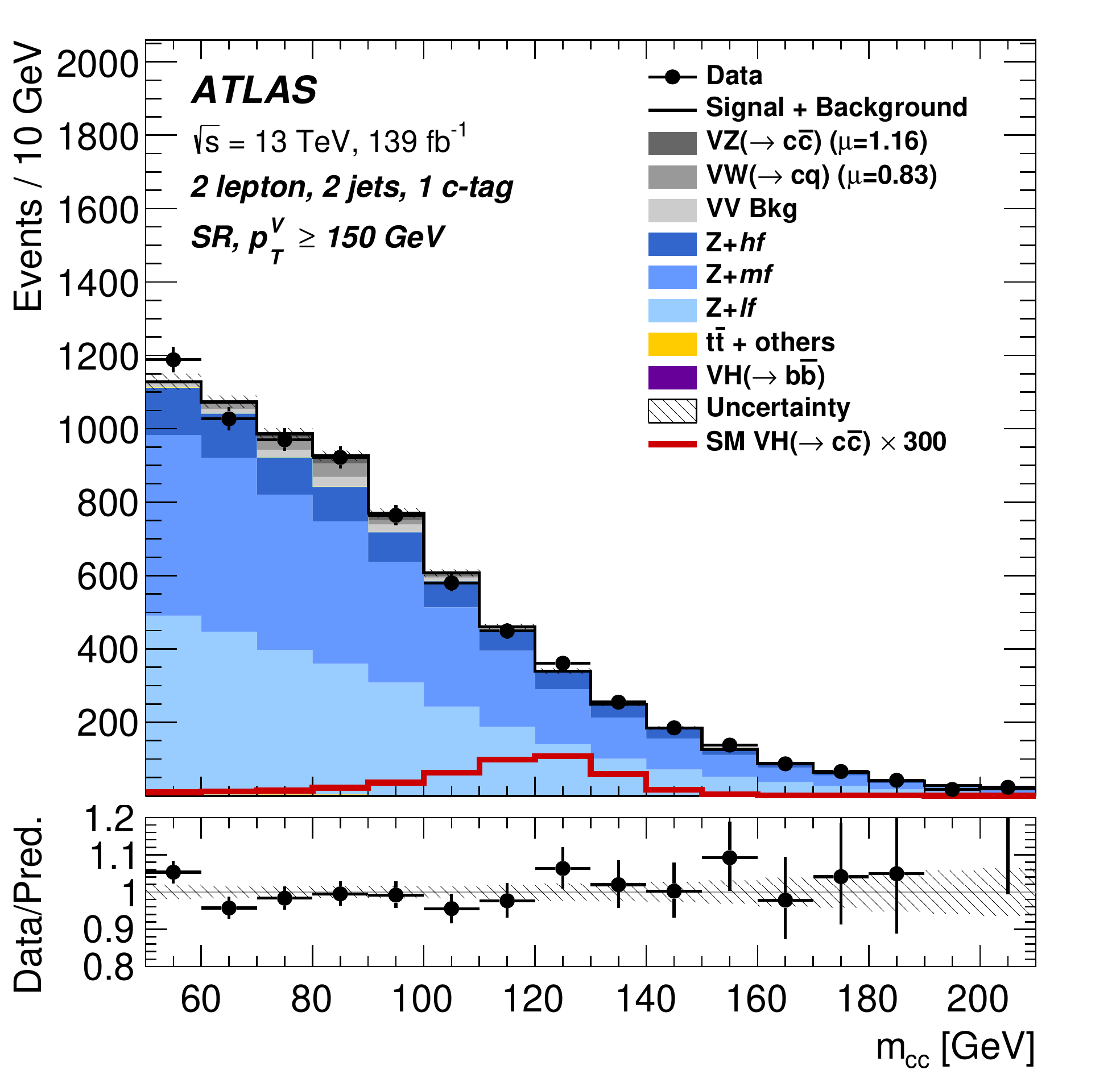}} \\
{\includegraphics[width=0.32\textwidth]{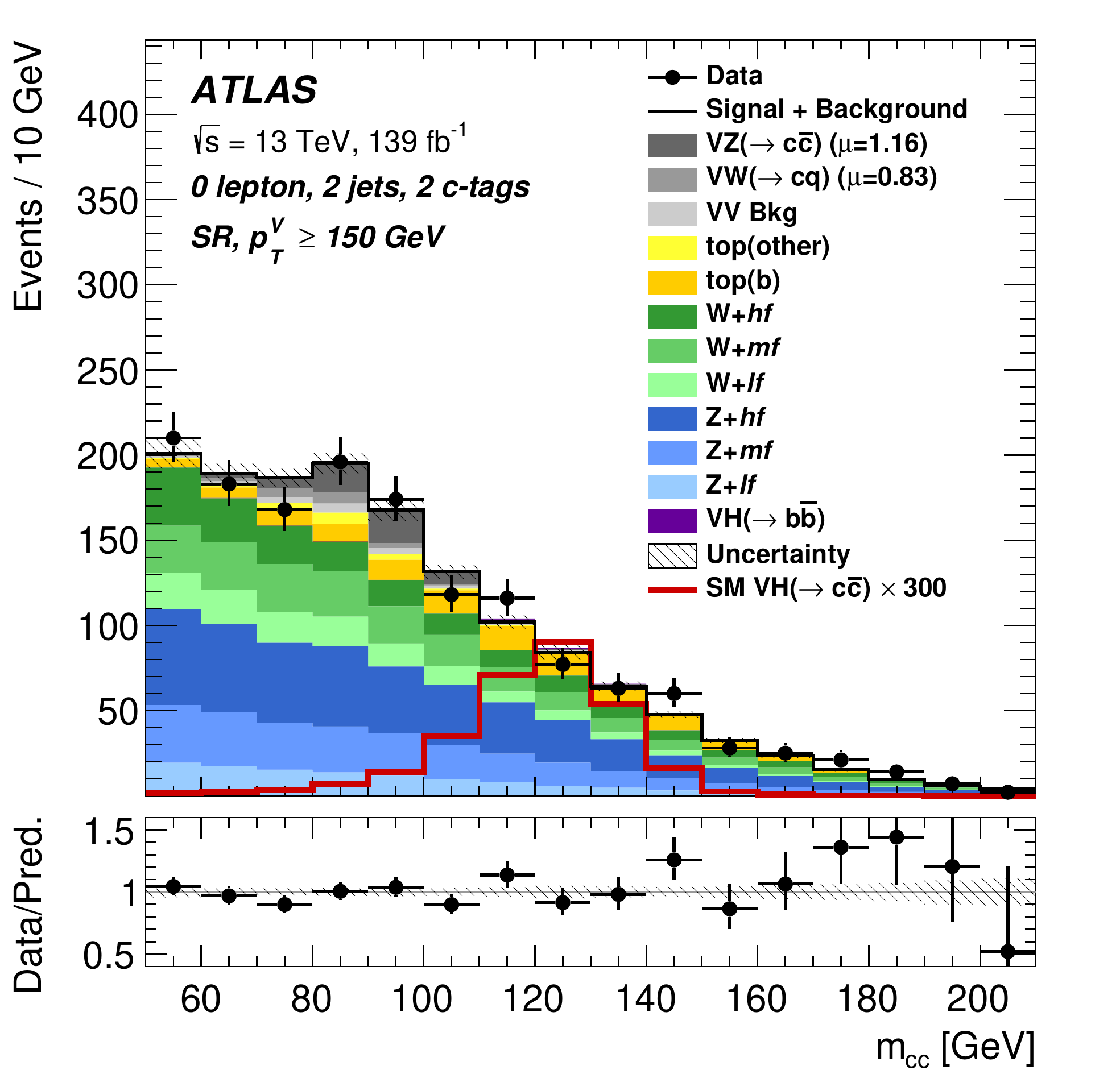}}
{\includegraphics[width=0.32\textwidth]{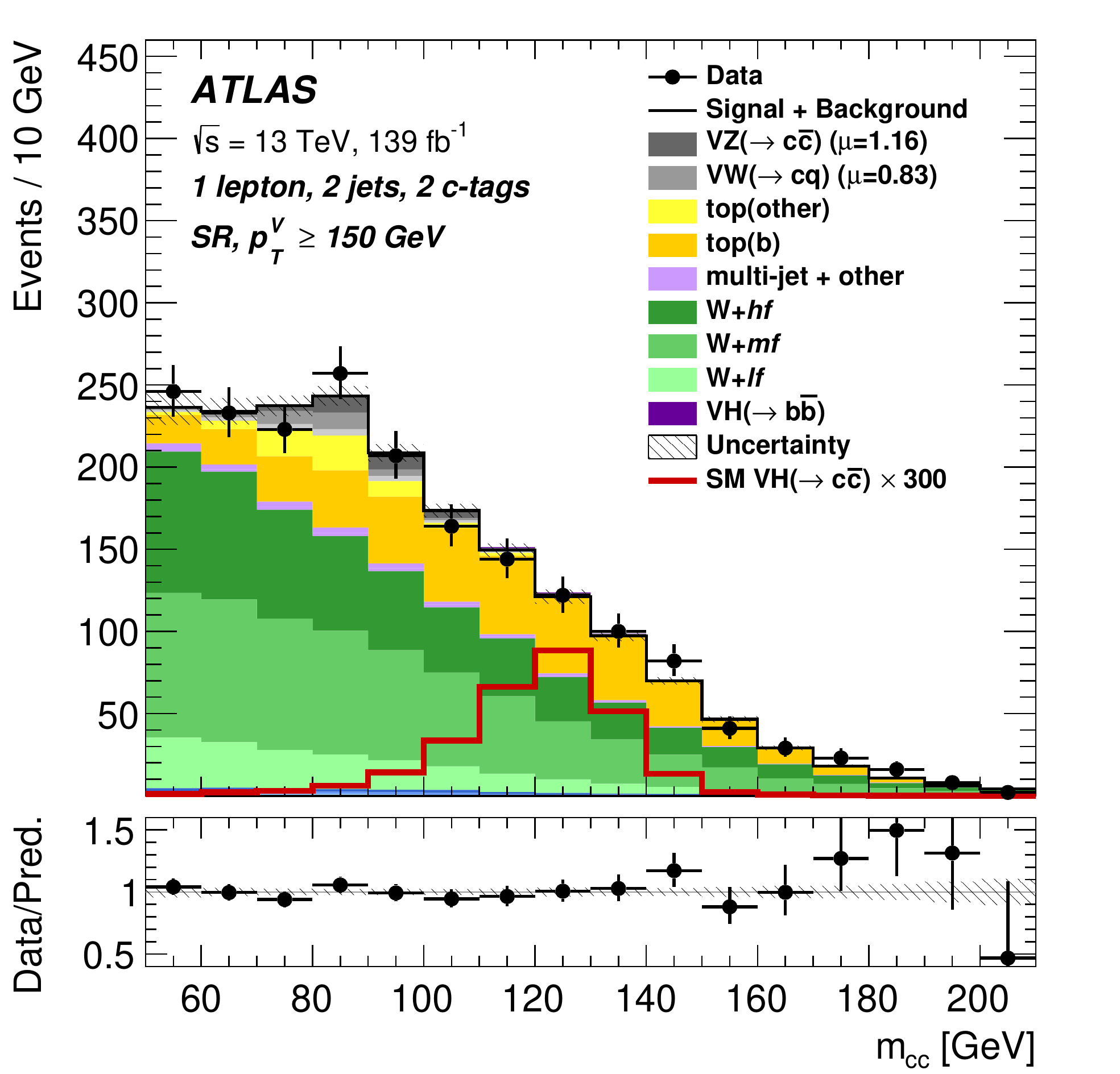}}
{\includegraphics[width=0.32\textwidth]{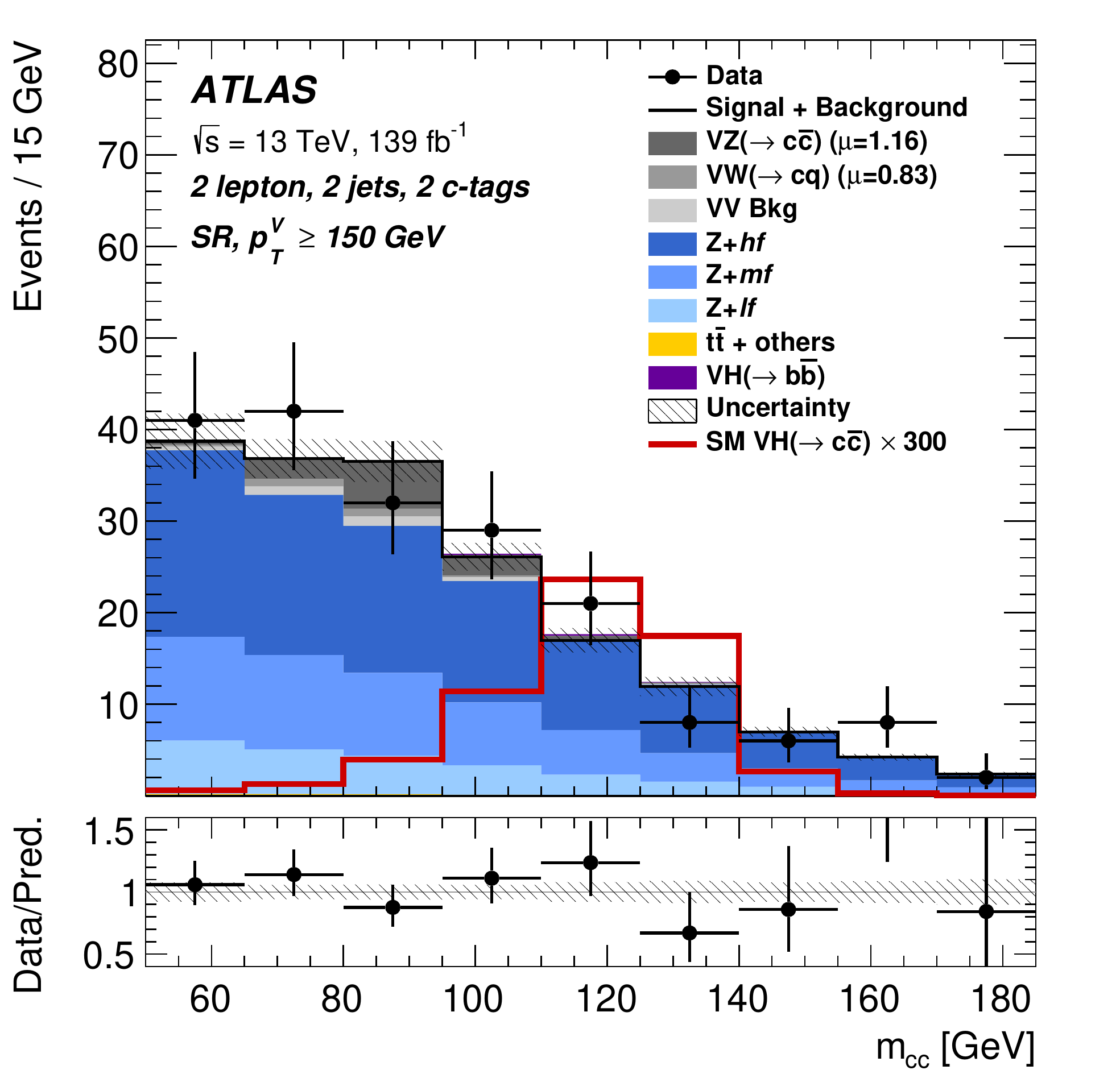}}
\caption{Post-fit distributions for six selected signal regions out of 44
analysis regions, with two jets and $\pt^V > 150~\GeV$ for the
0-lepton (left), 1-lepton (centre), and 2-lepton (right) channels, with one
$c$-tag (top) and two $c$-tags (bottom). The total signal-plus-background
prediction is shown by the solid black line and includes the $H\to c\bar{c}$
signal scaled to the best-fit value of $\mu_{VH(c\bar{c})} = -9$.
The $H\to c\bar{c}$ signal is also shown
as an unfilled histogram scaled to 300 times the SM prediction. The post-fit
uncertainty is shown as the hatched background. The ratio of the data to the
sum of the post-fit signal plus background is shown in the lower panel.
}
\label{fig:postfit}
\end{figure*}
 
Selected post-fit $m_{cc}$ distributions, where all normalisations and nuisance
parameters are adjusted by the likelihood fit, are shown in
Figure~\ref{fig:postfit} for the 0-, 1-, and 2-lepton channels. Post-fit
distributions for the remaining analysis regions can be found in the
\hyperref[sec:app_postfit_plots]{Appendix}. Table~\ref{tab:floating_norms} shows
the values of the free-floating background normalisation parameters obtained
from the likelihood fit to data. The fitted signal strengths are:
{\renewcommand{\arraystretch}{1.2}
\begin{center}
\begin{tabular}{r@{ = }r@{ $\pm$ }r@{ (stat.) $\pm$ }r@{ (syst.)}l}
$\mu_{VH(c\bar{c})}$ & $-9  $ & $10$ & $12$\\
$\mu_{VW(cq)}      $ & $0.83$ & $0.11$ & $0.21$  \\
$\mu_{VZ(c\bar{c})}$ & $1.16$ & $0.32$ & $0.36$  & . \\
\end{tabular}
\end{center}
}
 
The correlation between $\mu_{VH(c\bar{c})}$ and
$\mu_{VW(cq)}~(\mu_{VZ(c\bar{c})})$ is 17\% (16\%), while $\mu_{VW(cq)}$ and
$\mu_{VZ(c\bar{c})}$ are 17\% anti-correlated. The probability of compatibility with the SM,
defined as all three POIs being equal to unity, is 84\%. The observed (expected)
significances of the $VW(\rightarrow cq)$ and $VZ(\rightarrow c\bar{c})$ signals
are $3.8~(4.6)$ and $2.6~(2.2)$ standard deviations, respectively.
For the $\mu_{VH(c\bar{c})}$ signal strength, an upper limit of $26~(31^{+12}_{-8})$ is
observed (expected) at 95\% CL using a modified frequentist
$\mathrm{CL_{s}}$ method~\cite{Read:2002hq}, with the profile-likelihood ratio
as the test statistic~\cite{Cowan:2010js}, using the \textsc{RooFit/RooStats}
framework~\cite{RooFit,RooStats}. The limits for the three lepton channels are summarised
in Figure~\ref{fig:results:limit}.

\begin{table*}
\footnotesize
\centering
\caption{Values of the free-floating background normalisation parameters
obtained from the likelihood fit to data. The uncertainties represent the
combined statistical and systematic uncertainties. Unless otherwise stated,
normalisation parameters are correlated across all $\pt^V$ and number-of-jets
analysis regions.}
\begin{tabular}{lccc}
\toprule
Background & $\pt^V$ & Jets & Value \\
\midrule
Top($b$)                                         &                                & & $ 0.91 \pm 0.06 $ \\
Top(other)                                      &                                & & $ 0.94 \pm 0.08 $ \\
\cmidrule(lr){1-4}
\multirow{4}{*}{\ttbar (2-lepton)}                & \multirow{2}{*}{ \quad\quad $\pt^V > 150~\GeV$}  & 2 & $ 0.76 \pm 0.22 $ \\
&                                      & 3 & $ 0.96 \pm 0.13 $ \\
\cmidrule(lr){2-4}
&  \multirow{2}{*}{$75 < \pt^V < 150~\GeV$} & 2 & $ 1.08 \pm 0.08 $ \\
&   & 3 & $ 1.06 \pm 0.07 $ \\
\midrule
$W+\,$\textit{hf}                                &                                & & $ 1.16 \pm 0.35 $ \\
$W+\,$\textit{mf}                                &                                & & $ 1.28 \pm 0.14 $ \\
\cmidrule(lr){1-4}
\multirow{2}{*}{$W+\,$\textit{lf}}               &                                & 2 & $ 1.02 \pm 0.04 $ \\
&                                & 3 & $ 0.97 \pm 0.05 $ \\
\midrule
\multirow{2}{*}{$Z\,+\,$\textit{hf}}               & ~\quad\quad $\pt^V > 150~\GeV$             & & $ 1.19 \pm 0.22 $ \\ [0.2cm]
& $75 < \pt^V < 150~\GeV$        & & $ 1.25 \pm 0.25 $ \\
\cmidrule(lr){1-4}
\multirow{2}{*}{$Z\,+\,$\textit{mf}}               & ~\quad\quad $\pt^V > 150~\GeV$             & & $ 1.10 \pm 0.15 $ \\ [0.2cm]
& $75 < \pt^V < 150~\GeV$        & & $ 1.11 \pm 0.15 $ \\
\cmidrule(lr){1-4}
\multirow{4}{*}{$Z\,+\,$\textit{lf}}               & \multirow{2}{*}{ \quad\quad $\pt^V > 150~\GeV$}& 2 & $ 1.07 \pm 0.03 $ \\
&        & 3 & $ 1.08 \pm 0.05 $ \\
\cmidrule(lr){2-4}
& \multirow{2}{*}{$75 < \pt^V < 150~\GeV$}  & 2 & $ 1.12 \pm 0.04 $ \\
&  & 3 & $ 1.07 \pm 0.06 $ \\
\bottomrule
\end{tabular}
\label{tab:floating_norms}
\end{table*}
 
The effects of systematic uncertainties are summarised in
Table~\ref{tab:syst_breakdown}. For each POI, the statistical uncertainty is
obtained from a fit in which all nuisance parameters are fixed to their post-fit
values. The total systematic uncertainty is found by subtracting the squared statistical
uncertainty from the squared total uncertainty,
i.e.\ $\sigma_{\mathrm{syst.}} = (\sigma_{\mathrm{tot.}}^2 - \sigma_{\mathrm{stat.}}^2)^{1/2}$.
Similarly, the impact of a subset of the
systematic uncertainties is assessed by performing the fit with only their
nuisance parameters fixed to their post-fit values.
For each POI, the impact is
then computed as the square root of the decrease in the squared uncertainty of that
POI between the nominal fit and the fit with the nuisance parameters fixed.
Despite the additional uncertainties it introduces, the use of truth-flavour
tagging improves the expected limit on $\mu_{VH(c\bar{c})}$ by about
$10\%$ due to the improved statistical precision in the MC predictions.
 
The improvements in this analysis relative to the previous ATLAS search for
$ZH, H \to c\bar{c}$~\cite{HIGG-2017-01} are quantified by performing a fit in
the 2-lepton channel to the 2015--2016 data, corresponding to $36~\ifb$. Using
the same signal regions as the previous analysis a 36\% improvement in the expected
limit is found, with most of the improvement due to better flavour-tagging performance.
After also including the new 2-lepton signal and control regions
introduced in this analysis, a 43\% improvement in the expected limit is found.
Adding the full Run-2 dataset, along with the 0- and 1-lepton channels, the
expected limit is improved by a factor of five in this analysis,
relative to the previous ATLAS search.
 
\begin{figure*}
\centering
\includegraphics[width=0.65\textwidth]{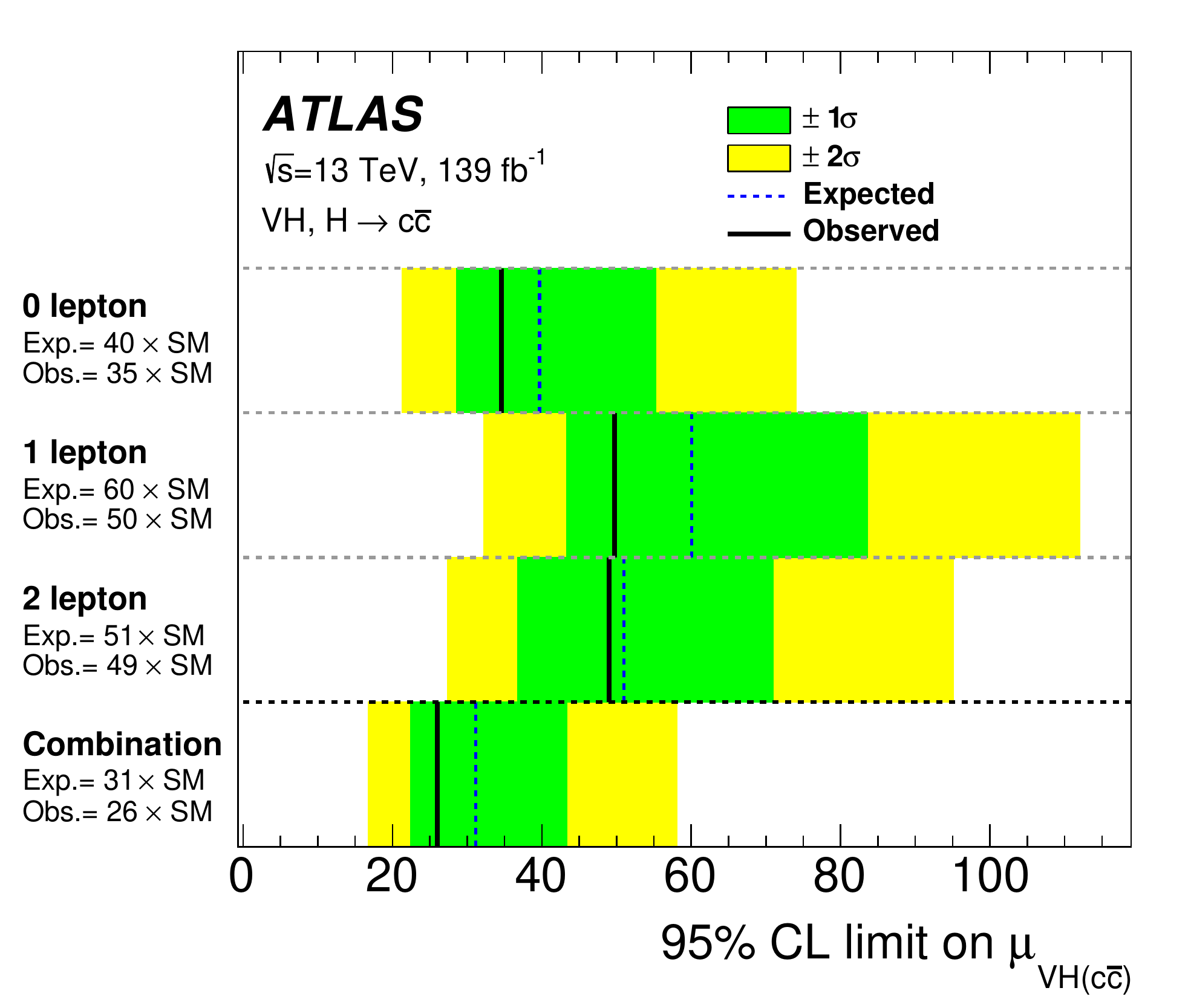}
\caption{The observed and expected 95\% CL upper limits on the cross-section
times branching fraction normalized to its SM prediction
in each lepton channel and for the combined fit.
The single-channel limits are obtained using a five-POI fit, in which each channel
has a separate $VH(\to c\bar{c})$ parameter of interest.}
\label{fig:results:limit}
\end{figure*}
 
\begin{table}
\centering
\caption{Breakdown of contributions to the uncertainty in the fitted values of
$\mu_{VH(c\bar{c})}$, $\mu_{VW(cq)}$ and $\mu_{VZ(c\bar{c})}$. The sum in
quadrature of uncertainties from different sources may differ from the total due
to correlations. In cases where the upward and downward systematic variations have
different values, the mean of the absolute values is shown.}
\small
\begin{tabular}{llrrr}
\toprule
\multicolumn{2}{l}{Source of uncertainty} & $\mu_{VH(c\bar{c})}$ & $\mu_{VW(cq)}$ & $\mu_{VZ(c\bar{c})}$ \\
\midrule
Total                 & & $  15.3 $ & $  0.24 $ & $  0.48 $ \\
Statistical           & & $  10.0 $ & $  0.11 $ & $  0.32 $ \\
Systematic            & & $  11.5 $ & $  0.21 $ & $  0.36 $ \\
\midrule
\multicolumn{4}{l}{Statistical uncertainties} \\
\midrule
\multicolumn{2}{l}{Signal normalisation}          & $  7.8 $  & $  0.05 $ & $  0.23 $ \\
\multicolumn{2}{l}{Other normalisations}       & $  5.1 $  & $  0.09 $ & $  0.22 $ \\
\midrule
\multicolumn{4}{l}{Theoretical and modelling uncertainties} \\
\midrule
$VH(\to c\bar{c})$            & & $  2.1 $  & $    <0.01 $ & $  0.01 $ \\
\Zjets                      & & $  7.0 $  & $  0.05 $ & $  0.17 $ \\
Top quark    	                & & $  3.9 $  & $  0.13 $ & $  0.09 $ \\
\Wjets                      & & $  3.0 $  & $  0.05 $ & $  0.11 $ \\
Diboson                       & & $  1.0 $  & $  0.09 $ & $  0.12 $ \\
$VH(\to b\bar{b})$            & & $  0.8 $  & $    <0.01 $ & $  0.01 $ \\
Multi-jet                     & & $  1.0 $  & $  0.03 $ & $  0.02 $ \\
\cmidrule(lr){1-5}
Simulation samples size       & & $  4.2 $  & $  0.09 $ & $  0.13 $ \\
\midrule
\multicolumn{4}{l}{Experimental uncertainties} \\
\midrule
Jets                          & & $  2.8 $  & $  0.06 $ & $  0.13 $ \\
Leptons                       & & $  0.5 $  & $  0.01 $ & $  0.01 $ \\
\met                          & & $  0.2 $  & $  0.01 $ & $  0.01 $ \\
\multicolumn{2}{l}{Pile-up and luminosity}
& $  0.3 $  & $  0.01 $ & $  0.01 $ \\
\cmidrule(lr){1-5}
\multirow{4}{*}{Flavour tagging}
& $c$-jets     & $  1.6 $  & $  0.05 $ & $  0.16 $ \\
& $b$-jets     & $  1.1 $  & $  0.01 $ & $  0.03 $ \\
& light-jets   & $  0.4 $  & $  0.01 $ & $  0.06 $ \\
& $\tau$-jets  & $  0.3 $  & $  0.01 $ & $  0.04 $ \\
\cmidrule(lr){1-5}
\multirow{2}{*}{Truth-flavour tagging}
& $\Delta R$ correction  & $  3.3 $  & $  0.03 $ & $  0.10 $ \\
& Residual non-closure   & $  1.7 $  & $  0.03 $ & $  0.10 $ \\
\bottomrule
\end{tabular}
\label{tab:syst_breakdown}
\end{table}
 
The $m_{cc}$ distributions for events with either one or two $c$-tagged jets, summed
over all channels and regions,
after background subtraction, and using the fitted signal strengths, are shown in
Figure~\ref{fig:mcc_weighted}.
 
\begin{figure*}
\centering
\subcaptionbox{\label{fig:mcc_weighted:1tag}}{\includegraphics[width=0.45\textwidth]{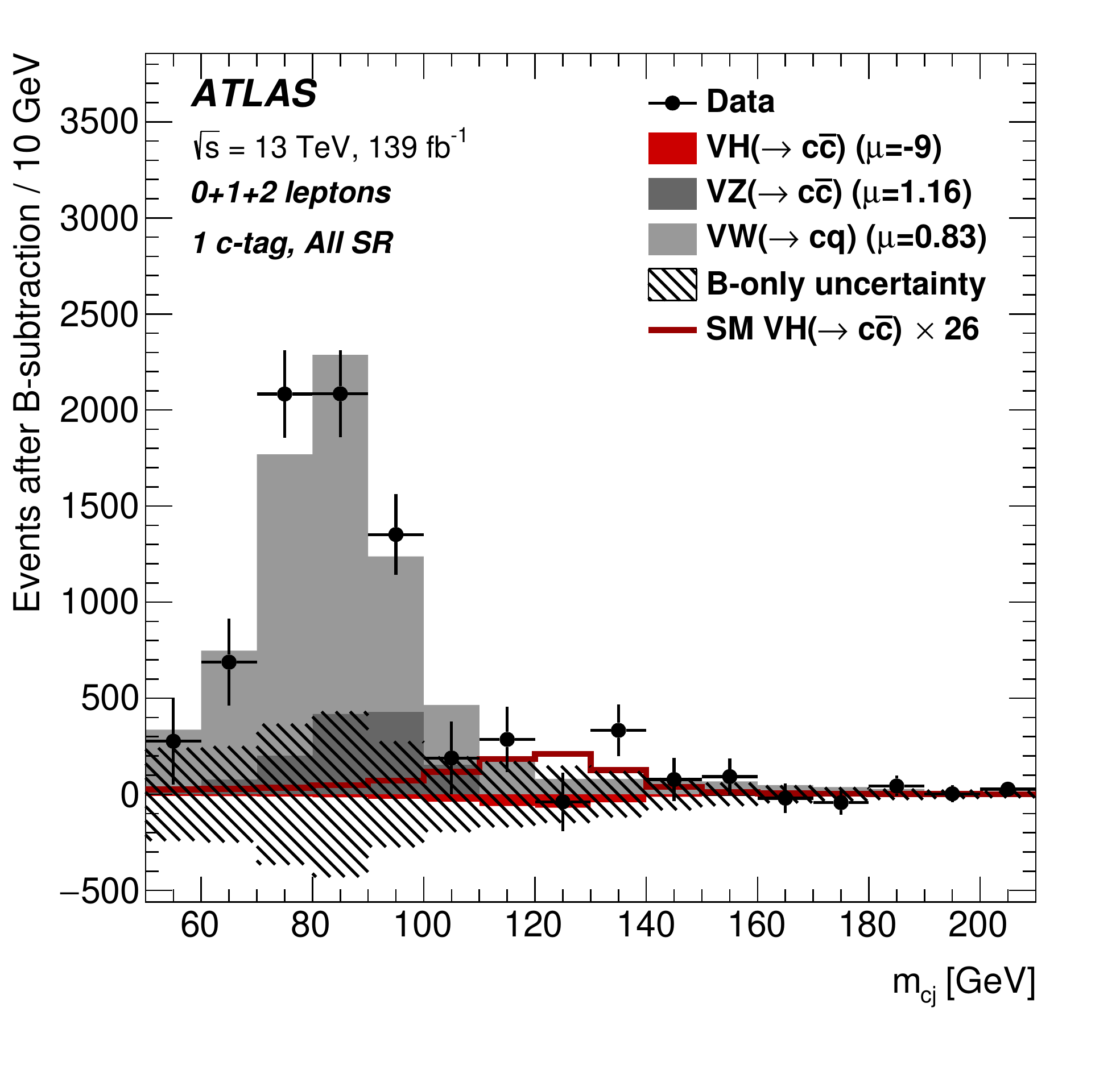}}
\subcaptionbox{\label{fig:mcc_weighted:2tag}}{\includegraphics[width=0.45\textwidth]{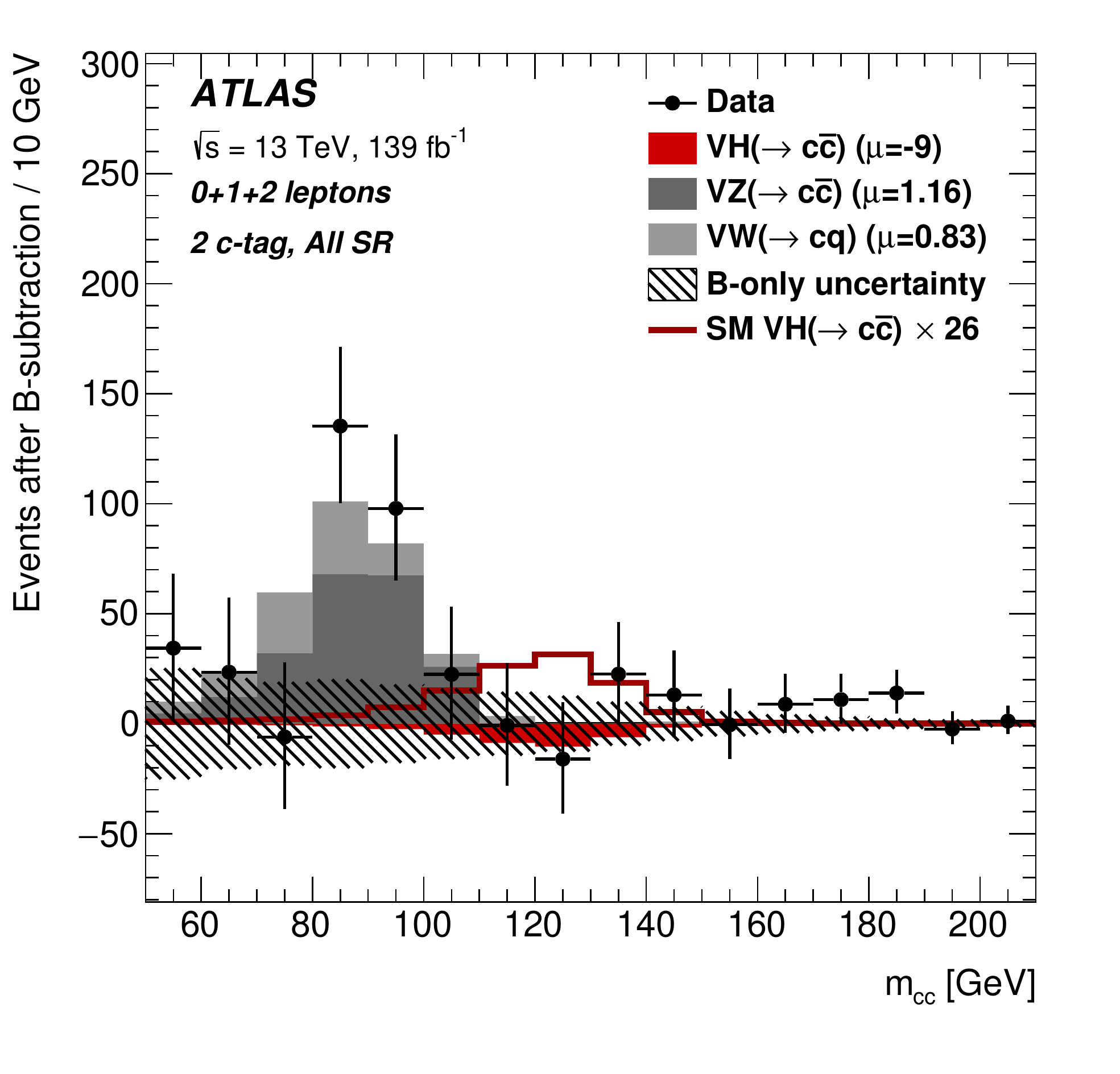}}
\caption{The post-fit $m_{cc}$ distribution summed over all signal regions
after subtracting backgrounds, leaving only the $VH(\to c\bar{c})$, $VW(\to
cq)$ and $VZ(\to c\bar{c})$ processes, for events with
\subref{fig:mcc_weighted:1tag} one $c$-tag and \subref{fig:mcc_weighted:2tag} two
$c$-tags. The red filled histogram corresponds to the $VH, H\to c\bar{c}$
signal for the fitted value of $\mu_{VH(c\bar{c})} = -9$, while the open red
histogram corresponds to the signal expected at the 95\% CL upper limit on
$\mu_{VH(c\bar{c})}~(\mu_{VH(c\bar{c})}=26)$.  The hatched band shows the
uncertainty of the fitted background.
}
\label{fig:mcc_weighted}
\end{figure*}

The best-fit value of the $VH(\to c\bar{c})$ signal strength is interpreted within
the kappa framework~\cite{Heinemeyer:2013tqa,deFlorian:2016spz}, by
reparameterising $\mu_{VH(c\bar{c})}$ in the likelihood function in terms of the Higgs--charm
coupling modifier, $\kappa_c$, assuming that the coupling modifier only affects the Higgs boson decays.
Including effects in both the partial and full width, considering only SM decays
and setting all other couplings to their SM predictions, $\mu_{VH(c\bar{c})}$
is parameterised as a function of $\kappa_c$
\begin{equation}\nonumber
\mu_{VH(c\bar{c})}(\kappa_c) = \frac{\kappa_c^2}{1 + B^{\mathrm{SM}}_{H \to c\bar{c}}(\kappa_c^2 - 1)}\,,
\end{equation}
where $B^{\mathrm{SM}}_{H \to c\bar{c}}$ is the $H \to c\bar{c}$
branching fraction predicted in the SM.
 
Constraints on $\kappa_c$ are set using the profile-likelihood ratio test
statistic and are shown at 95\% CL for each of the three channels and for the
combined likelihood fit in Figure~\ref{fig:kappa_likelihood}. The combination
allows an observed (expected) constraint of $|\kappa_c| < 8.5~(12.4)$ to be set
at the 95\% CL.
 
\begin{figure}
\centering
\includegraphics[width=0.65\textwidth]{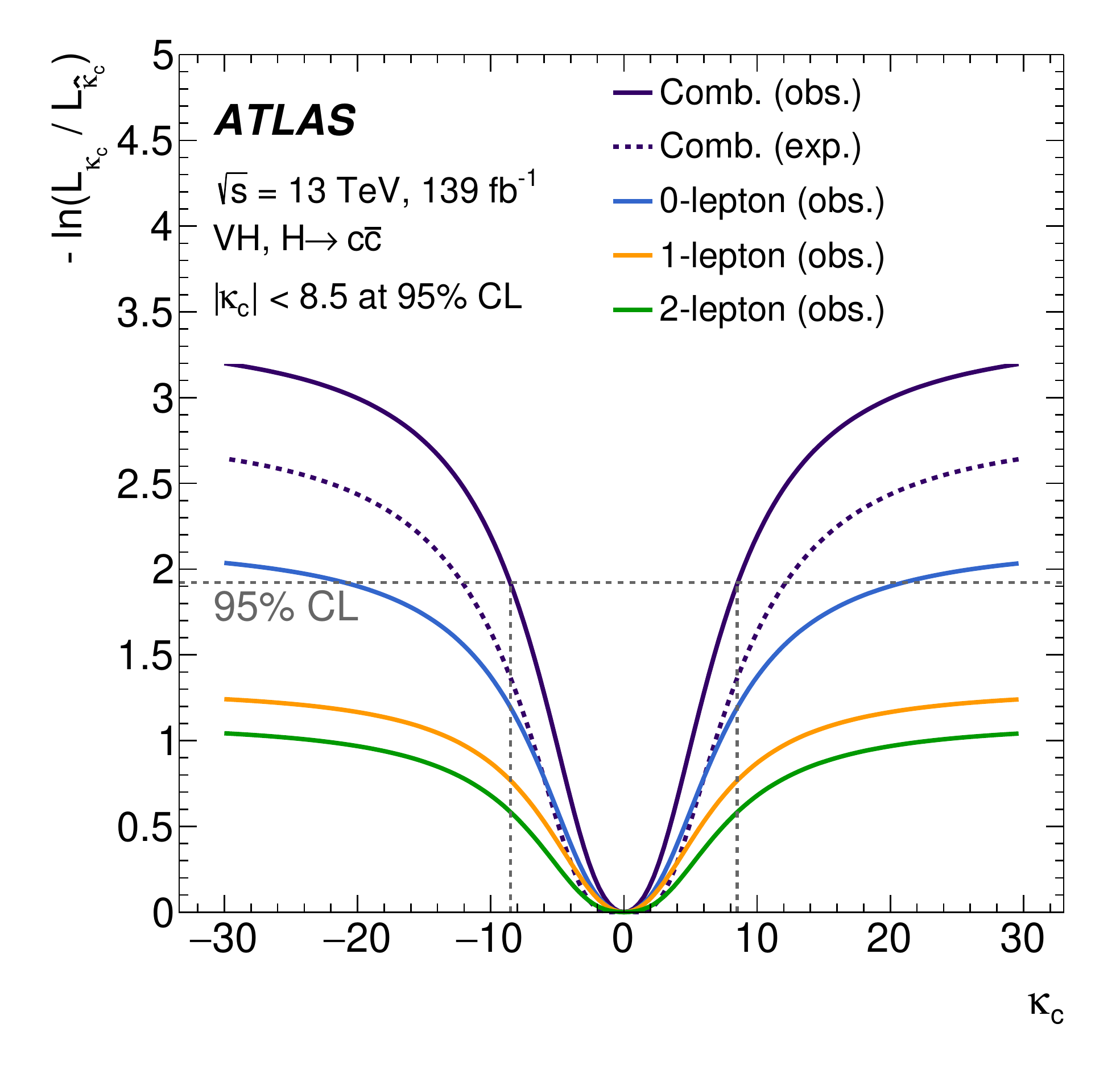}
\caption{Expected and observed values of the negative profile
log-likelihood ratio as a function of $\kappa_c$. The
single-channel likelihoods are obtained using a five-POI fit, in which each
channel has a separate $VH(\to c\bar{c})$ parameter of interest. The order of
the lines in the plot matches the order of the lines in the legend.}
\label{fig:kappa_likelihood}
\end{figure}
\FloatBarrier
\section{Combination with $VH, H\to b\bar{b}$}
\label{sec:combination}
 
A combination of the analysis presented in this paper with the ATLAS $VH, H\to
b\bar{b}$ measurement~\cite{HIGG-2018-51} is performed by creating a likelihood function
that is the product of the individual likelihood functions of the two analyses. Two
parameters of interest are used, $\mu_{VH(c\bar{c})}$ and $\mu_{VH(b\bar{b})}$
for the $VH, H \to c\bar{c}$ and $VH, H \to b\bar{b}$ signal strengths,
respectively, and are included in both of the input likelihood functions.
The importance of including both signal strengths in a combined likelihood was pointed out by
Refs.~\cite{PhysRevD.89.033014,PhysRevD.92.033016}.
Experimental systematic uncertainties that are common to both analyses, detailed in
Section~\ref{sec:exp_uncerts}, are considered correlated between the two
analyses, with the exception of the flavour-tagging systematic uncertainties due to
the different calibration procedures used for $b$- and $c$-tagging.
Background normalisations and modelling uncertainties are considered
uncorrelated between the two analyses.

The fitted signal strengths are
{\renewcommand{\arraystretch}{1.2}
\begin{center}
\begin{tabular}{r@{ = }r@{ $\pm$ }l}
$\mu_{VH(c\bar{c})}$ & $-9$  & 10 (stat.) $\pm~11$ (syst.) \\
$\mu_{VH(b\bar{b})}$ & 1.06  & 0.12 (stat.) $^{+0.15}_{-0.13}$ (syst.) \\
\end{tabular}
\end{center}
} with a correlation coefficient of $-12\%$. The fitted signal strengths are
consistent with those found in the individual analyses. The expected and
observed best-fit values and their 68\% and 95\% CL contours are shown in
Figure~\ref{fig:comb_llh}.
 
\begin{figure}
\centering
\includegraphics[width=0.7\textwidth]{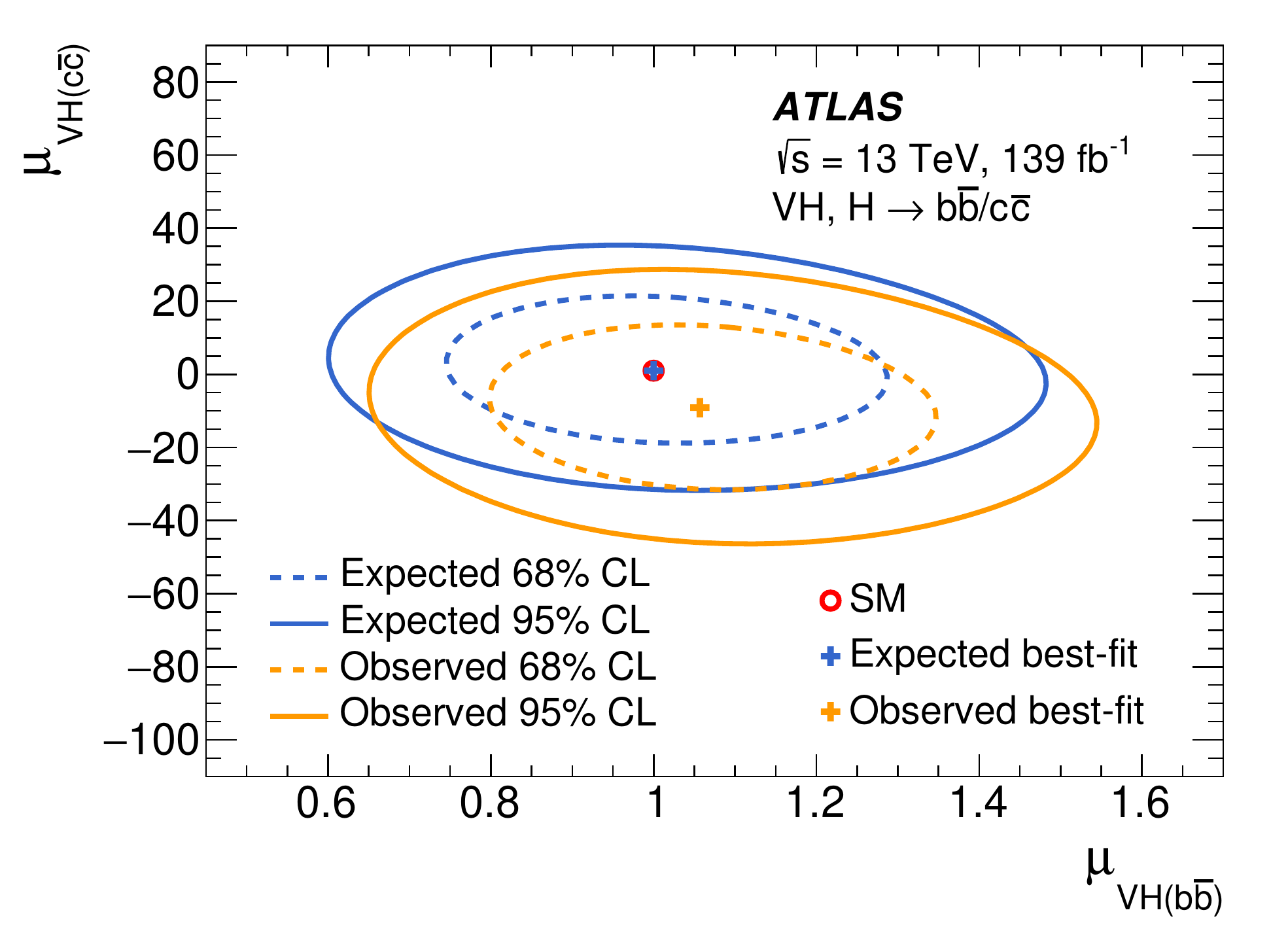}
\caption{The observed and expected best fit values of $\mu_{VH(c\bar{c})}$
and $\mu_{VH(b\bar{b})}$ with their 68\% and 95\% CL contours.}
\label{fig:comb_llh}
\end{figure}

Although the signal regions of the two analyses are orthogonal due to the
$b$-tagging veto used in the $c$-tagging definition, a small overlap of events
occurs in the control regions used in the two analyses. To test the impact of
this, events that appear in both analyses are removed from the $VH, H\to
c\bar{c}$ control regions. The results are unchanged. Treating the
normalisations of the backgrounds as correlated between the two analyses is also tested and
does not affect the expected sensitivity.
 
The best-fit values of $\mu_{VH(b\bar{b})}$ and $\mu_{VH(c\bar{c})}$ are
interpreted in the kappa framework by parameterising the likelihood function in terms of
both $\kappa_b$ and $\kappa_c$, while setting all other couplings to their SM
predictions and considering only SM Higgs boson decays.
Constraints on $\kappa_b$ and $\kappa_c$ are set using the profile-likelihood ratio test statistic.
The expected and
observed constraints
are shown in
Figures~\ref{fig:sub:kappa_bc_exp} and~\ref{fig:sub:kappa_bc_obs}, respectively.
The likelihood function is symmetric relative to the sign of $\kappa_c$ but
not to the sign of $\kappa_b$ due to the
inclusion of $\kappa_b$ in the parameterisation of $\sigma(gg \to
ZH)$~\cite{ATLAS-CONF-2020-027}, resulting in two minima in the expected
likelihood scan.
For most values of $\kappa_b$, a value of $\kappa_c$
is allowed at 95\% CL that compensates for
the effect of $\kappa_b$ via the width of the Higgs boson and vice versa.
The observed best-fit value is $(\kappa_b, \kappa_c) = (-1.02,
0)$. The difference in the value of the log-likelhood function between the
best-fit value and $(\kappa_b, \kappa_c) = (+1.02, 0)$ is 0.02. These
constraints complement those from measurements of the Higgs boson $\pt$
spectrum~\cite{HIGG-2018-29,HIGG-2019-13}.
 
The ratio $|\kappa_c/\kappa_b|$ is constrained to be less than 4.5 at
95\% CL (5.1 expected).  The observed value is smaller than the ratio
of the $b$- and $c$-quark masses, $m_b / m_c = 4.578 \pm
0.008$~\cite{PhysRevD.98.054517}, and therefore constrains the
coupling of the Higgs boson to the charm quark to be weaker than the
coupling of the Higgs boson to the bottom quark at 95\% CL.  The
profile likelihood scan, parameterised in terms of
$\kappa_c/\kappa_b$, with $\kappa_b$ as a free parameter, is shown in
Figure~\ref{fig:kappa_ratio_likelihood}.
 
\begin{figure}
\centering
\subcaptionbox{\label{fig:sub:kappa_bc_exp}}{\includegraphics[width=0.45\textwidth]{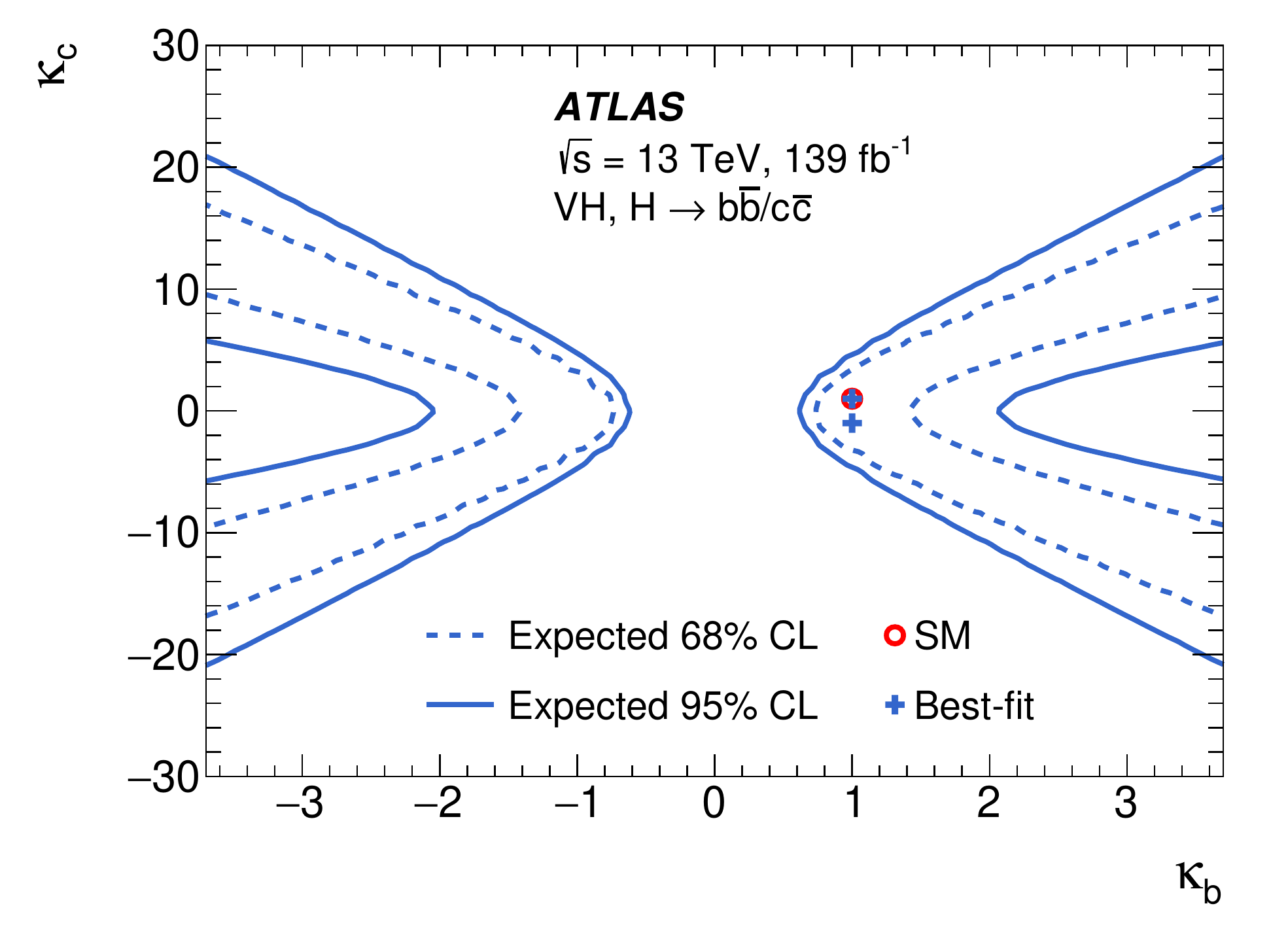}}
\subcaptionbox{\label{fig:sub:kappa_bc_obs}}{\includegraphics[width=0.45\textwidth]{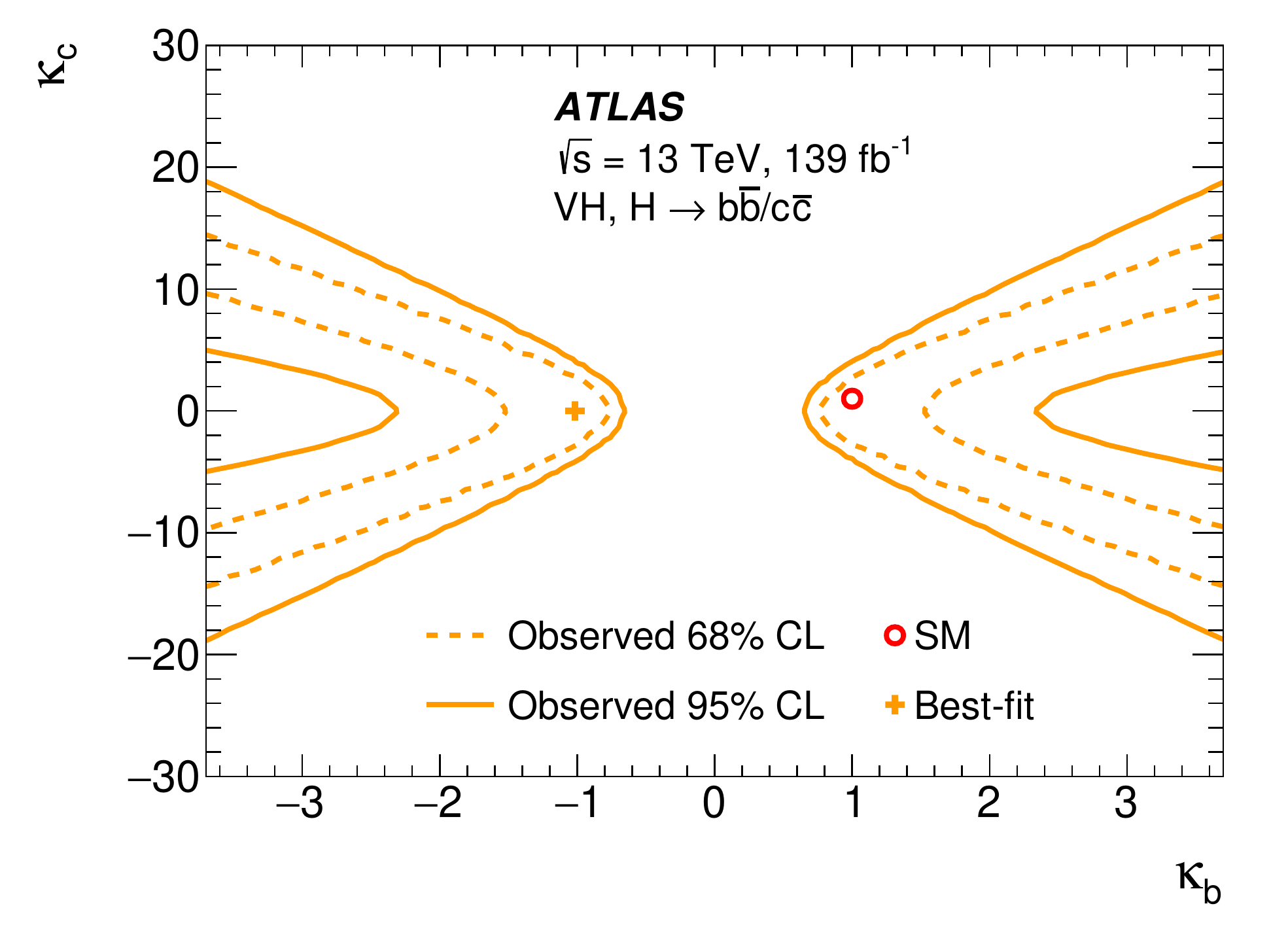}}
\caption{The \subref{fig:sub:kappa_bc_exp} expected and
\subref{fig:sub:kappa_bc_obs} observed constraints on $\kappa_c$ and
$\kappa_b$ at 68\% and 95\% confidence levels.}
\label{fig:kappa_bc}
\end{figure}
 
\begin{figure}
\centering
\includegraphics[width=0.45\textwidth]{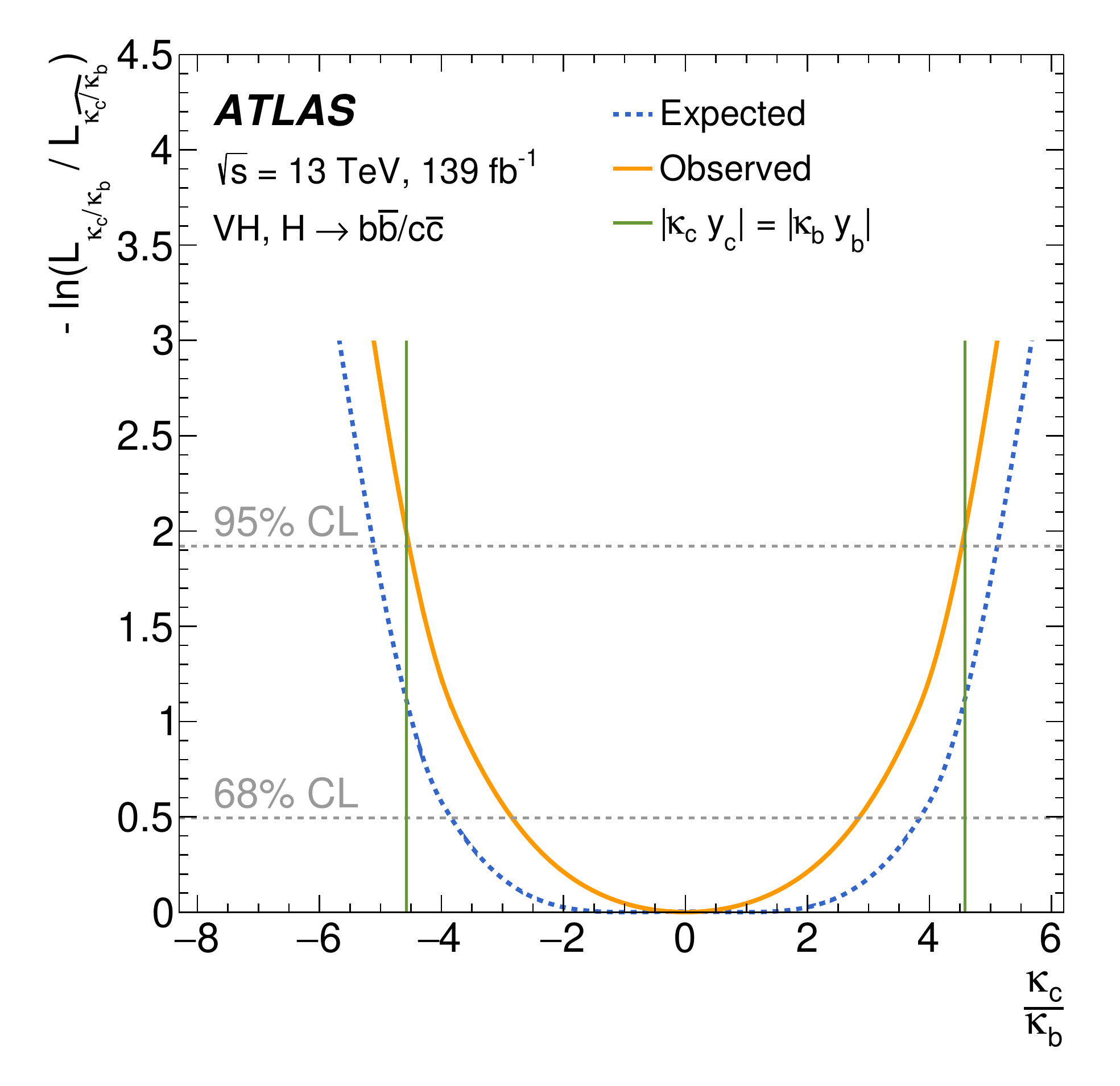}
\caption{Expected and observed values of the combined $VH, H\to c\bar{c}$ and $VH, H\to
b\bar{b}$ negative profile log-likelihood ratio as a function of $\kappa_c/\kappa_b$,
where $\kappa_b$ is a free parameter.
The vertical green lines correspond to the values of
$|\kappa_c / \kappa_b|$ for which the Higgs--charm and Higgs--bottom couplings are
equal, where each coupling strength $|\kappa_i y_i|$ is the product of the
$\kappa_i$ modifier and the Yukawa coupling, $y_i$, for $i = b,c$, and is
equal to $m_b / m_c = 4.578 \pm 0.008$~\cite{PhysRevD.98.054517}.}
\label{fig:kappa_ratio_likelihood}
\end{figure}
\FloatBarrier
\section{Conclusion}
A direct search for the decay of a Higgs boson to a charm
quark--antiquark pair has been performed using $139~\ifb$ of $pp$
collision data recorded at $\sqrt{s} = 13~\TeV$ by the ATLAS detector at the LHC.
The search uses three channels, $ZH \to \nu\nu c\bar{c}$, $WH \to
\ell\nu c\bar{c}$ and $ZH \to \ell\ell c\bar{c}$.
Signal events are identified using a multivariate charm tagging algorithm.
 
To enhance the signal sensitivity, events are categorised according to the \pt
of the reconstructed vector boson, the number of jets and the number of
$c$-tagged jets. The $m_{cc}$ observable is used as the main discriminant in the
likelihood fit to extract the signal. The analysis strategy is validated with the study of
diboson production, which is found to be consistent with the SM prediction, with
observed (expected) significances of $2.6$ ($2.2$) standard deviations for the
$VZ(\to c\bar{c})$ process and $3.8$ ($4.6$) standard deviations for the $VW(\to
cq)$ process.
 
The analysis yields an observed (expected) limit of $26~(31^{+12}_{-8})$ times
the predicted SM cross-section times branching fraction for a Higgs boson, with a
mass of $125~\GeV$, decaying into a charm quark--antiquark pair,
at the 95\% confidence level, the most stringent limit to date.
The expected
limit is a factor of five more stringent than in the previous ATLAS search
for $ZH, H\to c\bar{c}$, due to the larger dataset, improved $c$-tagging,
and inclusion of the 0- and 1-lepton channels and additional signal and control regions.
The result is interpreted in the kappa framework, considering effects on the
Higgs boson width and setting all other couplings to their SM values,
which results in an observed (expected) constraint on the
charm Yukawa coupling modifier strength $|\kappa_c| < 8.5~(12.4)$, at the 95\%
confidence level.

A combination with the ATLAS $H \to \bbbar$ measurement is
performed. Interpreted in the kappa framework the combination
constrains the observed ratio $|\kappa_c / \kappa_b|$ to be $< 4.5$ at
the 95\% confidence level.  This is less than the ratio of the $b$-
and $c$-quark masses, $m_b / m_c$, and thus constrains the coupling of
the Higgs boson to the charm quark to be weaker than the coupling of
the Higgs boson to the bottom quark at the 95\% confidence level.
 
\section*{Acknowledgements}
 

We thank CERN for the very successful operation of the LHC, as well as the
support staff from our institutions without whom ATLAS could not be
operated efficiently.
 
We acknowledge the support of
ANPCyT, Argentina;
YerPhI, Armenia;
ARC, Australia;
BMWFW and FWF, Austria;
ANAS, Azerbaijan;
SSTC, Belarus;
CNPq and FAPESP, Brazil;
NSERC, NRC and CFI, Canada;
CERN;
ANID, Chile;
CAS, MOST and NSFC, China;
Minciencias, Colombia;
MEYS CR, Czech Republic;
DNRF and DNSRC, Denmark;
IN2P3-CNRS and CEA-DRF/IRFU, France;
SRNSFG, Georgia;
BMBF, HGF and MPG, Germany;
GSRI, Greece;
RGC and Hong Kong SAR, China;
ISF and Benoziyo Center, Israel;
INFN, Italy;
MEXT and JSPS, Japan;
CNRST, Morocco;
NWO, Netherlands;
RCN, Norway;
MEiN, Poland;
FCT, Portugal;
MNE/IFA, Romania;
JINR;
MES of Russia and NRC KI, Russian Federation;
MESTD, Serbia;
MSSR, Slovakia;
ARRS and MIZ\v{S}, Slovenia;
DSI/NRF, South Africa;
MICINN, Spain;
SRC and Wallenberg Foundation, Sweden;
SERI, SNSF and Cantons of Bern and Geneva, Switzerland;
MOST, Taiwan;
TAEK, Turkey;
STFC, United Kingdom;
DOE and NSF, United States of America.
In addition, individual groups and members have received support from
BCKDF, CANARIE, Compute Canada and CRC, Canada;
COST, ERC, ERDF, Horizon 2020 and Marie Sk{\l}odowska-Curie Actions, European Union;
Investissements d'Avenir Labex, Investissements d'Avenir Idex and ANR, France;
DFG and AvH Foundation, Germany;
Herakleitos, Thales and Aristeia programmes co-financed by EU-ESF and the Greek NSRF, Greece;
BSF-NSF and GIF, Israel;
Norwegian Financial Mechanism 2014-2021, Norway;
NCN and NAWA, Poland;
La Caixa Banking Foundation, CERCA Programme Generalitat de Catalunya and PROMETEO and GenT Programmes Generalitat Valenciana, Spain;
G\"{o}ran Gustafssons Stiftelse, Sweden;
The Royal Society and Leverhulme Trust, United Kingdom.
 
The crucial computing support from all WLCG partners is acknowledged gratefully, in particular from CERN, the ATLAS Tier-1 facilities at TRIUMF (Canada), NDGF (Denmark, Norway, Sweden), CC-IN2P3 (France), KIT/GridKA (Germany), INFN-CNAF (Italy), NL-T1 (Netherlands), PIC (Spain), ASGC (Taiwan), RAL (UK) and BNL (USA), the Tier-2 facilities worldwide and large non-WLCG resource providers. Major contributors of computing resources are listed in Ref.~\cite{ATL-SOFT-PUB-2021-003}.
 

\clearpage
\appendix
\part*{Appendix}
\addcontentsline{toc}{part}{Appendix}
\FloatBarrier
\label{sec:app_postfit_plots}
 
Post-fit $m_{cc}$ distributions for 34 of the 44 analysis regions used in the
statistical analysis of the $VH, H\to \ccbar$~search are shown in
Figures~\ref{fig:postfit_0lep}~to~\ref{fig:top_CRs}. The 0-, 1- and
2-lepton signal regions are shown in
Figures~\ref{fig:postfit_0lep},~\ref{fig:postfit_1lep}
and~\ref{fig:postfit_2lep}, respectively, with the corresponding high-$\Delta R$
control regions shown in
Figures~\ref{fig:postfit_0lep_dR},~\ref{fig:postfit_1lep_dR}
and~\ref{fig:postfit_2lep_dR}. The 0- and 1-lepton top-quark control regions are
shown in Figure~\ref{fig:top_CRs}.
Figures~\ref{fig:background_comp_0lep},~\ref{fig:background_comp_1lep}
and~\ref{fig:background_comp_2lep} show the post-fit background composition,
including in control regions, for the 0-, 1- and 2-lepton channels, respectively.
 
\begin{figure*}
\centering
 
\includegraphics[width=0.47\textwidth]{fig_01a.pdf}
\includegraphics[width=0.47\textwidth]{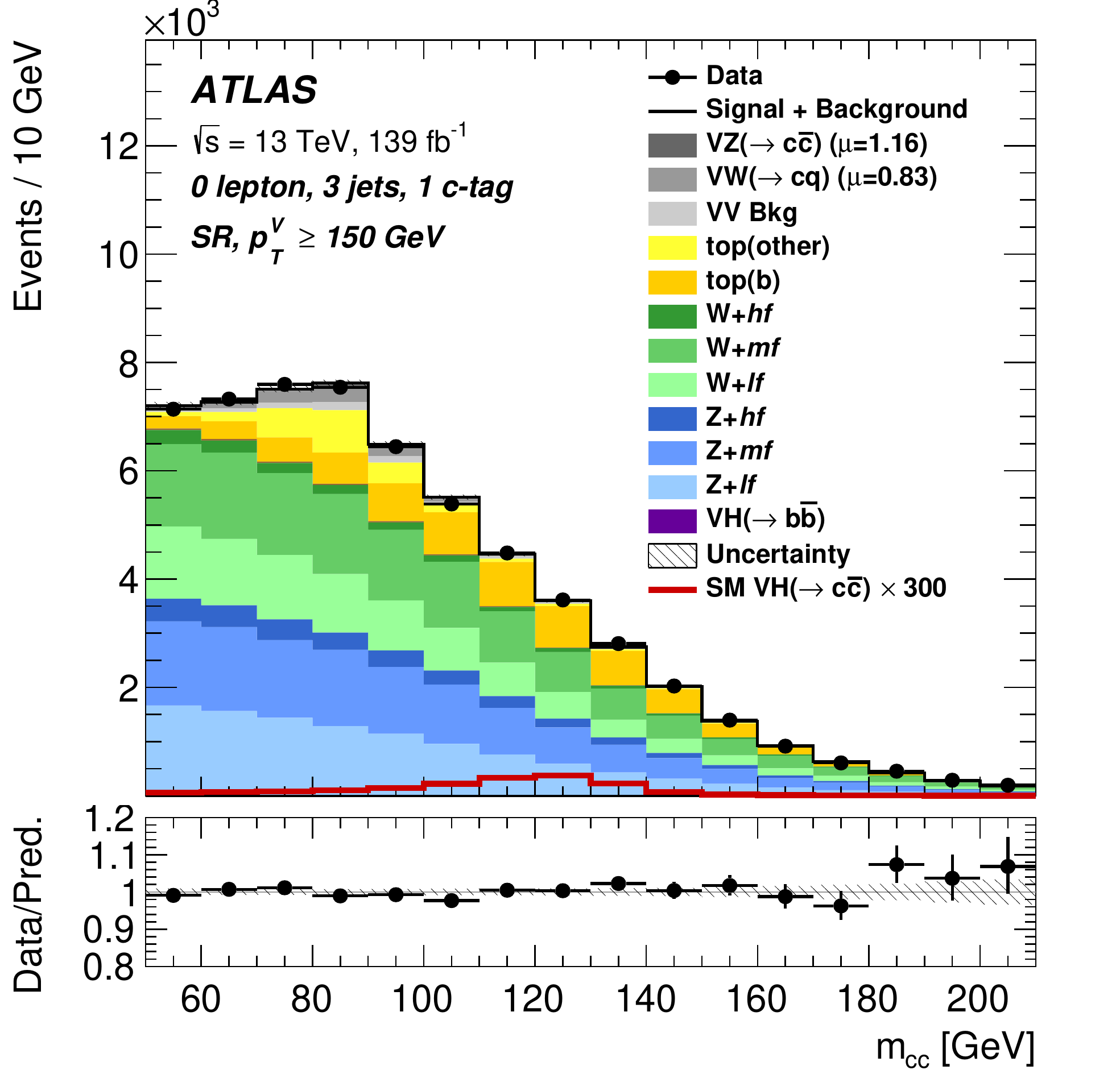} \\
\includegraphics[width=0.47\textwidth]{fig_01d.pdf}
\includegraphics[width=0.47\textwidth]{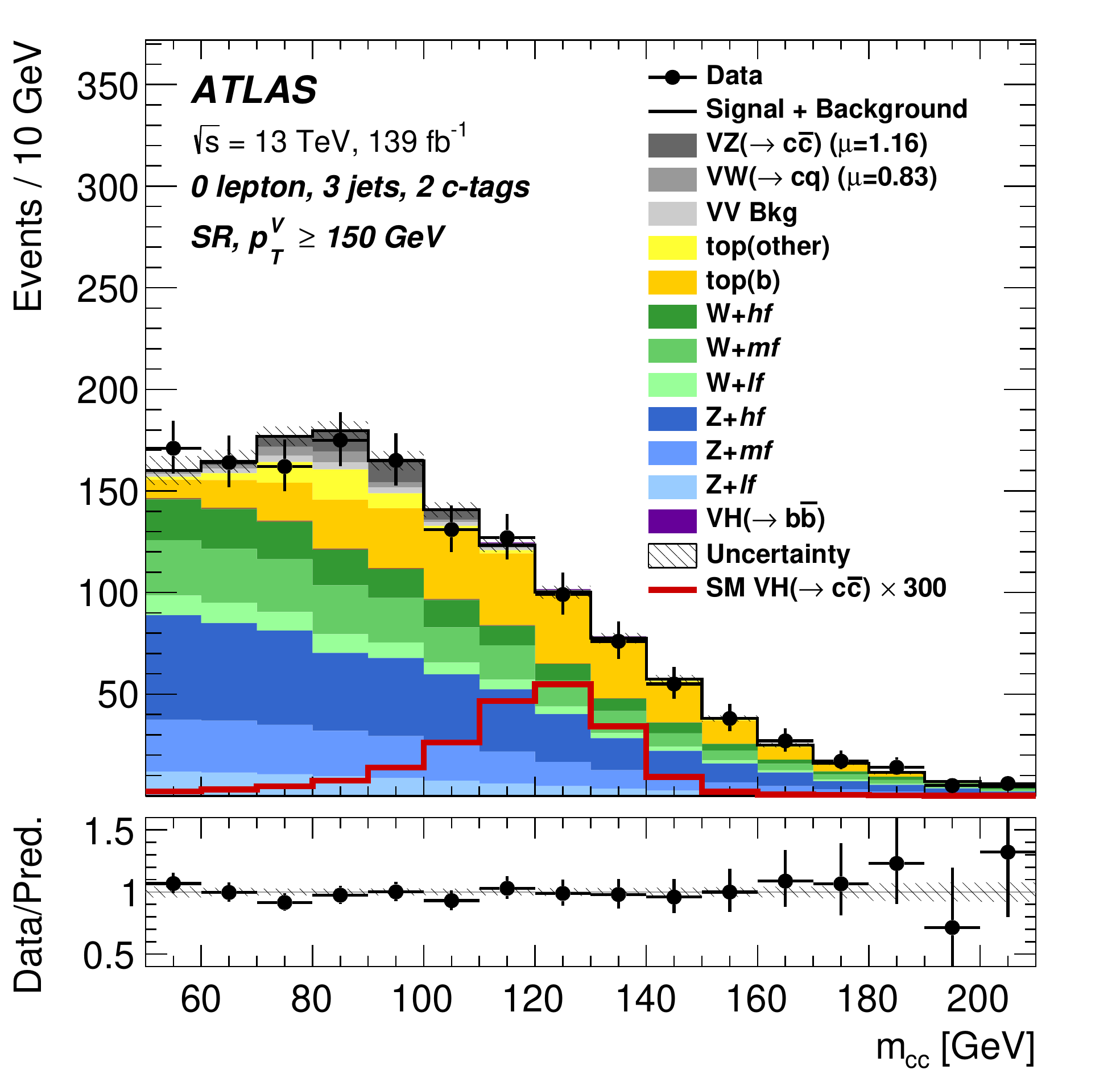}
\caption{Post-fit distributions of the four 0-lepton signal regions. The total
signal-plus-background prediction is shown by the solid black line and
includes the $H\to c\bar{c}$ signal scaled to the best-fit value of
$\mu_{VH(c\bar{c})} = -9$. The $H\to c\bar{c}$ signal is also shown as an
unfilled histogram scaled to 300 times the SM prediction. The post-fit
uncertainty is shown as the hatched background including correlations between
uncertainties. The ratio of the data to the sum of the post-fit signal plus
background is shown in the lower panel.}
\label{fig:postfit_0lep}
\end{figure*}
 
\begin{figure*}
\centering
\includegraphics[width=0.47\textwidth]{fig_01b.pdf}
\includegraphics[width=0.47\textwidth]{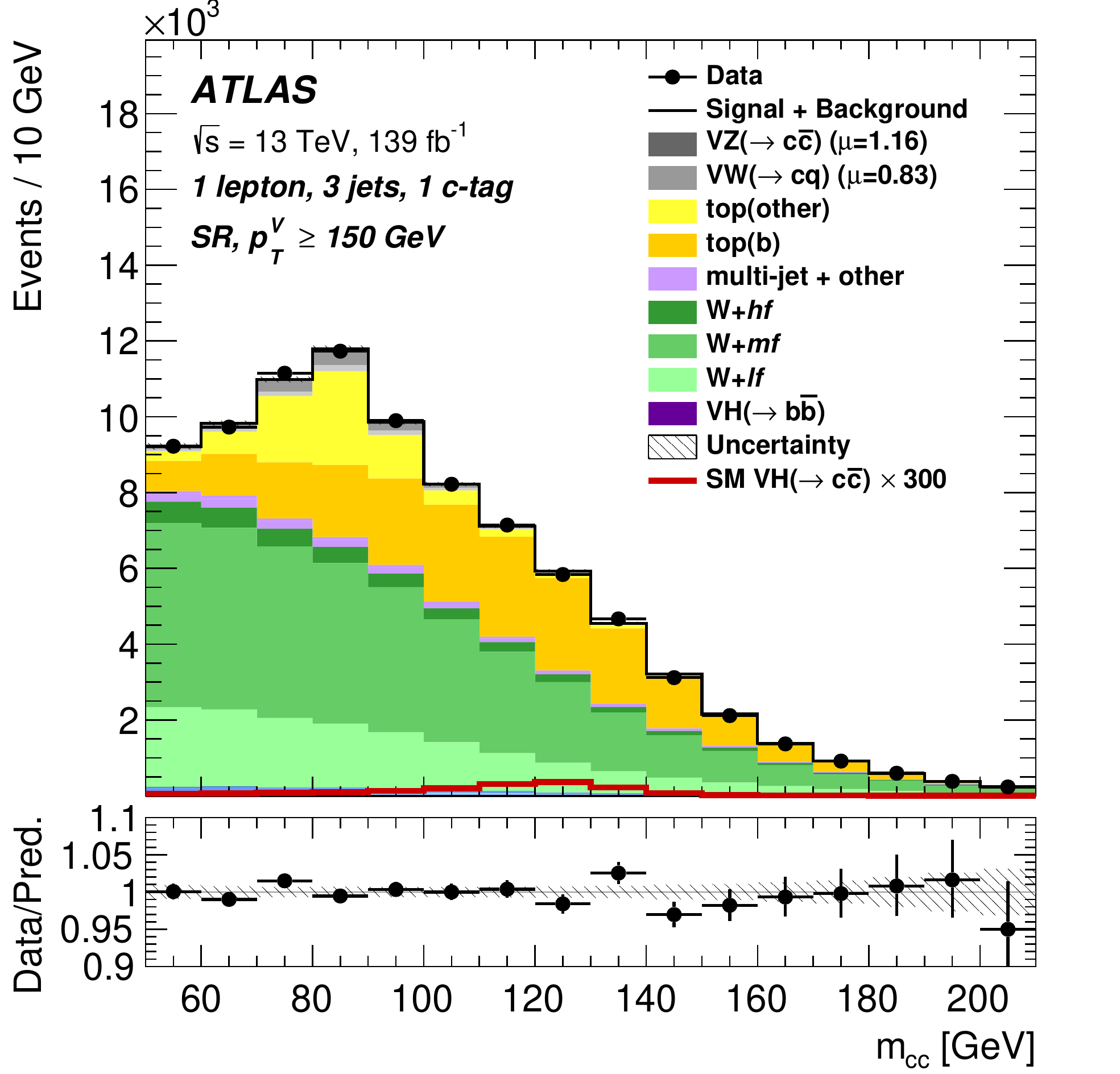} \\
\includegraphics[width=0.47\textwidth]{fig_01e.pdf}
\includegraphics[width=0.47\textwidth]{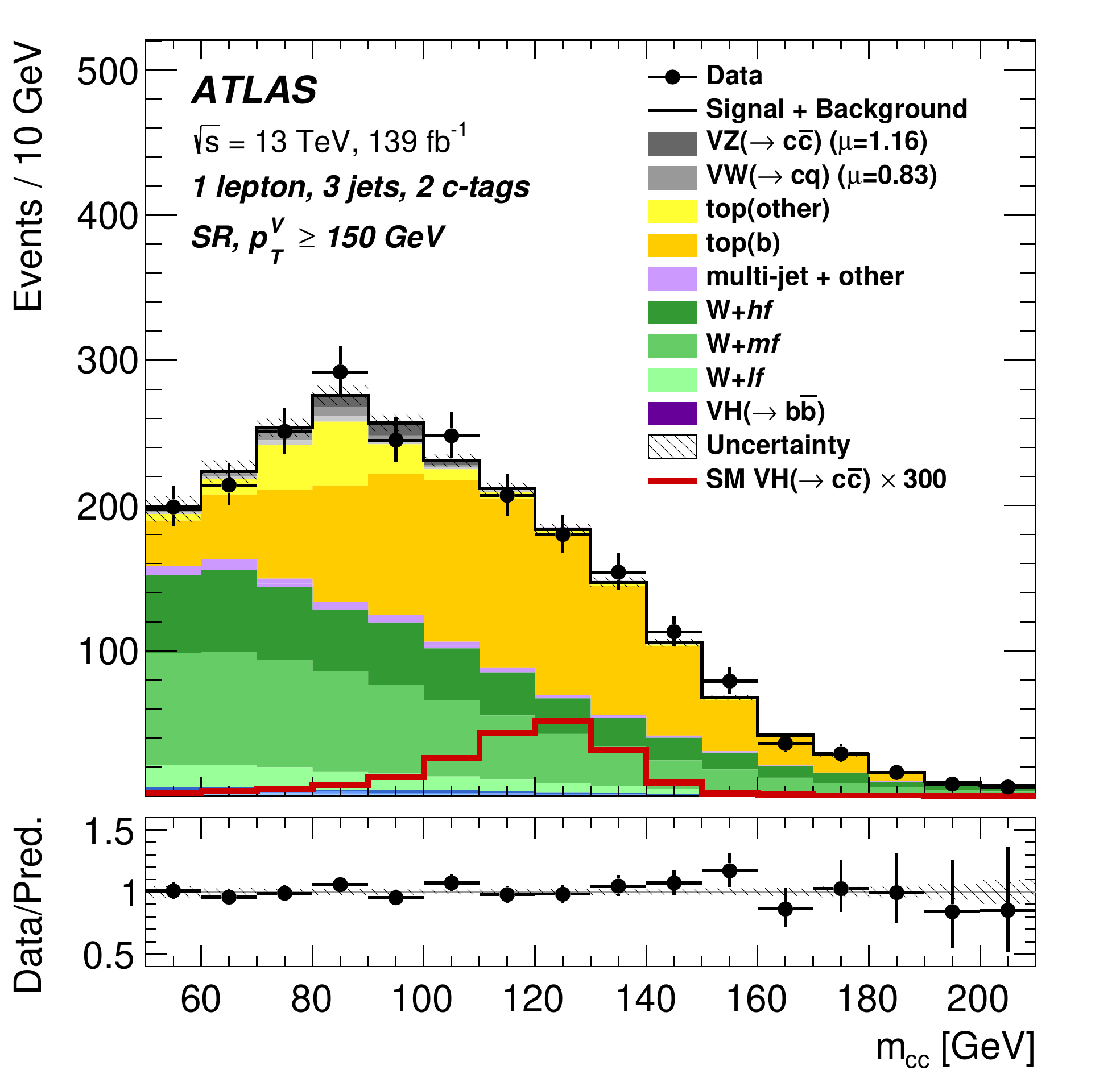}
\caption{Post-fit distributions of the four 1-lepton signal regions. The total
signal-plus-background prediction is shown by the solid black line and includes
the $H\to c\bar{c}$ signal scaled to the best-fit value of $\mu_{VH(c\bar{c})} =
-9$. The $H\to c\bar{c}$ signal is also shown as an unfilled histogram scaled to
300 times the SM prediction. The post-fit uncertainty is shown as the hatched
background including correlations between uncertainties. The ratio of the data
to the sum of the post-fit signal plus background is shown in the lower panel.}
\label{fig:postfit_1lep}
\end{figure*}
 
\begin{figure*}
\centering
\includegraphics[width=0.32\textwidth]{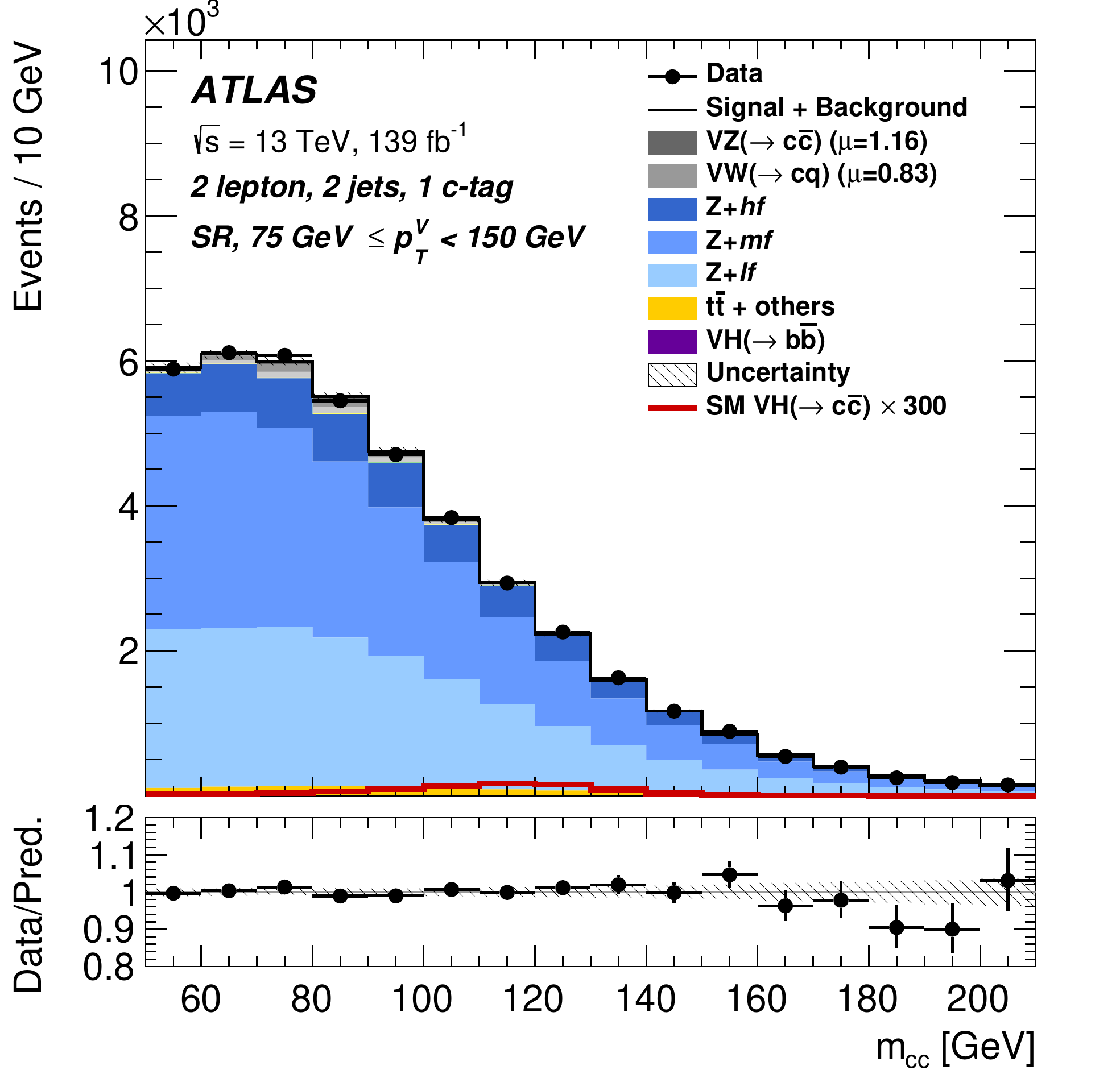}
\includegraphics[width=0.32\textwidth]{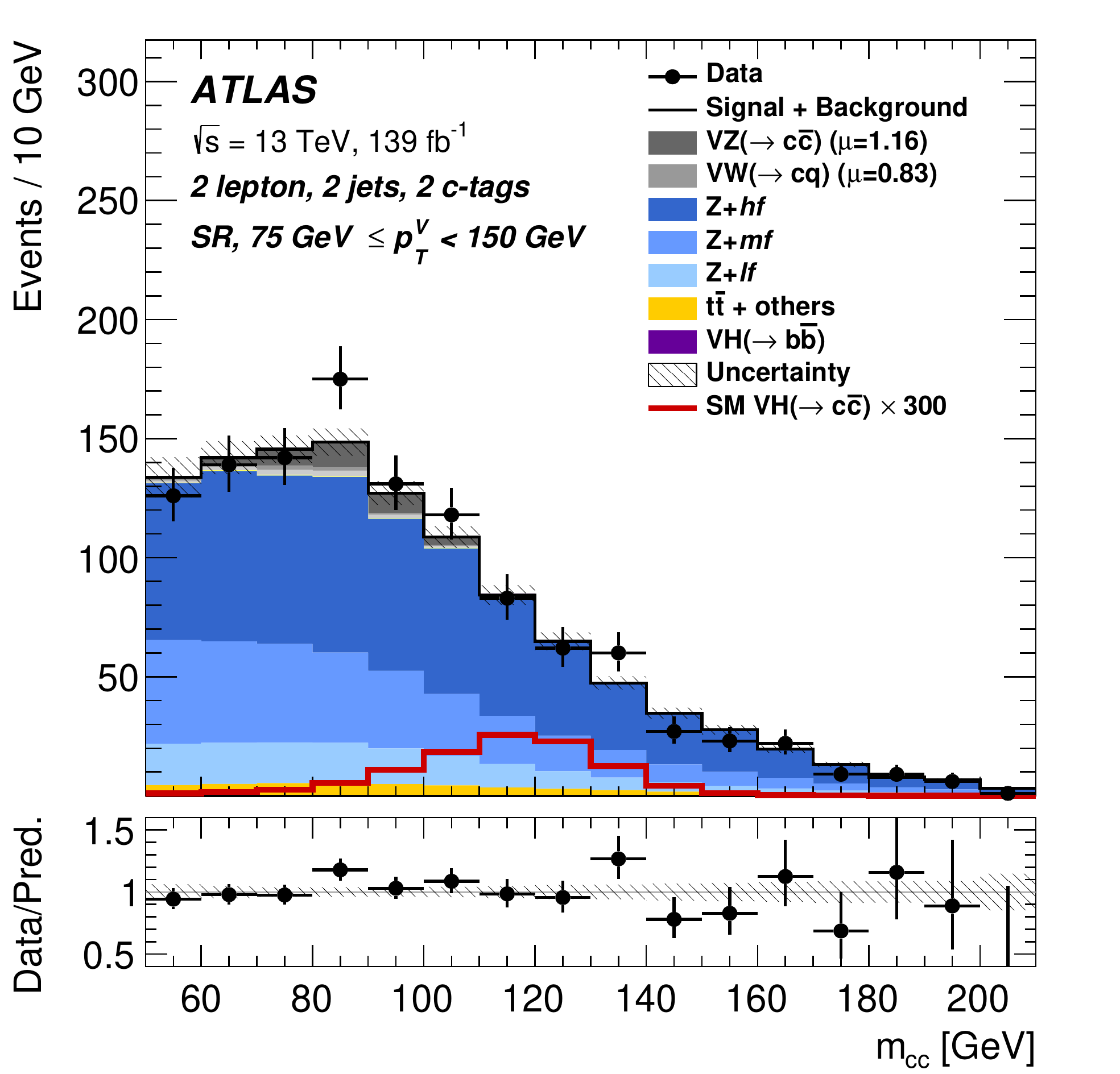}
\includegraphics[width=0.32\textwidth]{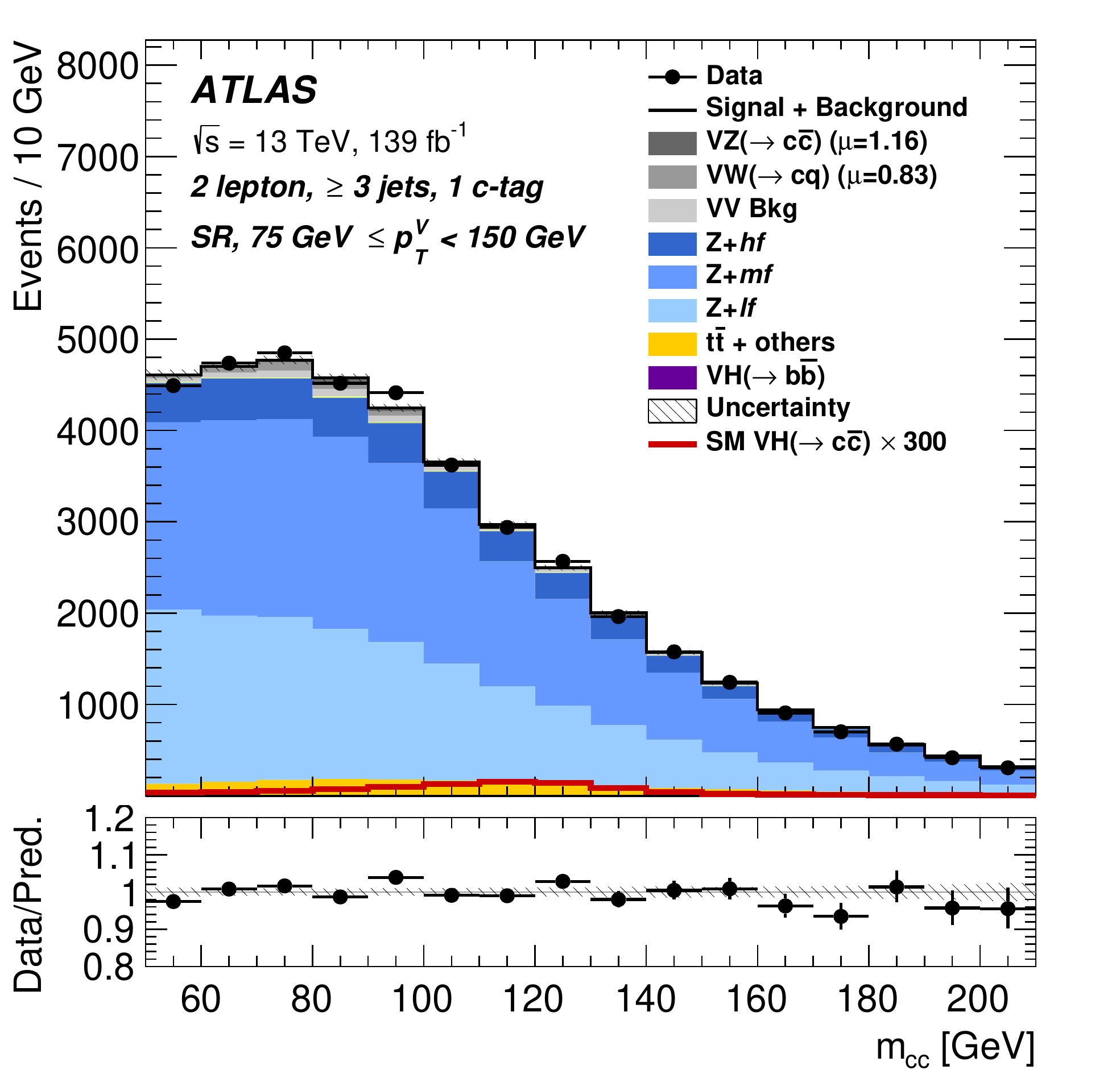}
\includegraphics[width=0.32\textwidth]{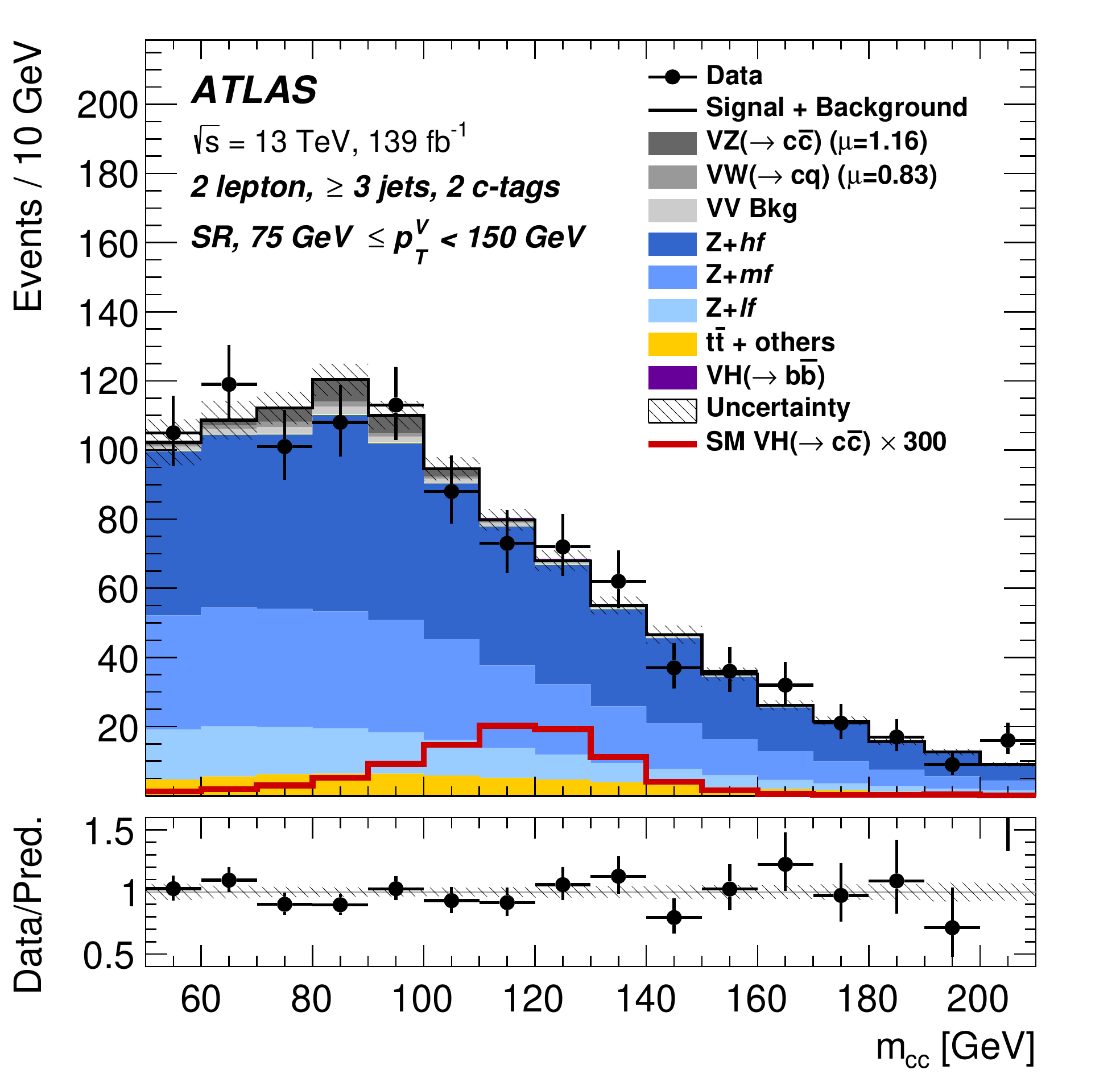}
\includegraphics[width=0.32\textwidth]{fig_01c.pdf}
\includegraphics[width=0.32\textwidth]{fig_01f.pdf}
\includegraphics[width=0.32\textwidth]{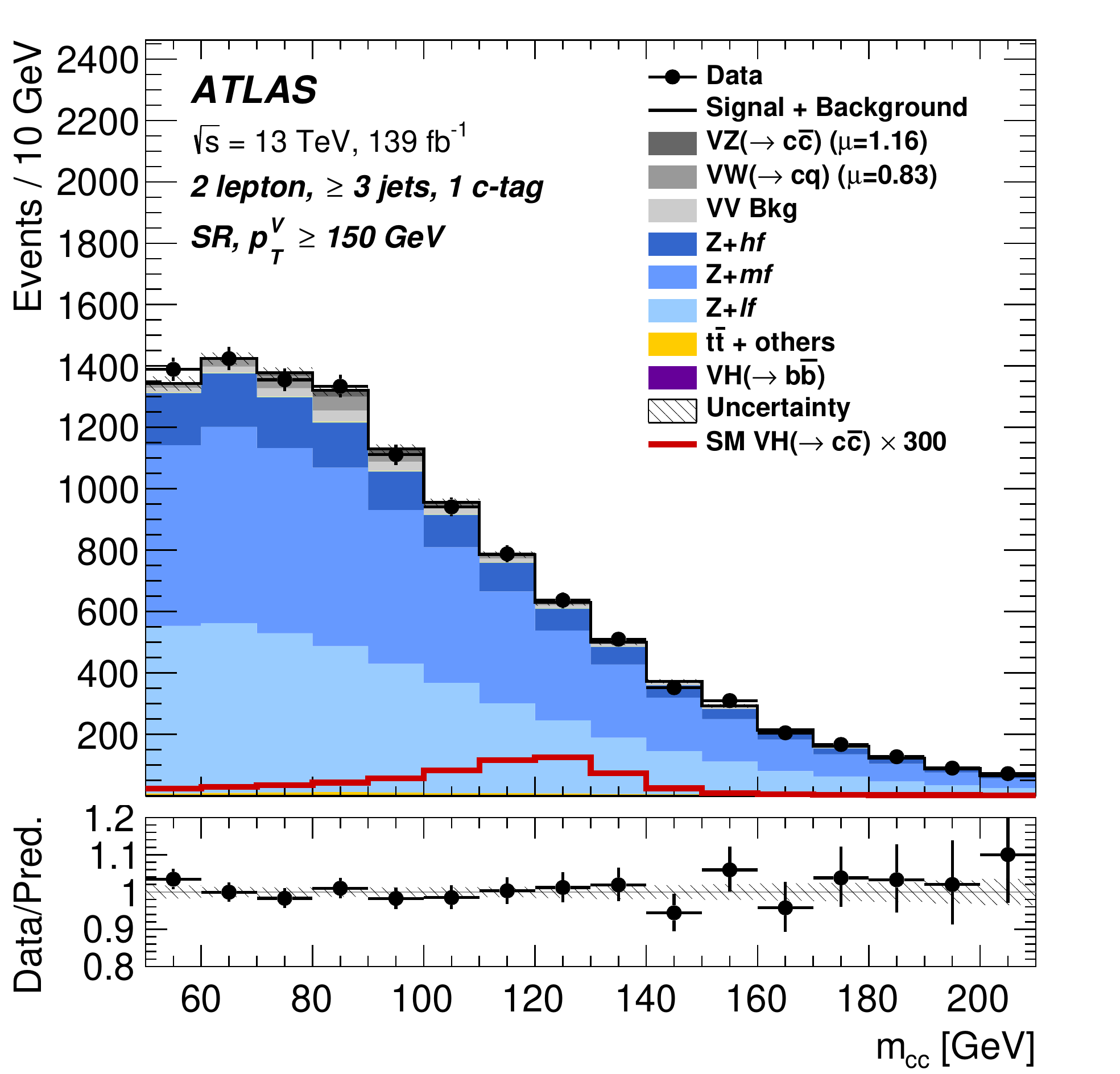}
\includegraphics[width=0.32\textwidth]{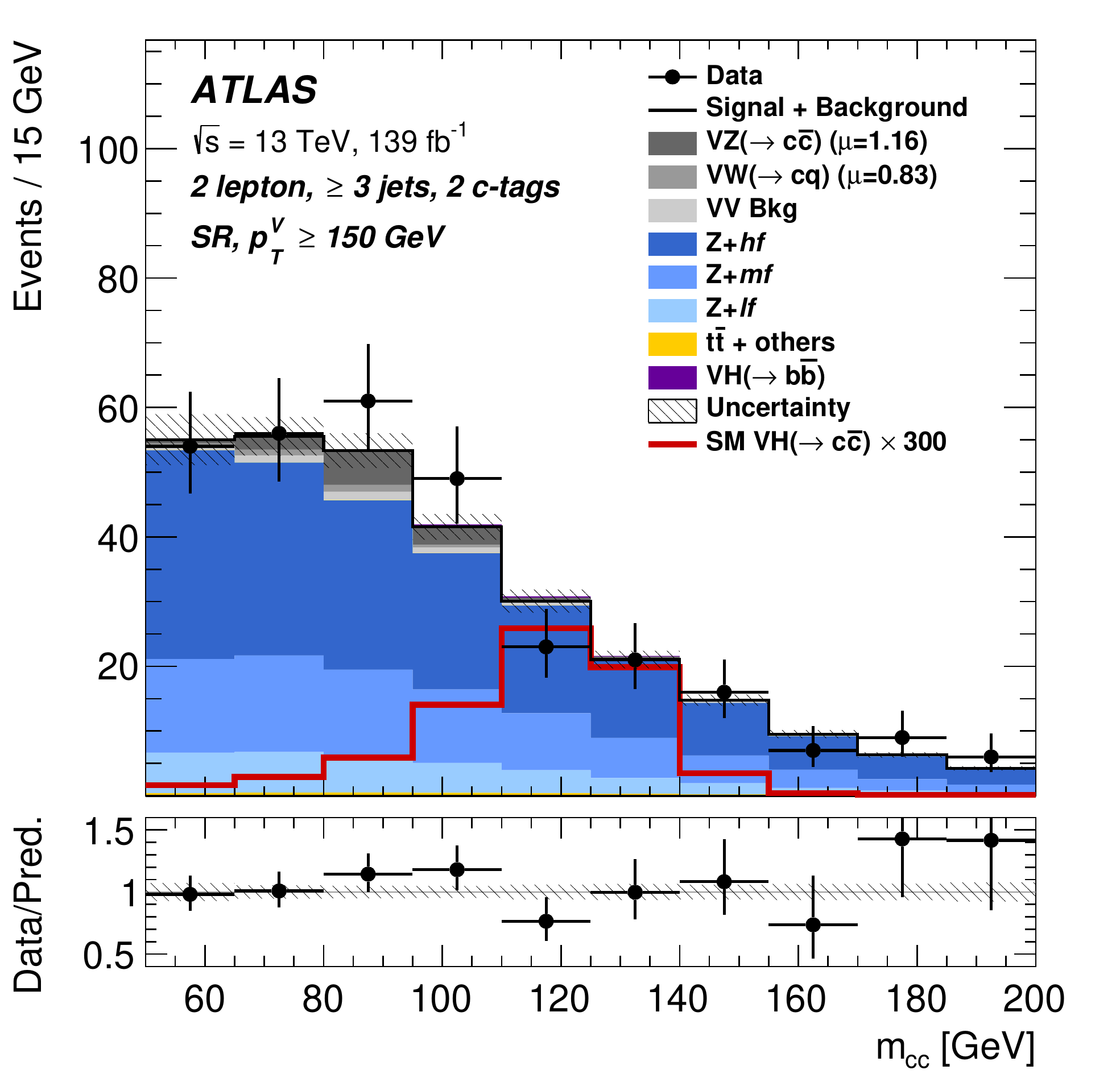}
\caption{Post-fit distributions of the eight 2-lepton signal regions. The total
signal-plus-background prediction is shown by the solid black line and includes
the $H\to c\bar{c}$ signal scaled to the best-fit value of $\mu_{VH(c\bar{c})} =
-9$. The $H\to c\bar{c}$ signal is also shown as an unfilled histogram scaled to
300 times the SM prediction. The post-fit uncertainty is shown as the hatched
background including correlations between uncertainties. The ratio of the
data to the sum of the post-fit signal plus background is shown in the lower
panel.}
\label{fig:postfit_2lep}
\end{figure*}

\begin{figure*}
\centering
 
\includegraphics[width=0.47\textwidth]{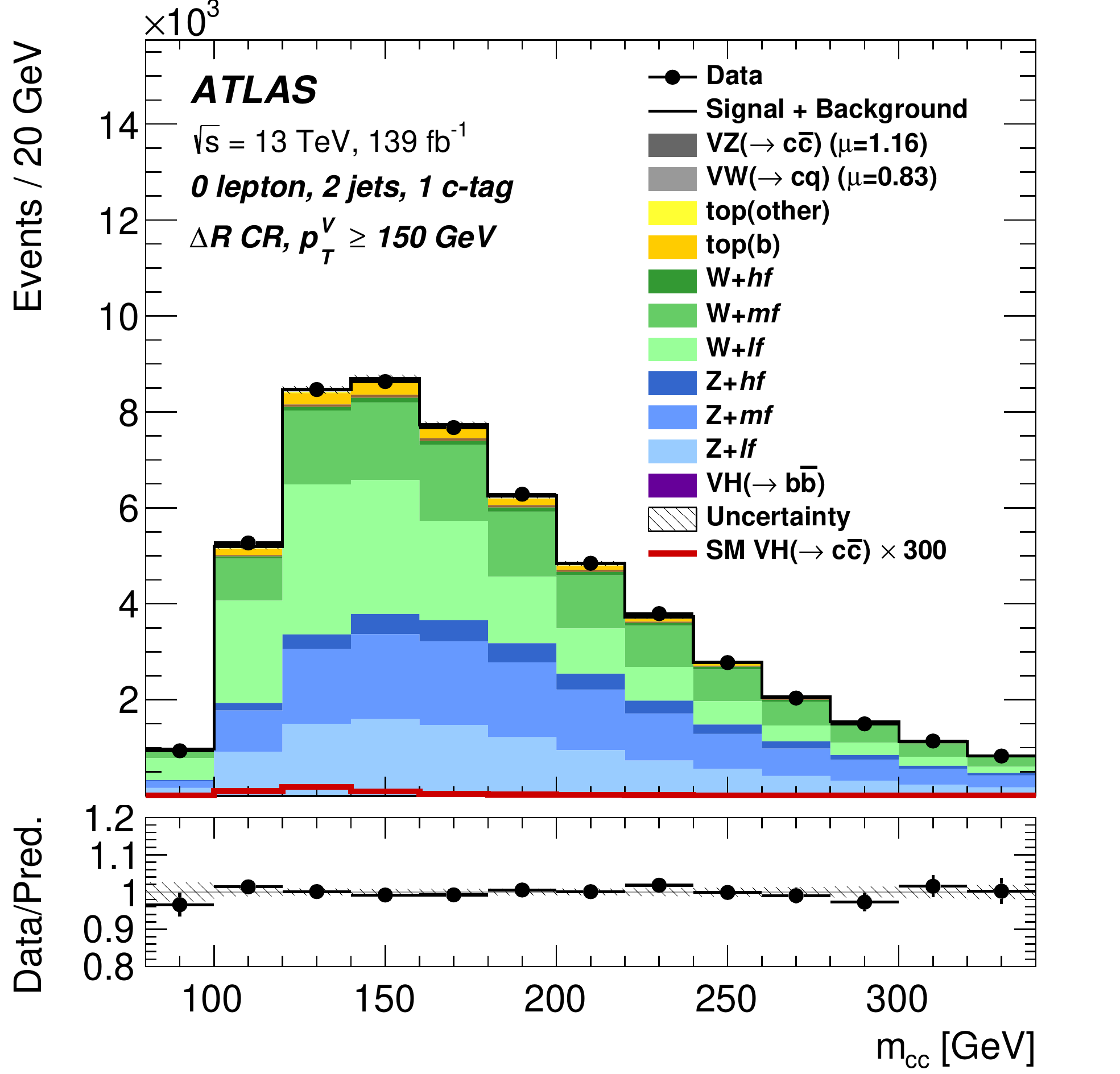}
\includegraphics[width=0.47\textwidth]{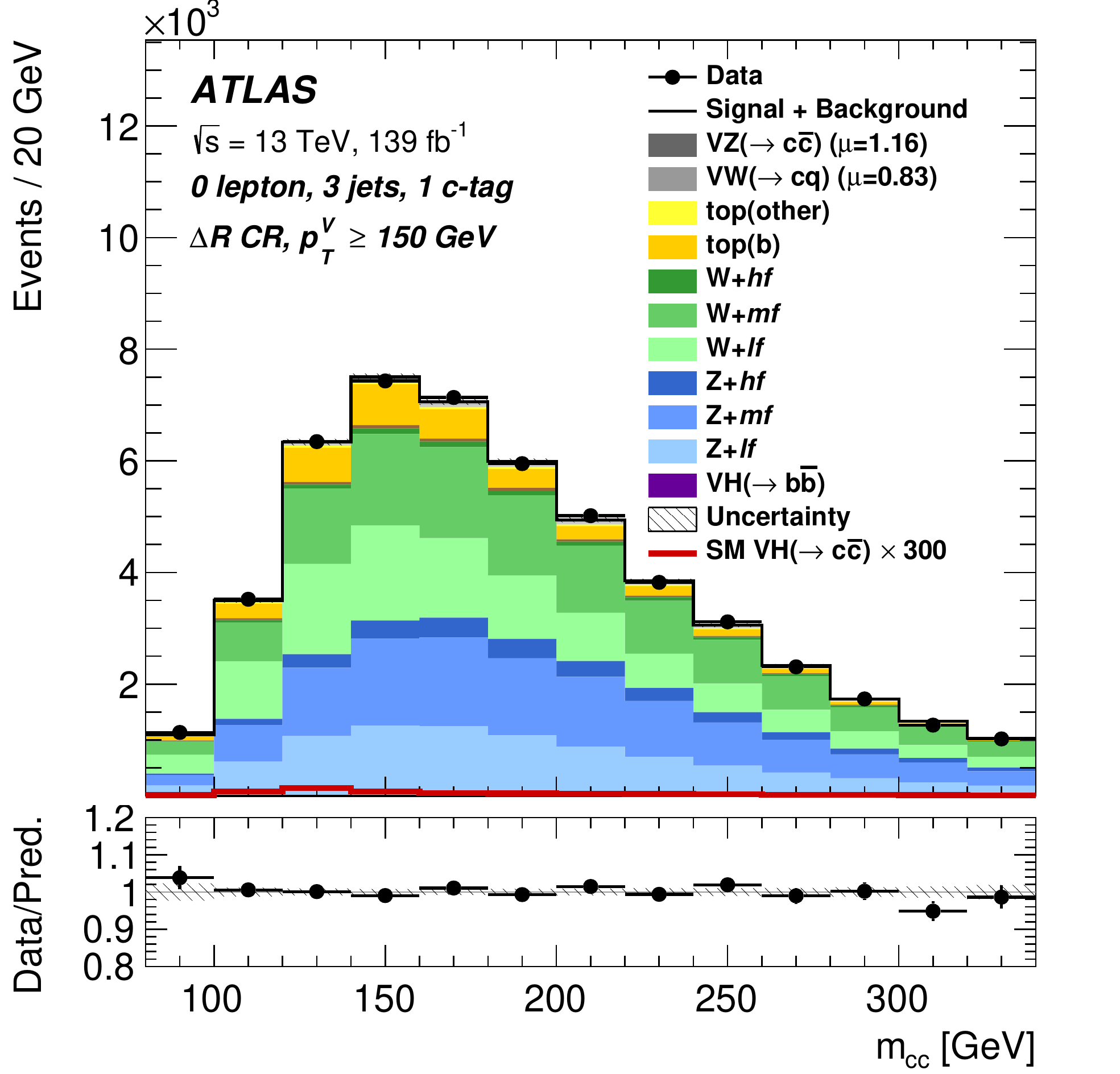} \\
\includegraphics[width=0.47\textwidth]{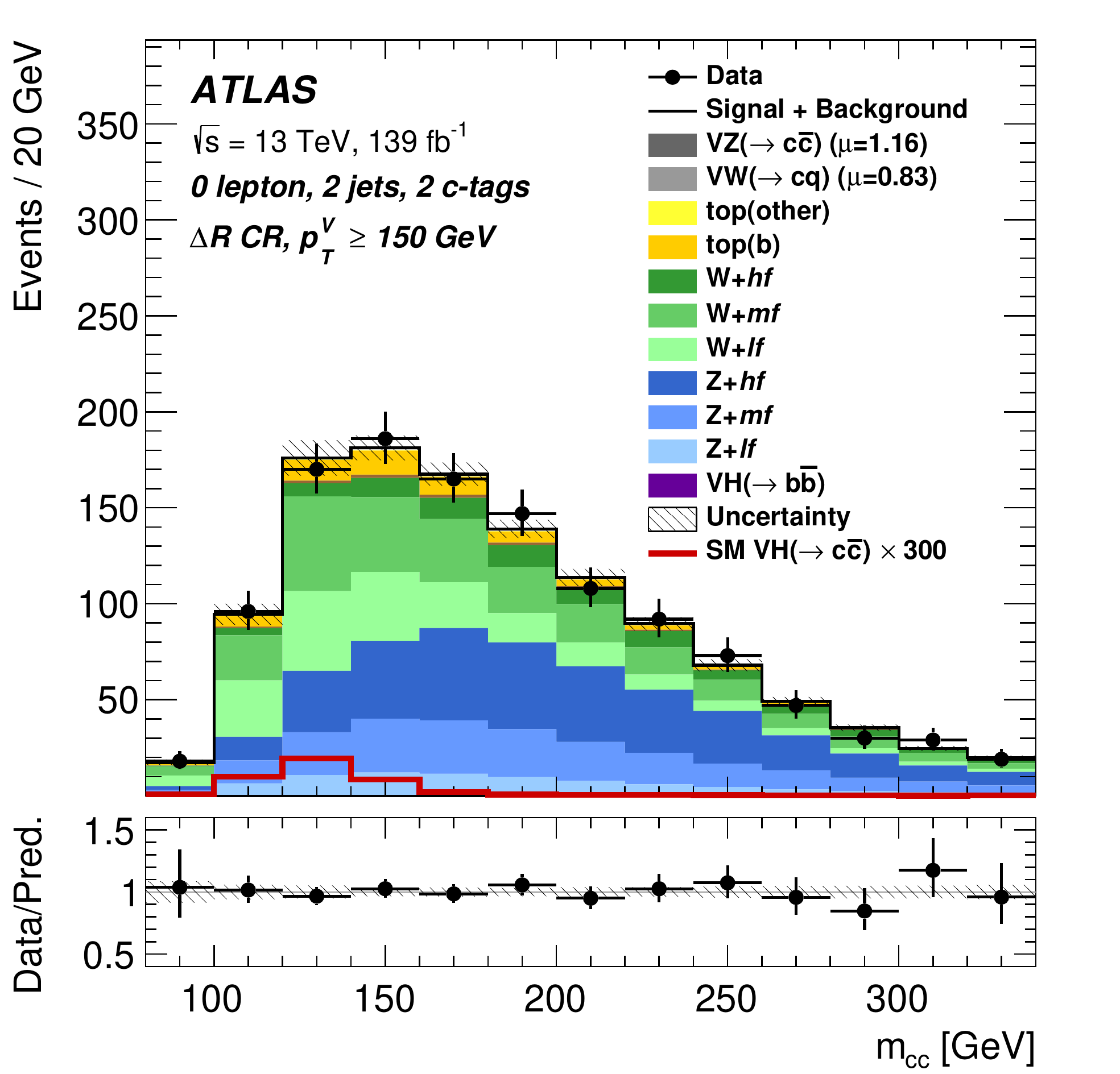}
\includegraphics[width=0.47\textwidth]{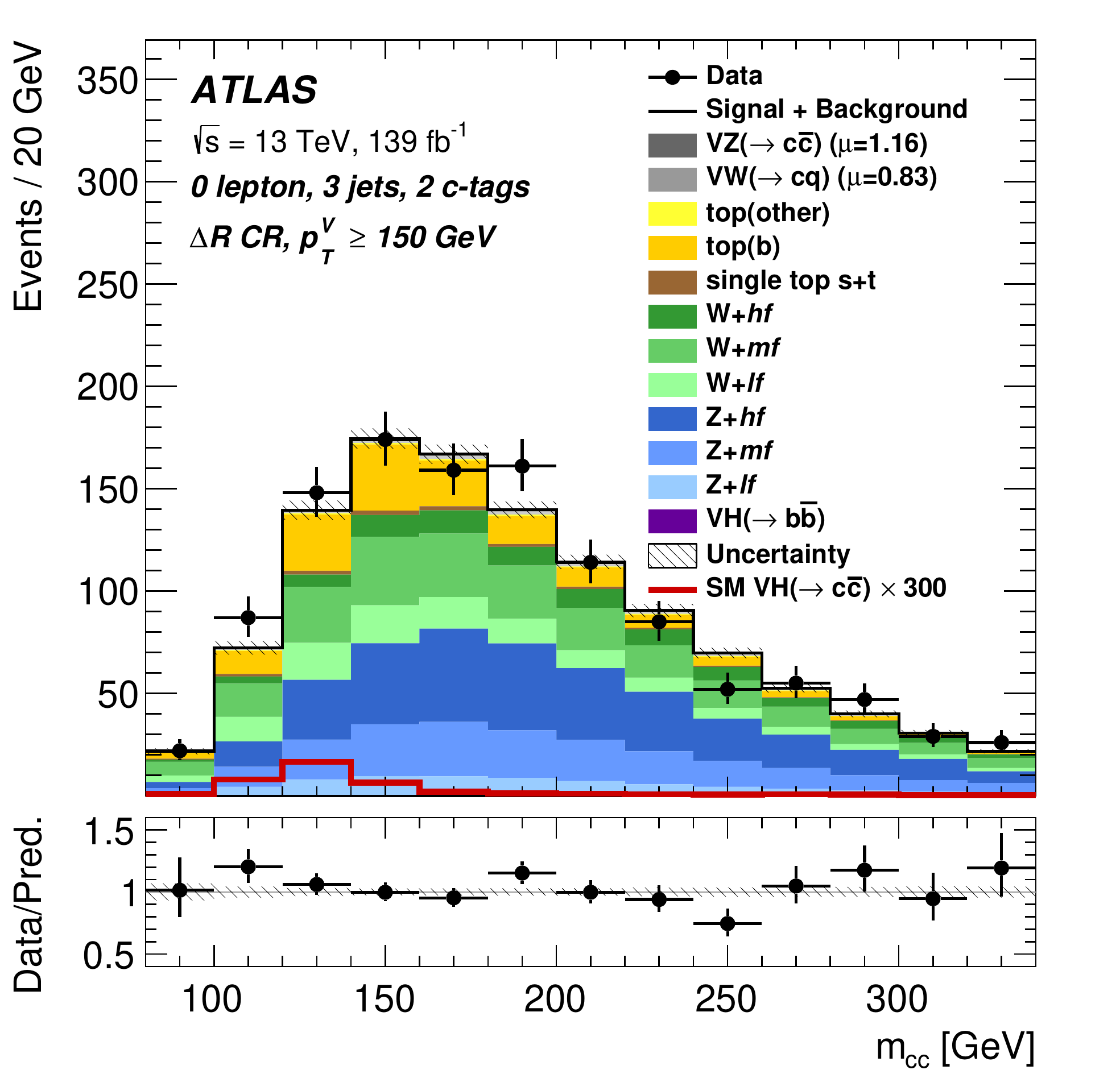}
\caption{Post-fit distributions of the four 0-lepton high-$\Delta R$ control
regions. The total signal-plus-background prediction is shown by the solid
black line and includes the $H\to c\bar{c}$ signal scaled to the best-fit
value of $\mu_{VH(c\bar{c})} = -9$. The $H\to c\bar{c}$ signal is also shown
as an unfilled histogram scaled to 300 times the SM prediction. The post-fit
uncertainty is shown as the hatched background including correlations
between uncertainties. The ratio of the data to the sum of the post-fit
signal plus background is shown in the lower panel.}
\label{fig:postfit_0lep_dR}
\end{figure*}
 
\begin{figure*}
\centering
\includegraphics[width=0.47\textwidth]{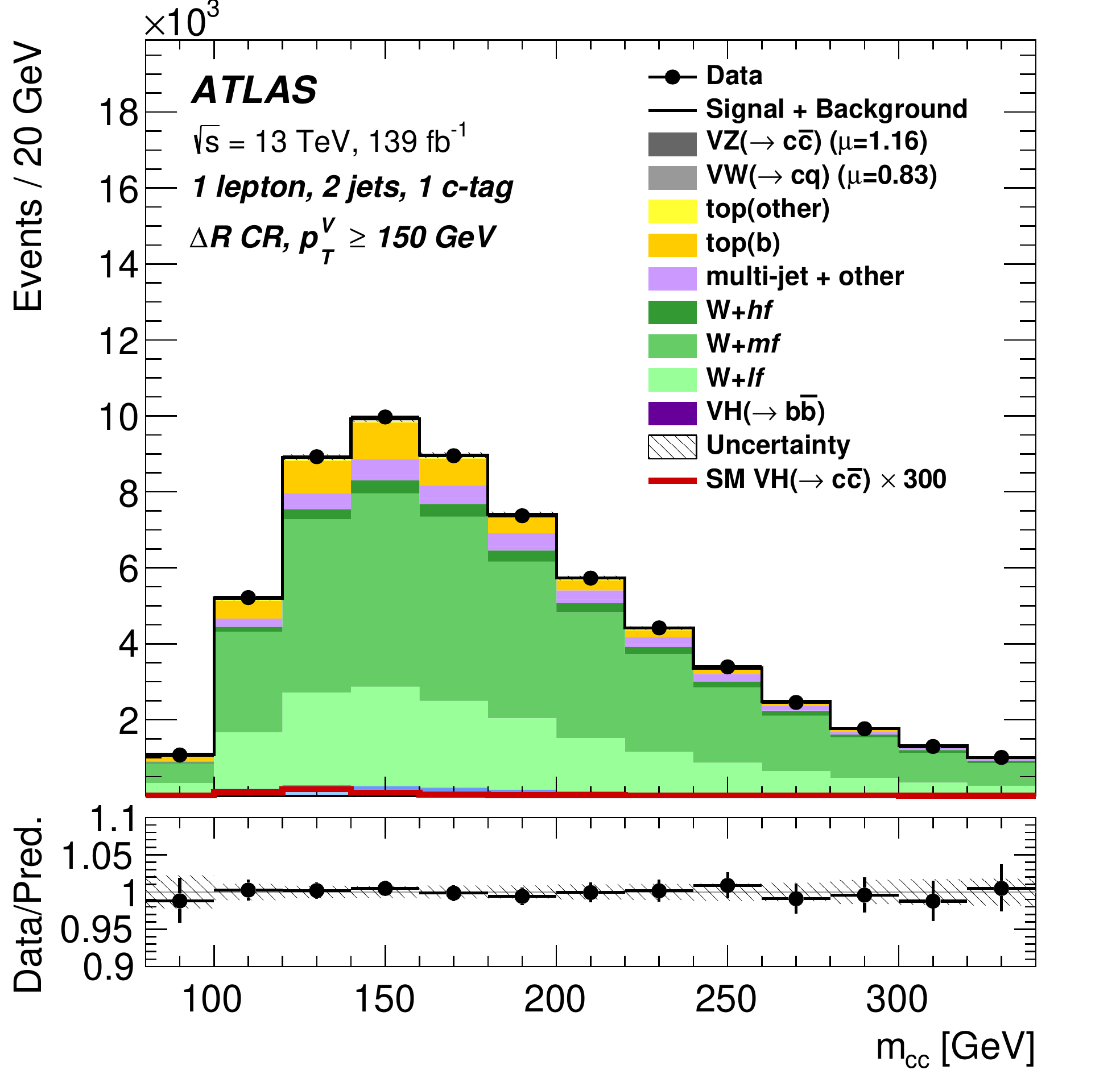}
\includegraphics[width=0.47\textwidth]{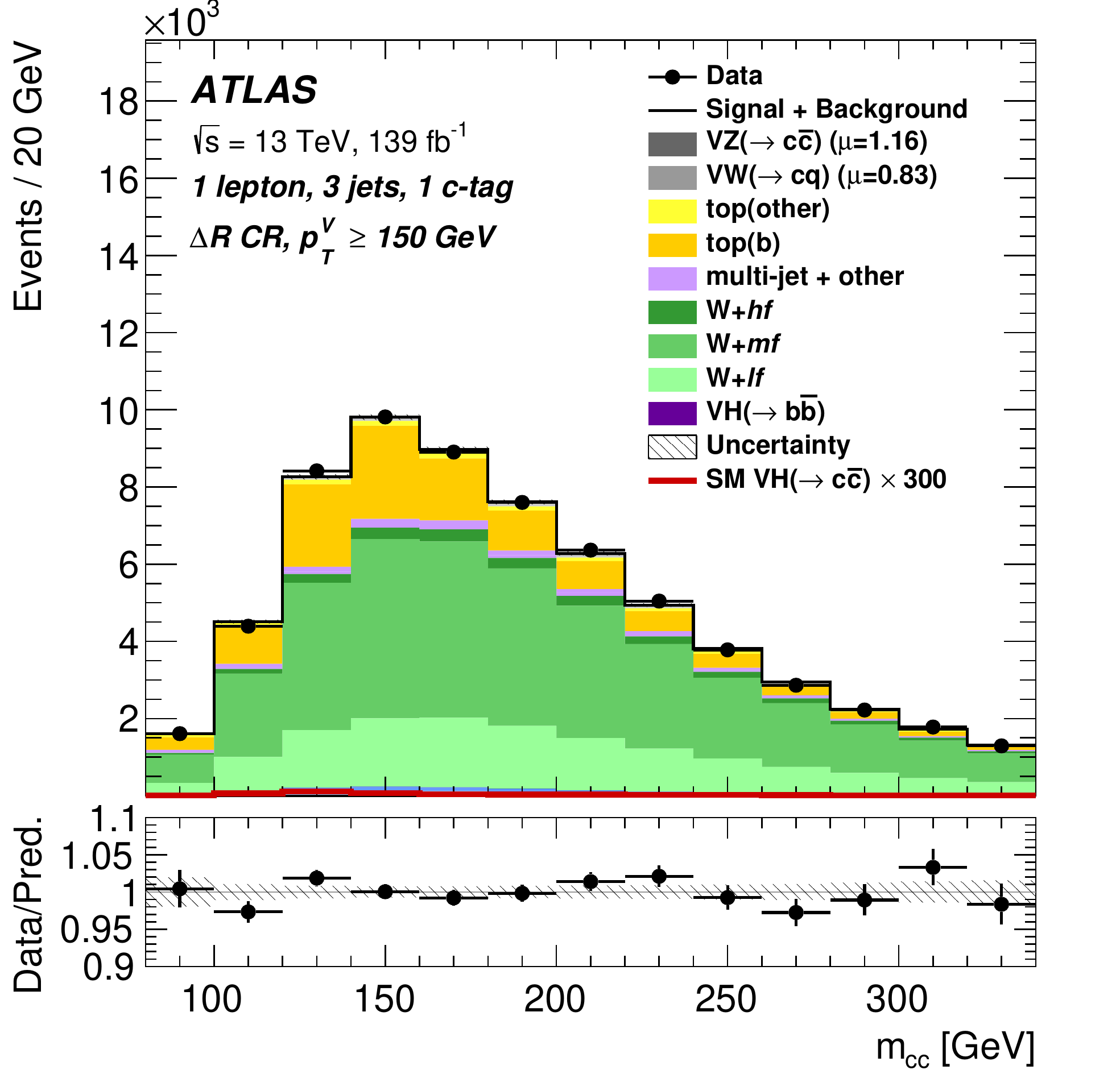} \\
\includegraphics[width=0.47\textwidth]{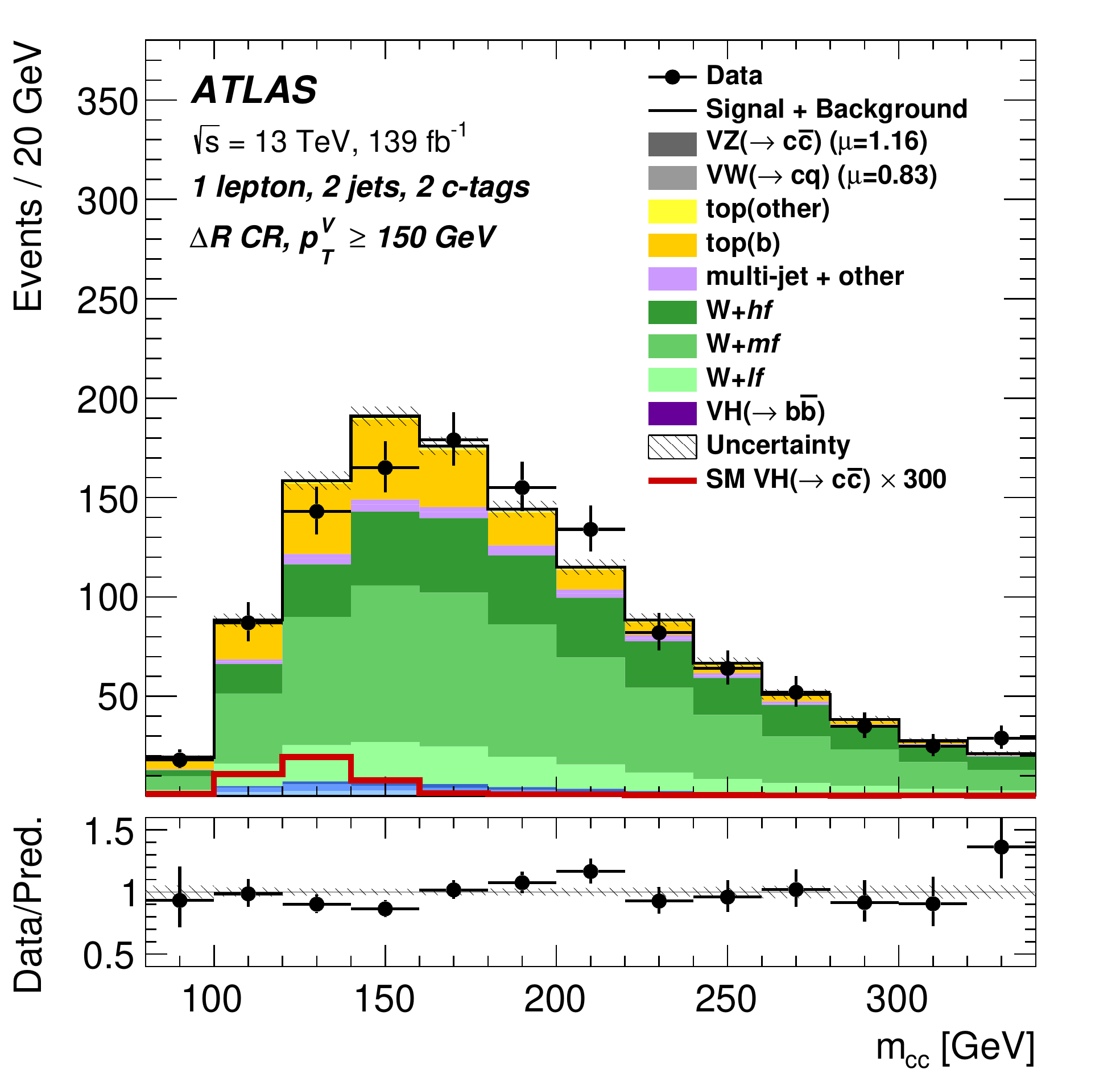}
\includegraphics[width=0.47\textwidth]{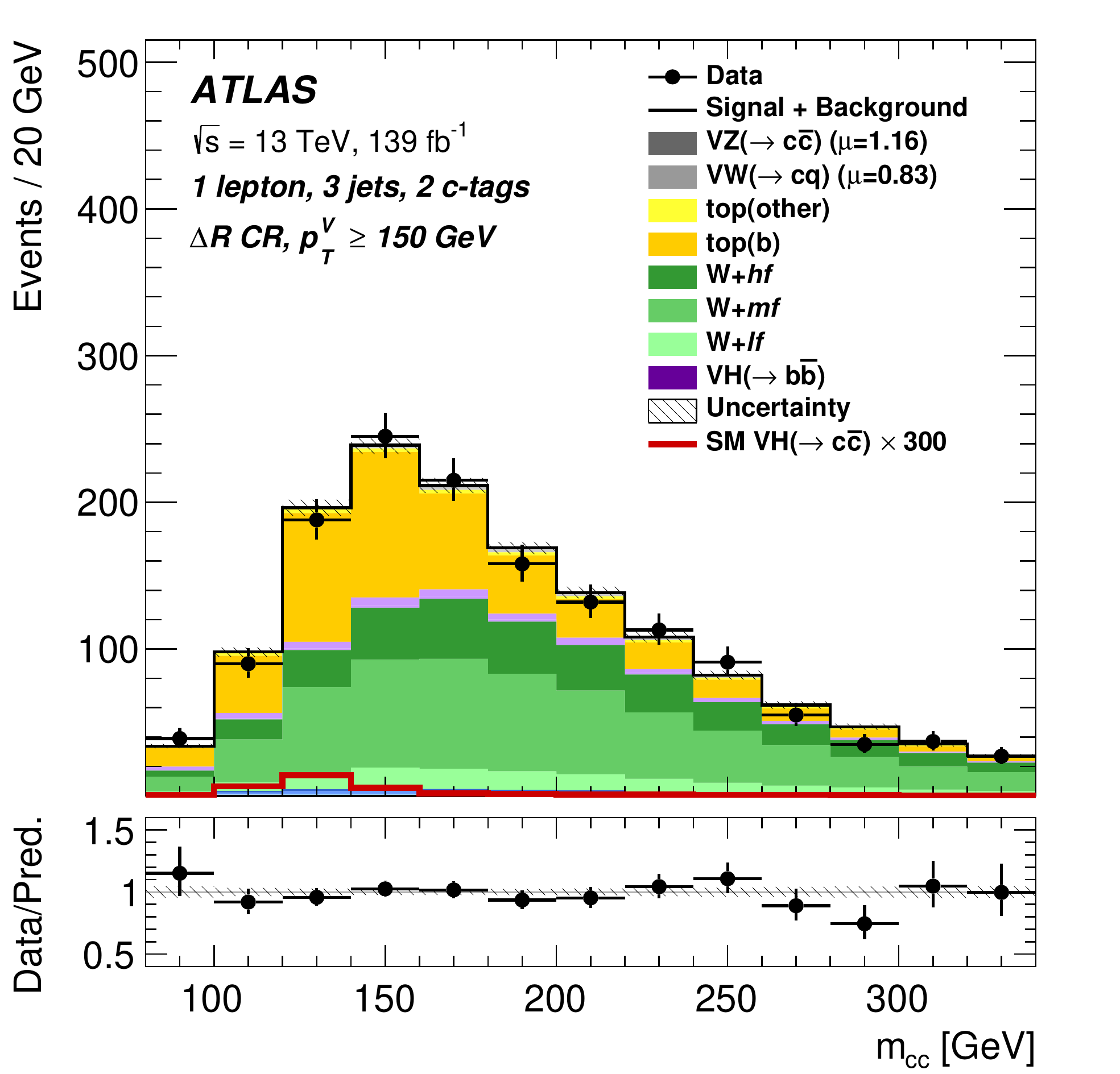}
\caption{Post-fit distributions of the four 1-lepton high-$\Delta R$ control
regions. The total signal-plus-background prediction is shown by the solid
black line and includes the $H\to c\bar{c}$ signal scaled to the best-fit
value of $\mu_{VH(c\bar{c})} = -9$. The $H\to c\bar{c}$ signal is also shown
as an unfilled histogram scaled to 300 times the SM prediction. The post-fit
uncertainty is shown as the hatched background including correlations between
uncertainties. The ratio of the data to the sum of the post-fit signal plus
background is shown in the lower panel.}
\label{fig:postfit_1lep_dR}
\end{figure*}
 
\begin{figure*}
\centering
\includegraphics[width=0.32\textwidth]{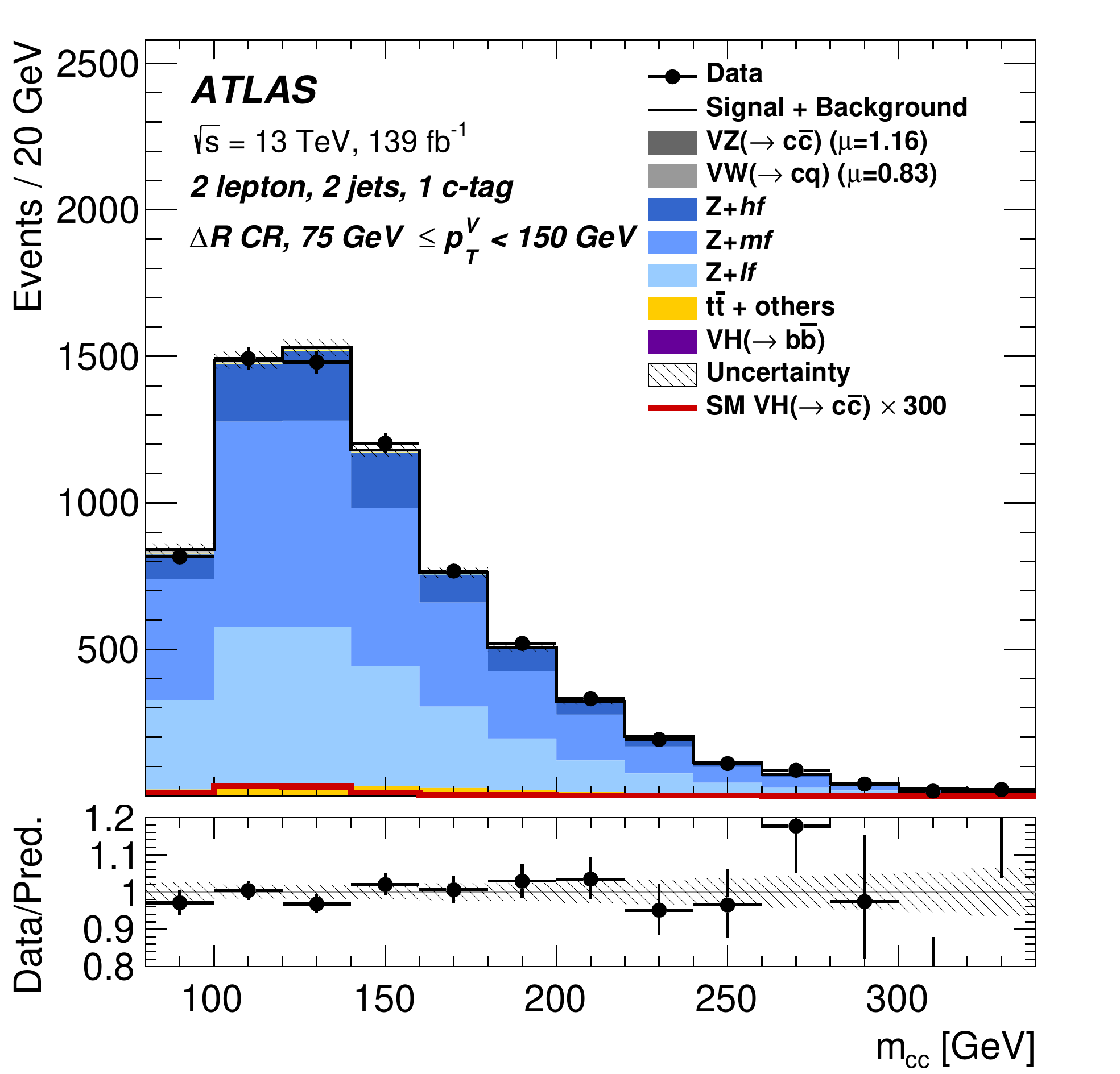}
\includegraphics[width=0.32\textwidth]{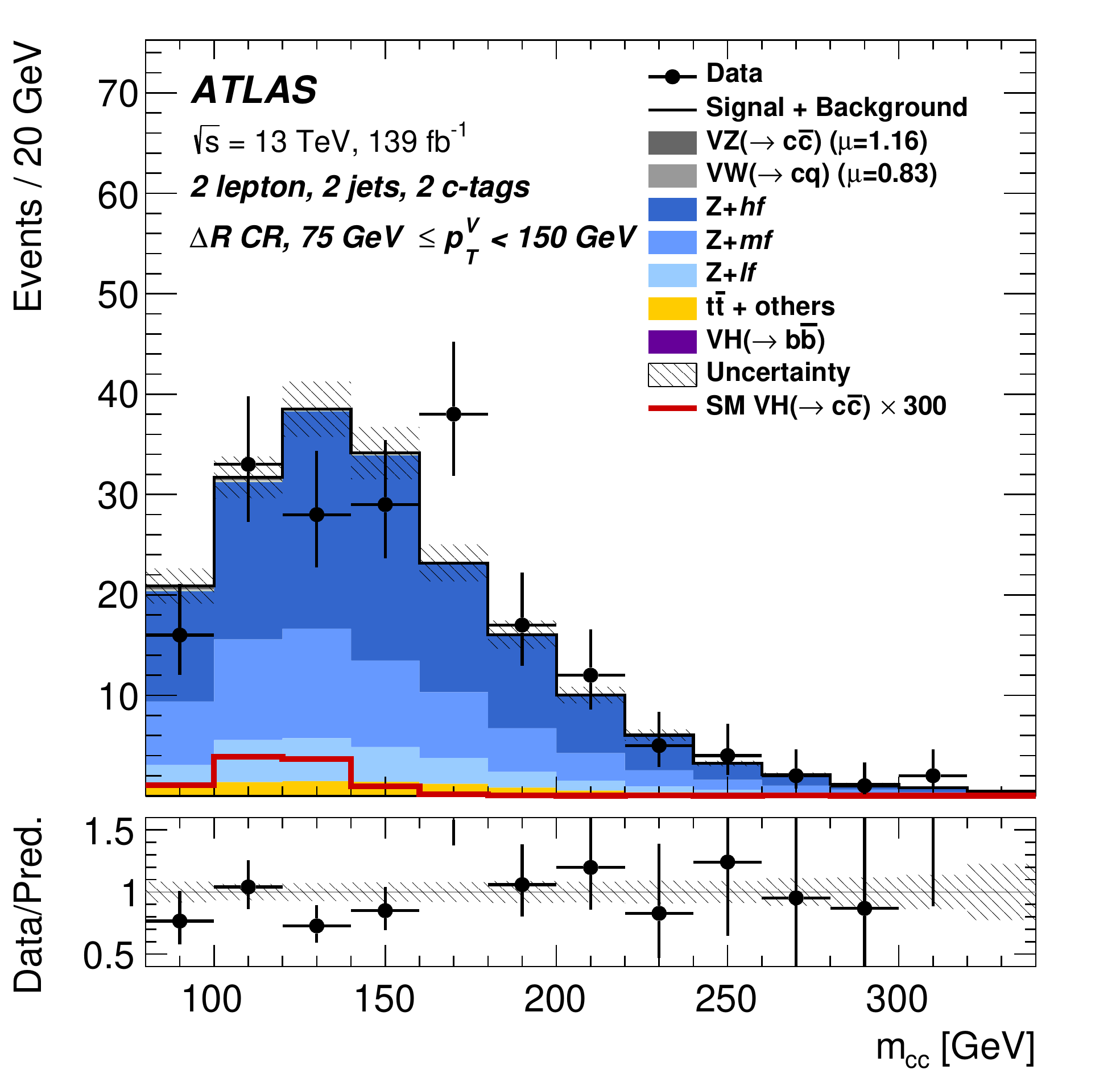}
\includegraphics[width=0.32\textwidth]{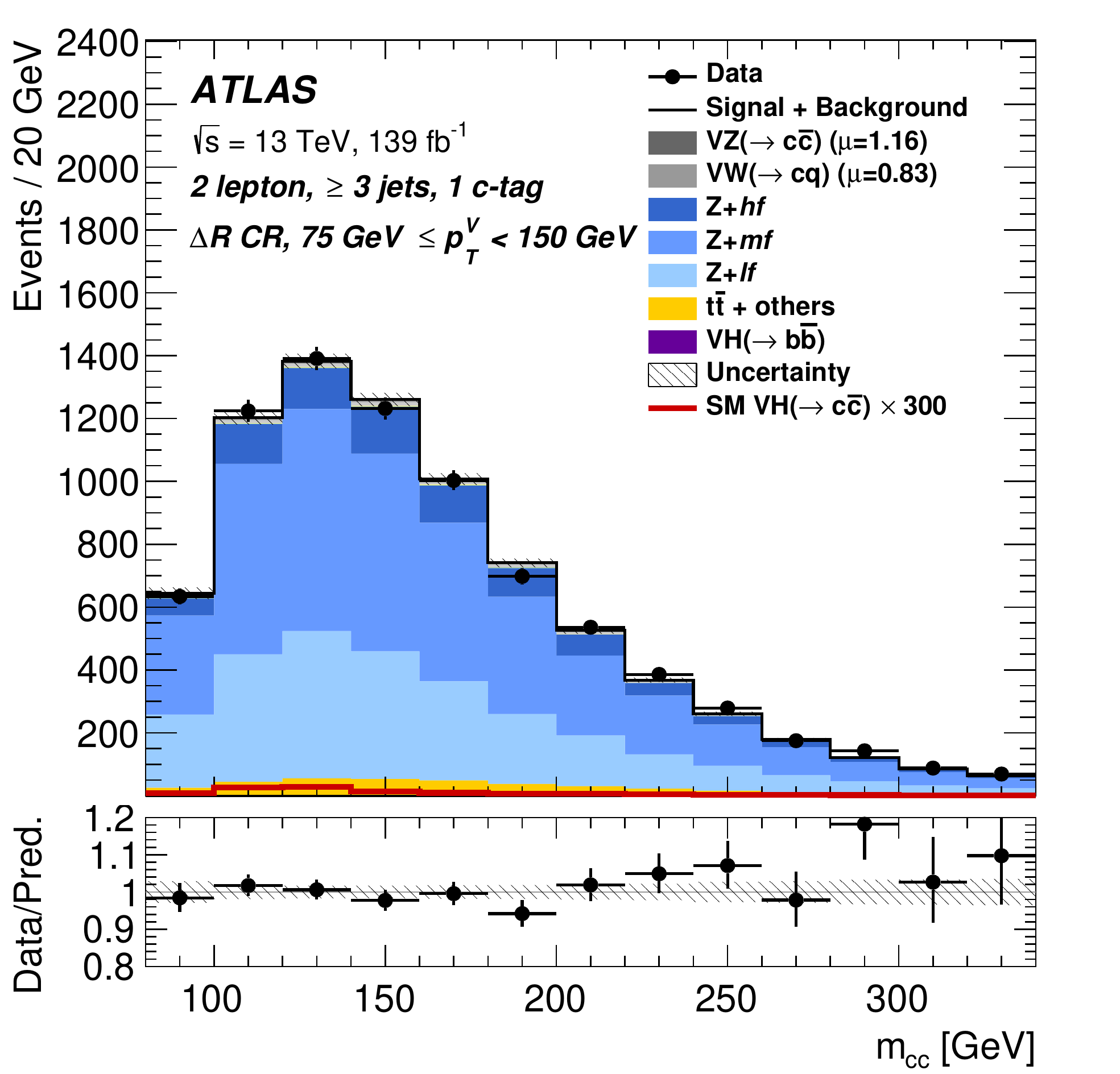}
\includegraphics[width=0.32\textwidth]{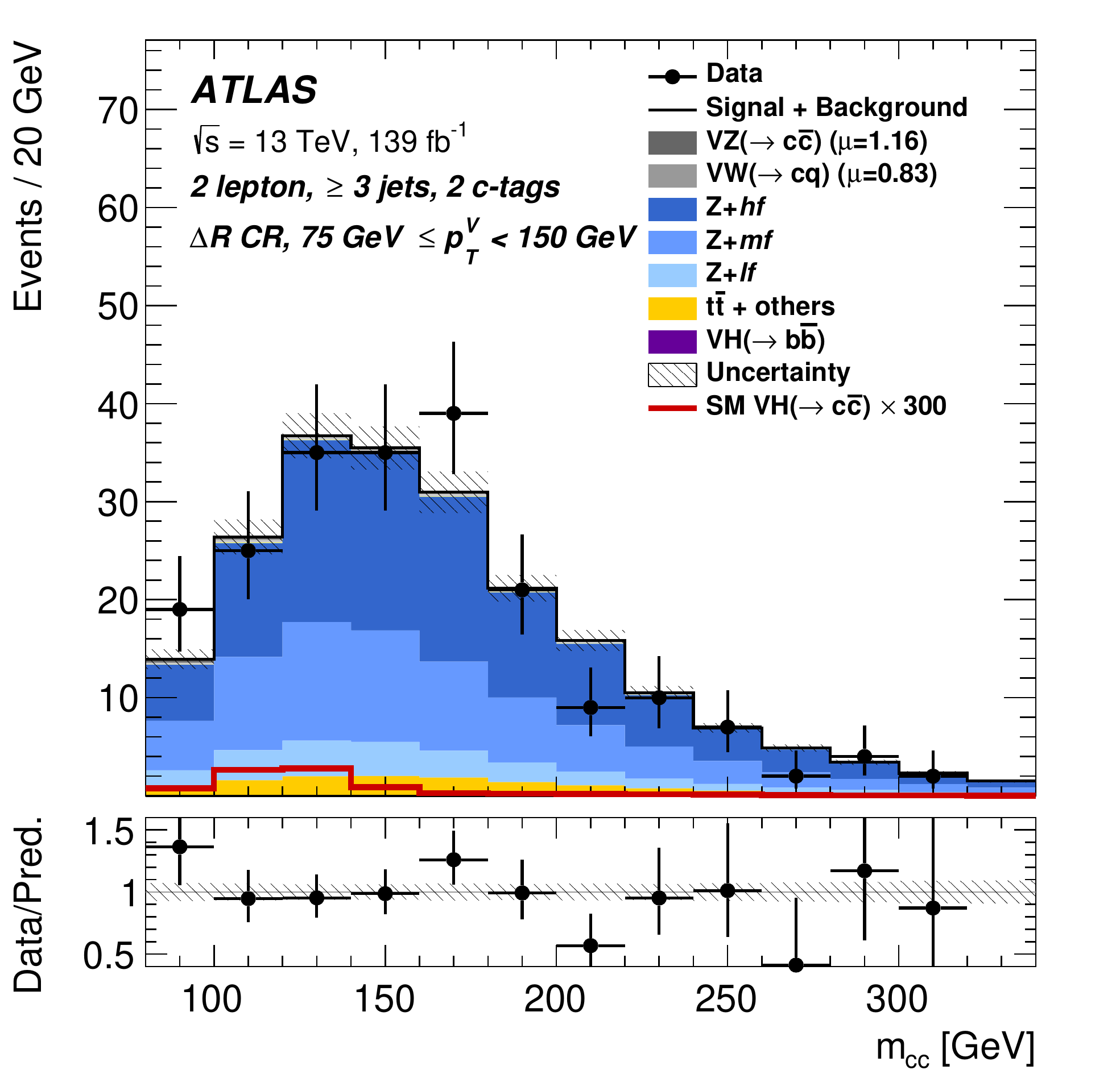}
\includegraphics[width=0.32\textwidth]{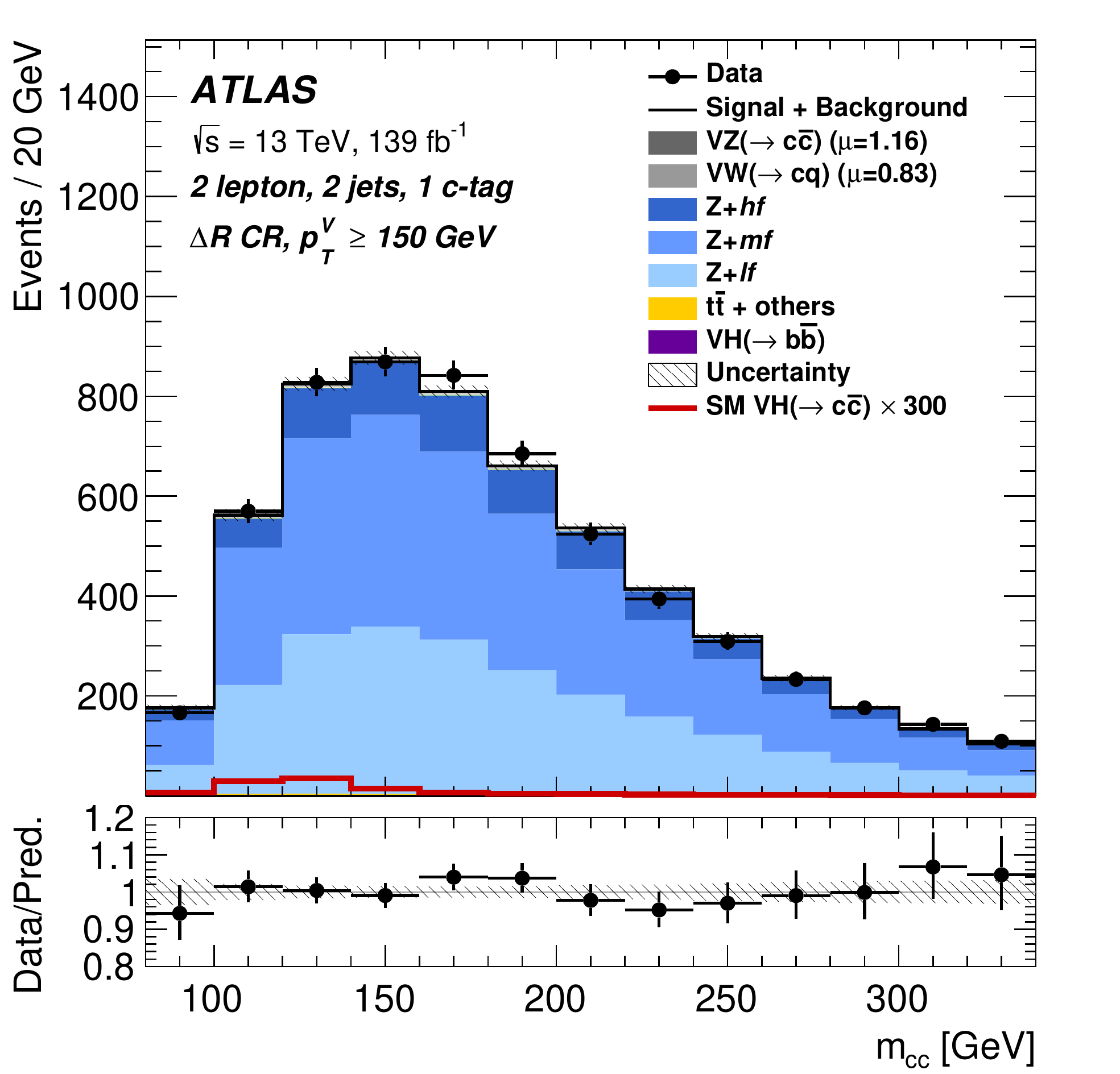}
\includegraphics[width=0.32\textwidth]{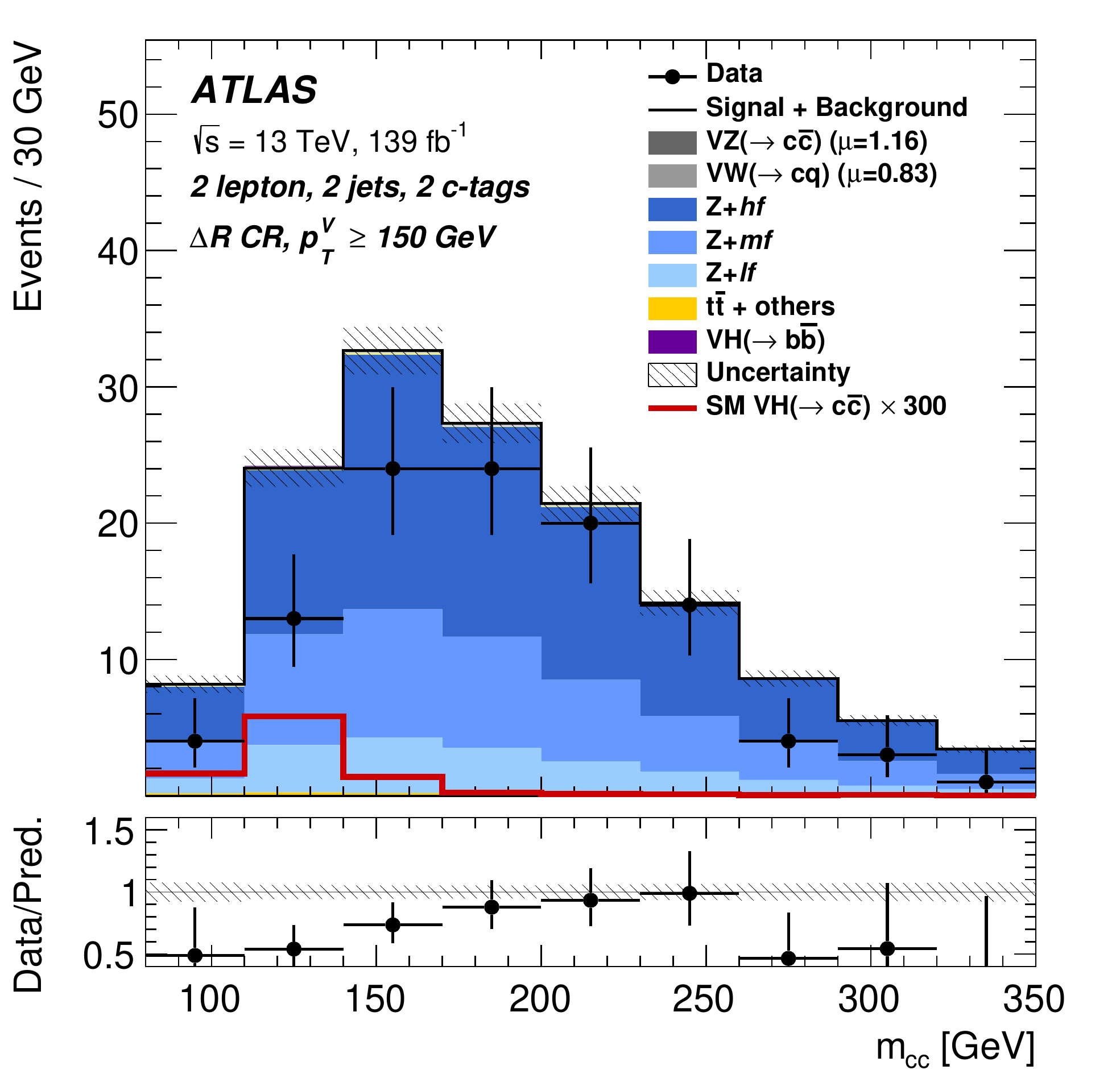}
\includegraphics[width=0.32\textwidth]{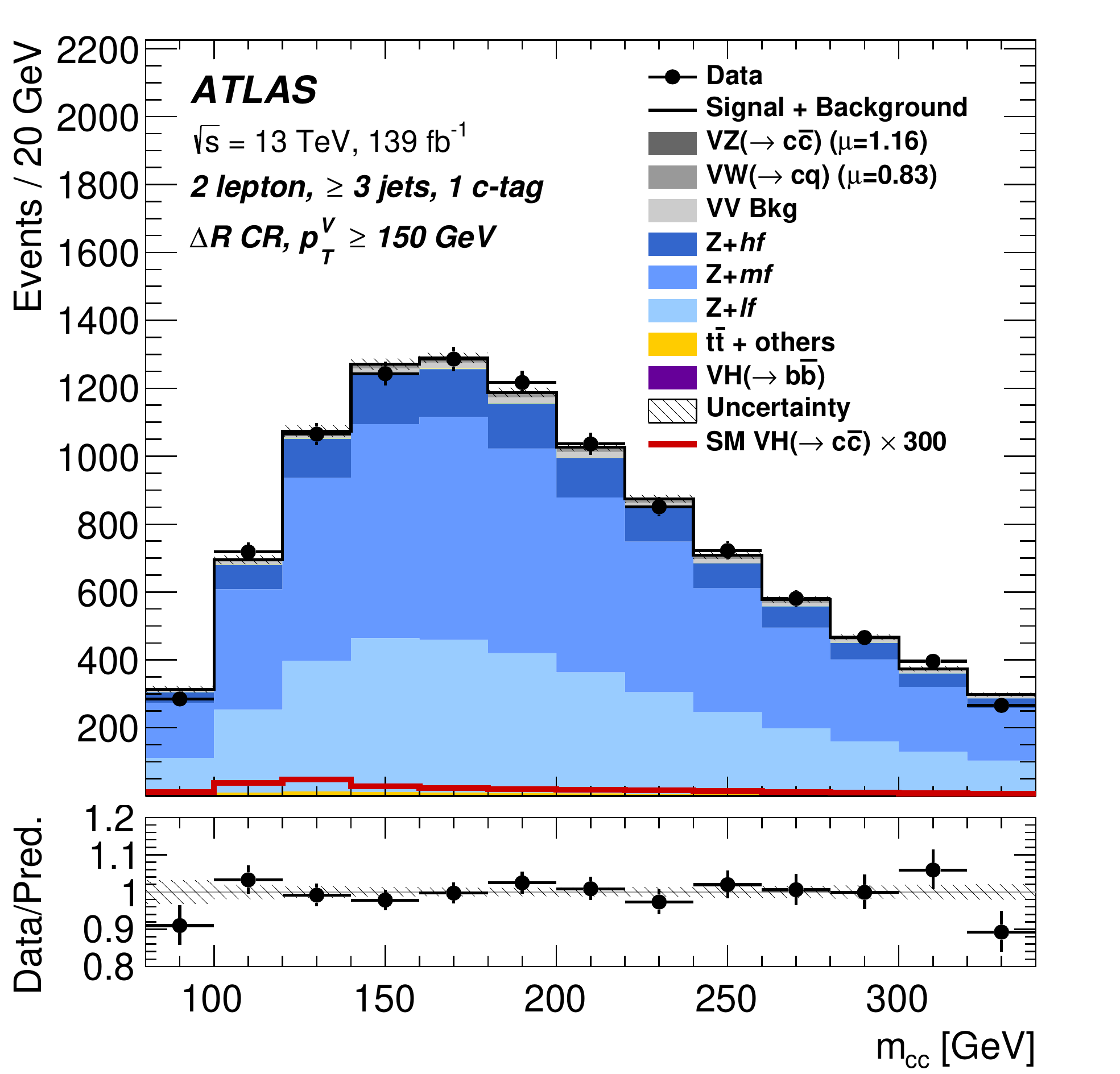}
\includegraphics[width=0.32\textwidth]{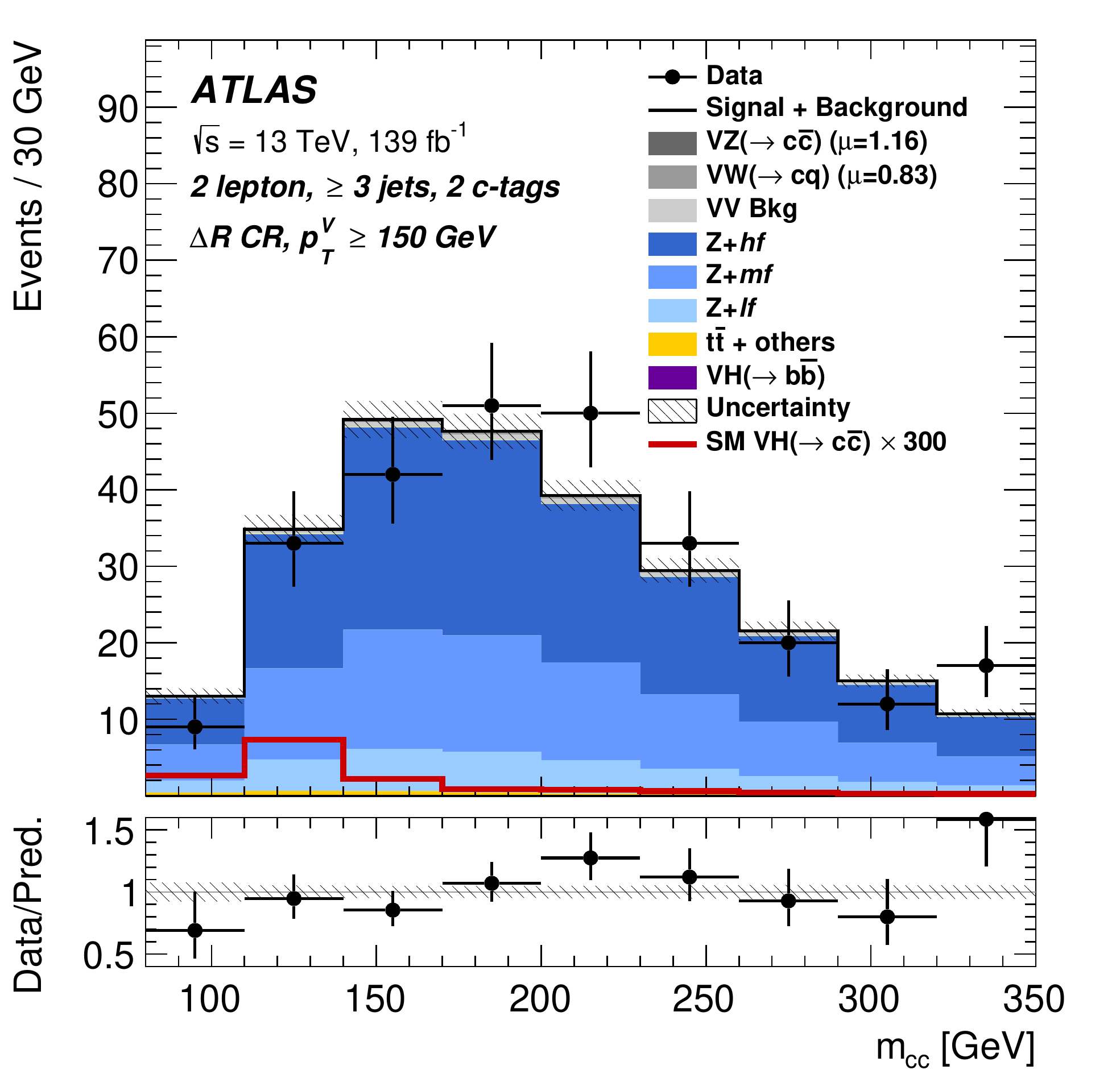}
\caption{Post-fit distributions of the eight 2-lepton high-$\Delta R$ control
regions. The total signal-plus-background prediction is shown by the solid
black line and includes the $H\to c\bar{c}$ signal scaled to the best-fit
value of $\mu_{VH(c\bar{c})} = -9$. The $H\to c\bar{c}$ signal is also shown
as an unfilled histogram scaled to 300 times the SM prediction. The post-fit
uncertainty is shown as the hatched background including correlations between
uncertainties. The ratio of the data to the sum of the post-fit signal plus
background is shown in the lower panel.}
\label{fig:postfit_2lep_dR}
\end{figure*}

\begin{figure*}
\centering
\includegraphics[width=0.47\textwidth]{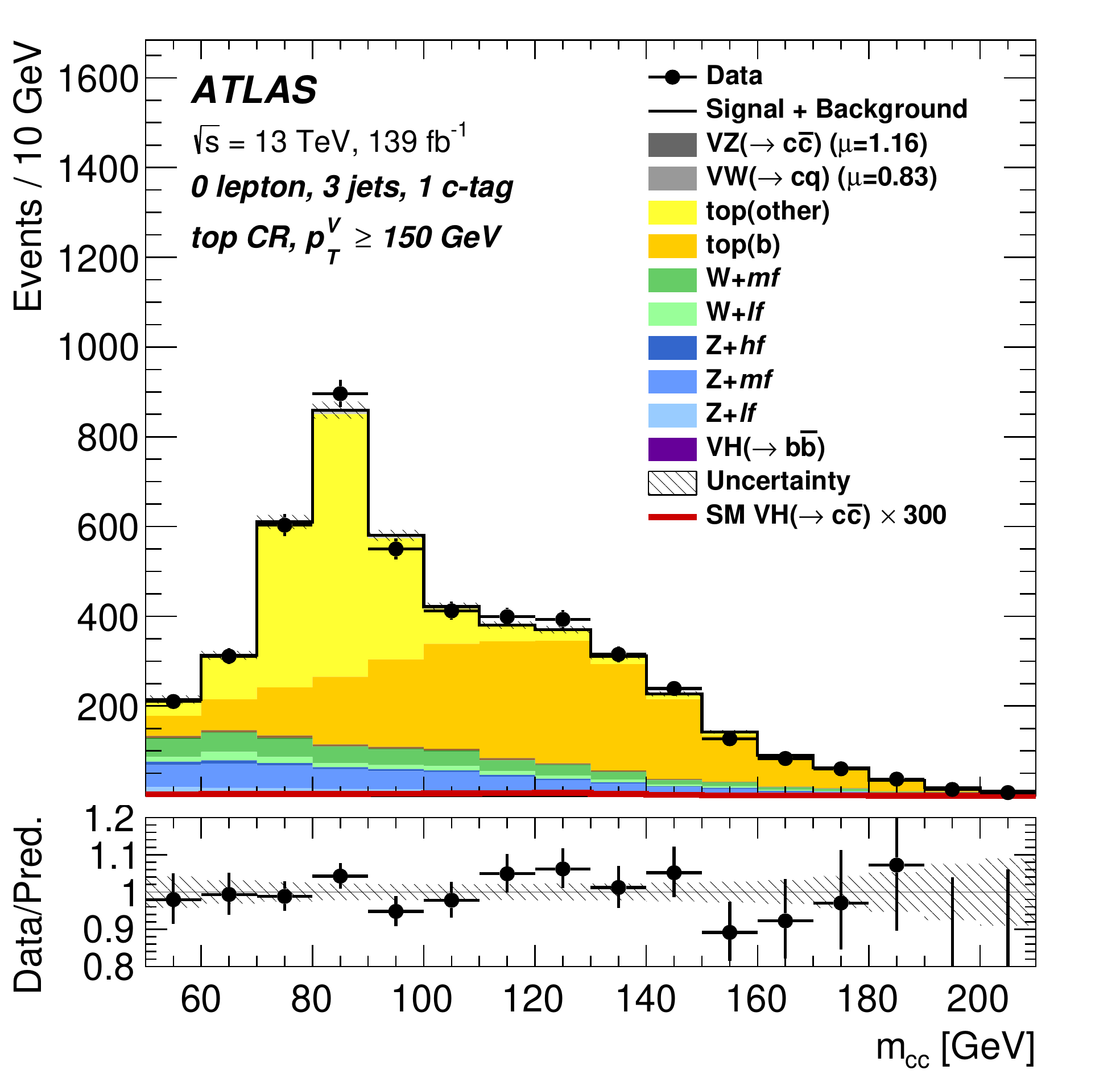}
\includegraphics[width=0.47\textwidth]{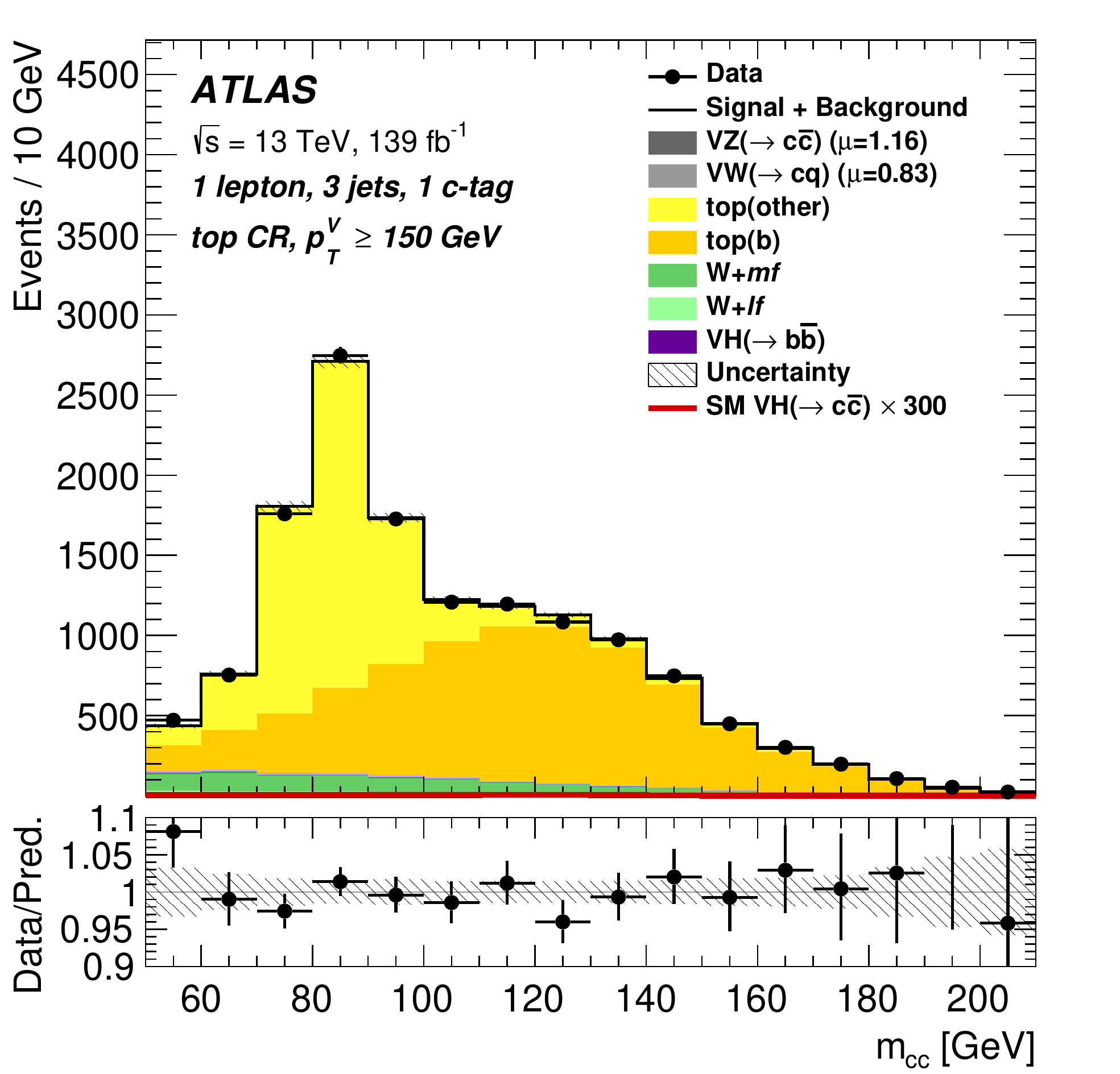} \\
\caption{Post-fit distributions of the 0- and 1-lepton top control regions.
The total signal-plus-background prediction is shown by the solid black line
and includes the $H\to c\bar{c}$ signal scaled to the best-fit value of
$\mu_{VH(c\bar{c})} = -9$. The $H\to c\bar{c}$ signal is also shown as an
unfilled histogram scaled to 300 times the SM prediction. The post-fit
uncertainty is shown as the hatched background including correlations between
uncertainties. The ratio of the data to the sum of the post-fit signal plus
background is shown in the lower panel.}
\label{fig:top_CRs}
\end{figure*}

\begin{figure*}
\centering
\includegraphics[width=\textwidth]{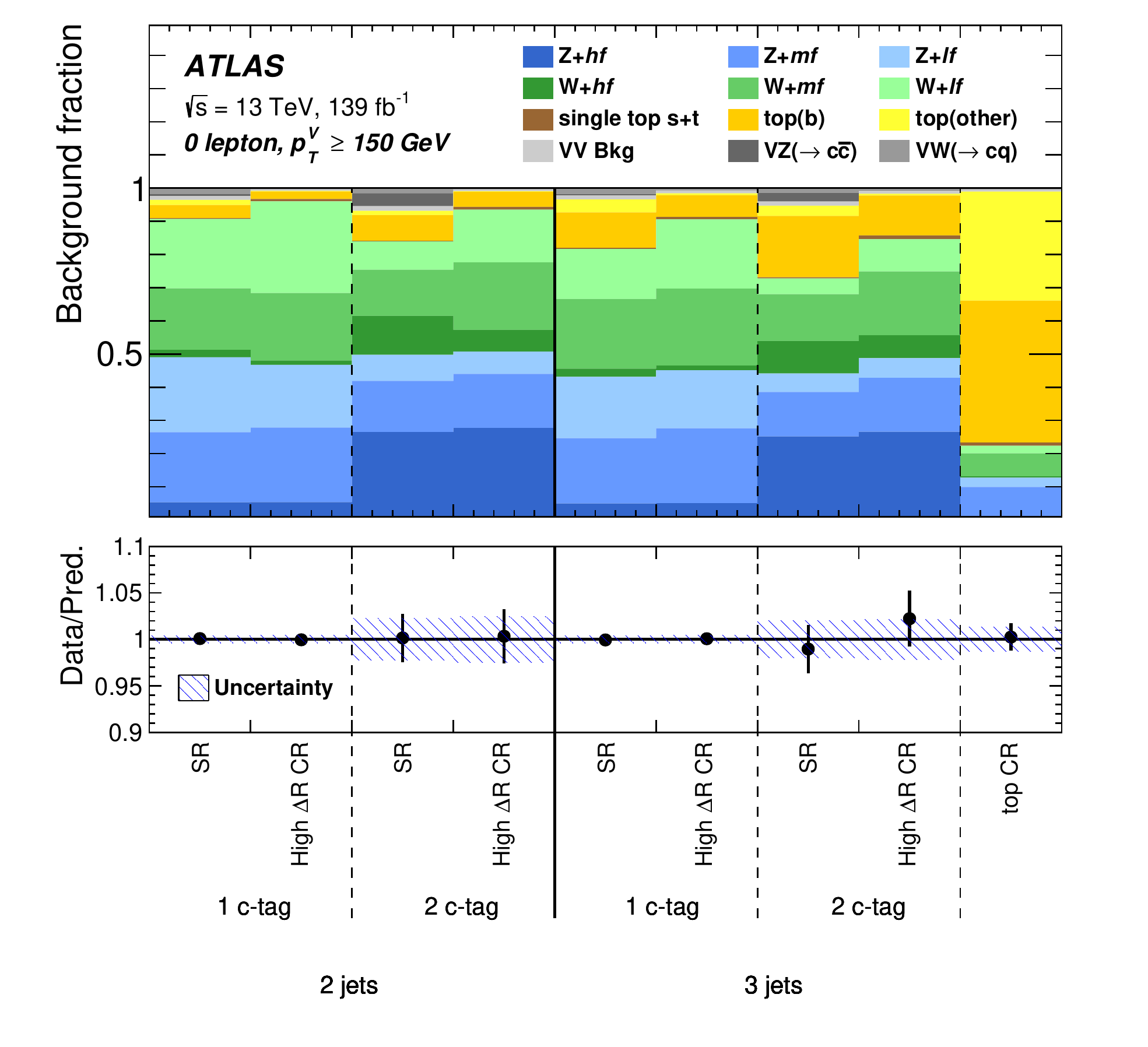}
\caption{The background composition in all 0-lepton signal and control
regions.}
\label{fig:background_comp_0lep}
\end{figure*}
 
\begin{figure*}
\centering
\includegraphics[width=\textwidth]{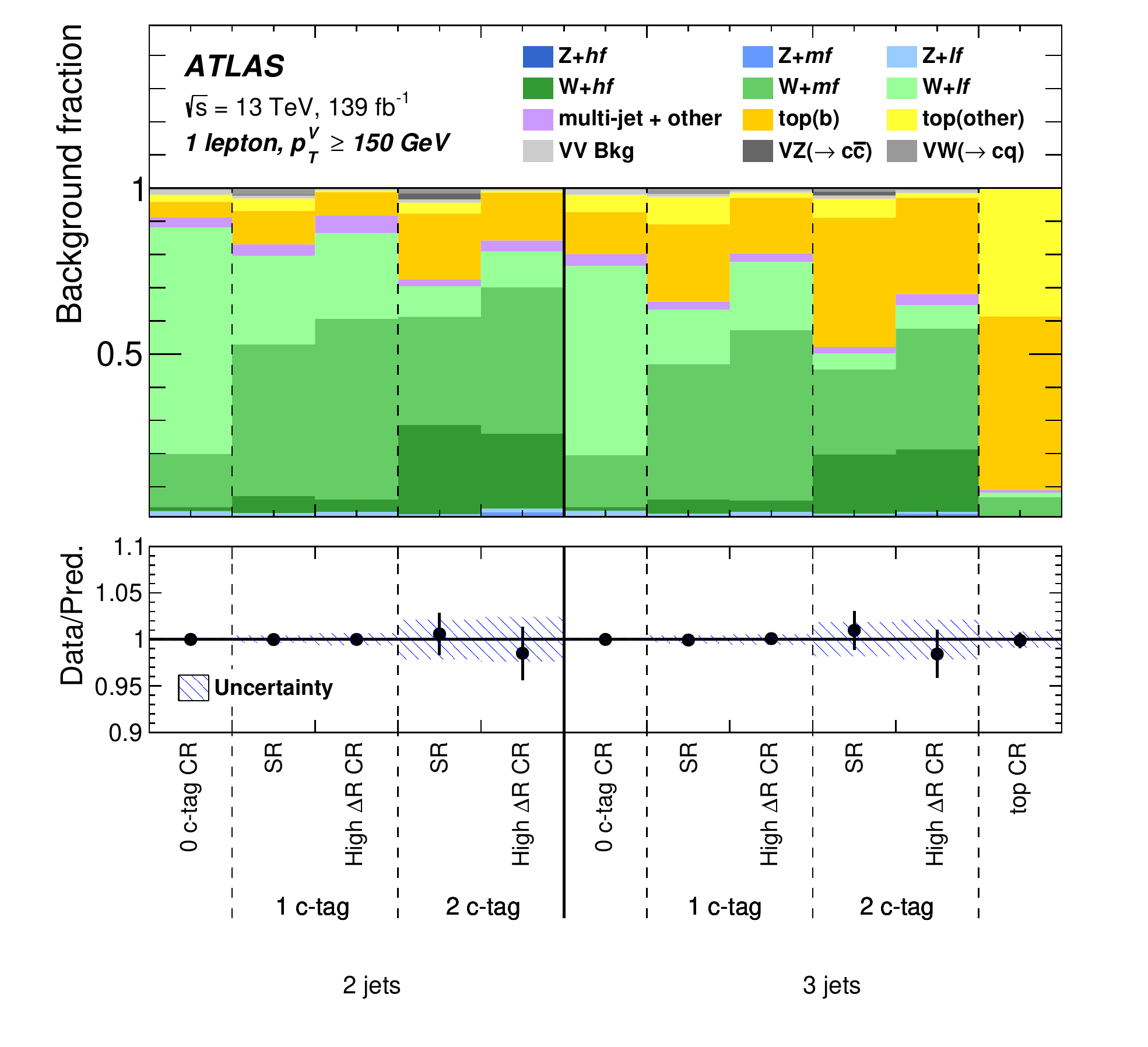}
\caption{The background composition in all 1-lepton signal and control
regions.}
\label{fig:background_comp_1lep}
\end{figure*}
 
\begin{figure*}
\centering
\includegraphics[height=0.46\textheight]{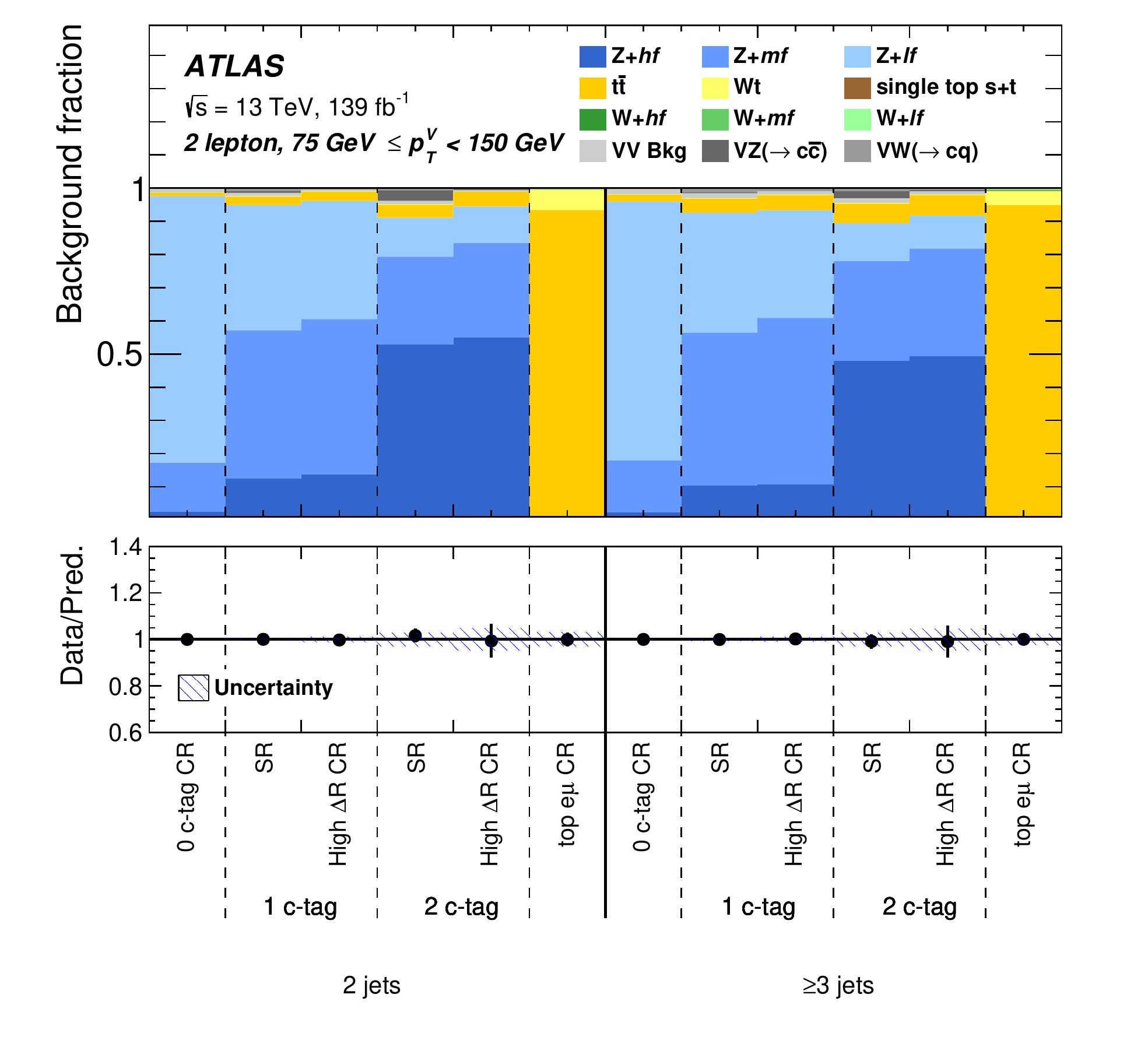}
\includegraphics[height=0.46\textheight]{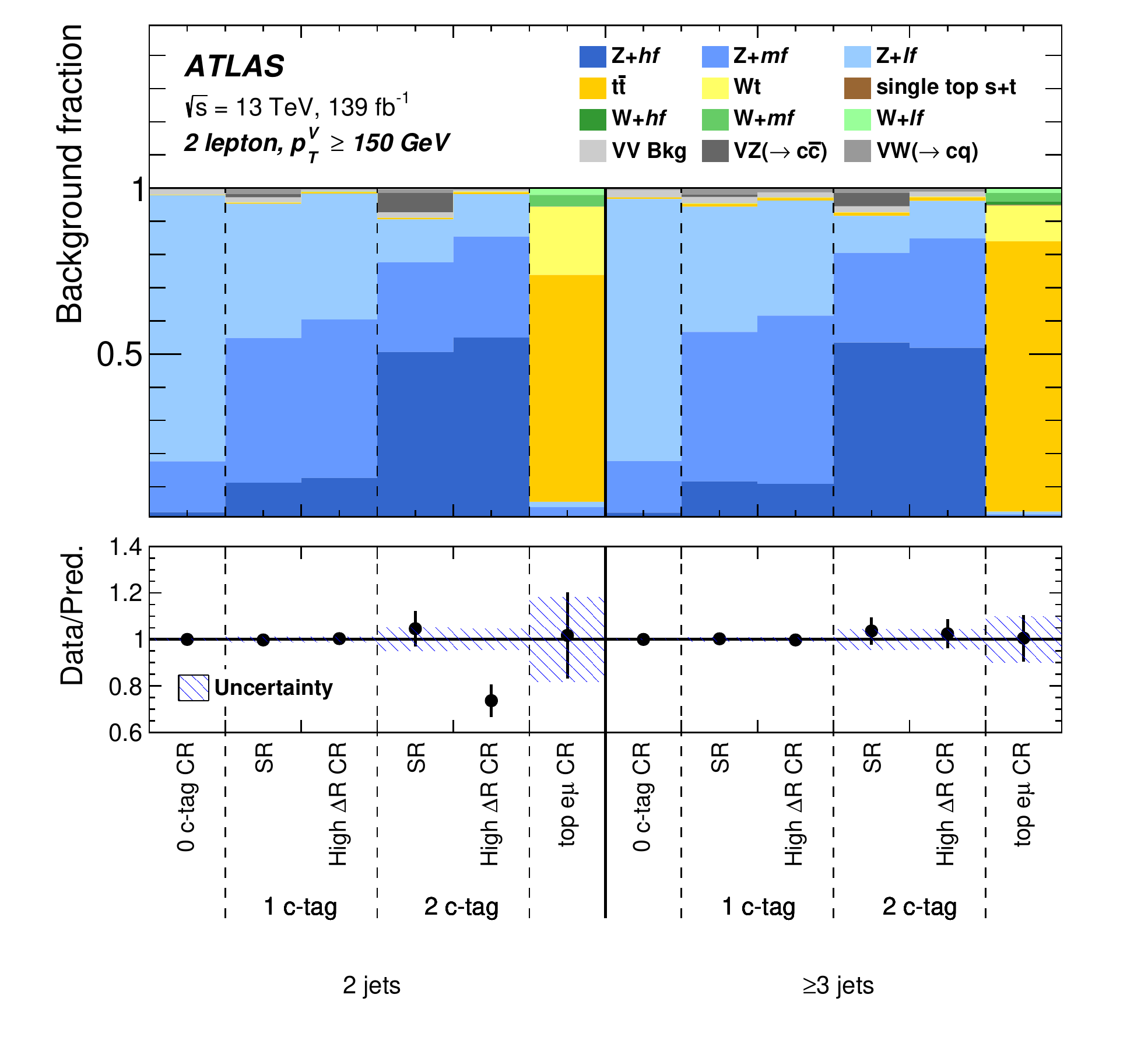}
\caption{The background composition in all 2-lepton signal and control
regions, for events with $75 < \pt^V < 150~\GeV$ (top) and $\pt^V > 150~\GeV$
(bottom).}
\label{fig:background_comp_2lep}
\end{figure*}
\FloatBarrier
 
\printbibliography
\clearpage \input{atlas_authlist}

\end{document}

%% file: atlas_authlist.tex
 
\begin{flushleft}
\hypersetup{urlcolor=black}
{\Large The ATLAS Collaboration}

\bigskip

\AtlasOrcid[0000-0002-6665-4934]{G.~Aad}$^\textrm{\scriptsize 100}$,    
\AtlasOrcid[0000-0002-5888-2734]{B.~Abbott}$^\textrm{\scriptsize 126}$,    
\AtlasOrcid[0000-0002-7248-3203]{D.C.~Abbott}$^\textrm{\scriptsize 101}$,    
\AtlasOrcid[0000-0002-2788-3822]{A.~Abed~Abud}$^\textrm{\scriptsize 36}$,    
\AtlasOrcid[0000-0002-1002-1652]{K.~Abeling}$^\textrm{\scriptsize 53}$,    
\AtlasOrcid[0000-0002-2987-4006]{D.K.~Abhayasinghe}$^\textrm{\scriptsize 93}$,    
\AtlasOrcid[0000-0002-8496-9294]{S.H.~Abidi}$^\textrm{\scriptsize 29}$,    
\AtlasOrcid[0000-0002-9987-2292]{A.~Aboulhorma}$^\textrm{\scriptsize 35e}$,    
\AtlasOrcid[0000-0001-5329-6640]{H.~Abramowicz}$^\textrm{\scriptsize 158}$,    
\AtlasOrcid[0000-0002-1599-2896]{H.~Abreu}$^\textrm{\scriptsize 157}$
% ...
\bigskip
\\

% ...
$^{29}$Physics Department, Brookhaven National Laboratory, Upton NY; United States of America.\\
% ...
$^{35}$$^{(a)}$Facult\'e des Sciences Ain Chock, R\'eseau Universitaire de Physique des Hautes Energies - Universit\'e Hassan II, Casablanca;$^{(b)}$Facult\'{e} des Sciences, Universit\'{e} Ibn-Tofail, K\'{e}nitra;$^{(c)}$Facult\'e des Sciences Semlalia, Universit\'e Cadi Ayyad, LPHEA-Marrakech;$^{(d)}$LPMR, Facult\'e des Sciences, Universit\'e Mohamed Premier, Oujda;$^{(e)}$Facult\'e des sciences, Universit\'e Mohammed V, Rabat;$^{(f)}$Mohammed VI Polytechnic University, Ben Guerir; Morocco.\\
$^{36}$CERN, Geneva; Switzerland.\\
% ...
$^{53}$II. Physikalisches Institut, Georg-August-Universit\"{a}t G\"ottingen, G\"ottingen; Germany.\\
% ...
$^{93}$Department of Physics, Royal Holloway University of London, Egham; United Kingdom.\\
% ...
$^{100}$CPPM, Aix-Marseille Universit\'e, CNRS/IN2P3, Marseille; France.\\
$^{101}$Department of Physics, University of Massachusetts, Amherst MA; United States of America.\\
% ...
$^{126}$Homer L. Dodge Department of Physics and Astronomy, University of Oklahoma, Norman OK; United States of America.\\
% ...
$^{157}$Department of Physics, Technion, Israel Institute of Technology, Haifa; Israel.\\
$^{158}$Raymond and Beverly Sackler School of Physics and Astronomy, Tel Aviv University, Tel Aviv; Israel.\\
% ...
$^{a}$ Also at Borough of Manhattan Community College, City University of New York, New York NY; United States of America.\\
$^{b}$ Also at Bruno Kessler Foundation, Trento; Italy.\\
$^{c}$ Also at Center for High Energy Physics, Peking University; China.\\
$^{d}$ Also at Centro Studi e Ricerche Enrico Fermi; Italy.\\
$^{e}$ Also at CERN, Geneva; Switzerland.\\
$^{f}$ Also at D\'epartement de Physique Nucl\'eaire et Corpusculaire, Universit\'e de Gen\`eve, Gen\`eve; Switzerland.\\
% ...

\end{flushleft}
